\documentclass[lettersize,journal]{IEEEtran}

\usepackage[T1]{fontenc}
\usepackage[utf8]{inputenc}
\usepackage[ngerman,english]{babel}
\usepackage{amsmath, amssymb, ﻿amsfonts}   
\usepackage{cuted} 
\usepackage{graphicx}         
\usepackage{algorithmic}
\usepackage{algorithm}
\usepackage{stfloats}
\usepackage{url}
\usepackage{verbatim}
\usepackage{graphicx}
\usepackage{cite}

\usepackage{upgreek}			  
\usepackage{mathtools}		  
\usepackage{url}					
\usepackage{enumitem}		
\usepackage{accents}			
\usepackage{bbm}
\usepackage{xcolor}
\usepackage{lipsum}
\usepackage{dirtytalk}
\usepackage{xcolor}
\usepackage{float}
\usepackage{accents}
\usepackage[export]{adjustbox}

\ifCLASSOPTIONcompsoc
	\usepackage[caption=false,font=normalsize,labelfont=sf,textfont=sf]{subfig}
\else
	\usepackage[caption=false,font=footnotesize]{subfig}
\fi

\usepackage{pifont} 
\usepackage{caption}

\usepackage{booktabs, tabularx}
\usepackage{array}

\usepackage{amsthm}
\usepackage{enumitem}

\usepackage{tikz}
\usetikzlibrary{decorations.markings,chains,matrix}
\usetikzlibrary{intersections,calc}
\usepackage{tikz-3dplot} 
\usepackage{pgfplots}
\usepackage{pgfplotstable}
\pgfplotsset{compat = newest}
\usepackage{datatool}
\usepackage{textcomp}
\usetikzlibrary{shapes,arrows}
\usetikzlibrary{decorations.pathreplacing,angles,quotes}
\newtheorem{thm}{Theorem}

\newtheorem{defn}{Definition}

\newtheorem{rmk}{Remark}

\DeclareMathOperator{\FDP}{FDP}
\DeclareMathOperator{\FDR}{FDR}
\DeclareMathOperator{\TPP}{TPP}
\DeclareMathOperator{\TPR}{TPR}

\DeclareMathOperator{\NHG}{NHG}
\DeclareMathOperator{\Var}{Var}

\newtheorem{innercustomgeneric}{\customgenericname}
\providecommand{\customgenericname}{}
\newcommand{\newcustomtheorem}[2]{%
  \newenvironment{#1}[1]
  {%
   \renewcommand\customgenericname{#2}%
   \renewcommand\theinnercustomgeneric{##1}%
   \innercustomgeneric
  }
  {\endinnercustomgeneric}
}

\newcustomtheorem{customthm}{Theorem}
\newcustomtheorem{customlemma}{Lemma}
\newcustomtheorem{customcor}{Corollary}

\newcommand{\y}{\boldsymbol{y}}
\newcommand{\x}{\boldsymbol{x}}

\newcommand{\X}{\boldsymbol{X}}

\newcommand{\bbeta}{\boldsymbol{\beta}}
\newcommand{\bepsilon}{\boldsymbol{\epsilon}}
\newcommand{\hatbbeta}{\boldsymbol{\hat{\beta}}}

\newcommand{\A}{\mathcal{A}}

\newcommand{\XK}{\boldsymbol{\protect\accentset{\circ}{X}}}
\newcommand{\xK}{\boldsymbol{\protect\accentset{\circ}{x}}}
\newcommand{\xk}{\protect\accentset{\circ}{x}}
\newcommand{\XWK}{\boldsymbol{\widetilde{X}}}
\newcommand{\XKvar}{\protect\accentset{\circ}{X}}
\newcommand{\GKvar}{\protect\accentset{\circ}{G}}
\newcommand{\QKvar}{\protect\accentset{\circ}{Q}}

\newcommand{\C}{\mathcal{C}}
\newcommand{\Z}{\mathcal{Z}}

\newcommand{\Dum}{\mathcal{D}}

\makeatletter
\let\save@mathaccent\mathaccent
\newcommand*\if@single[3]{%
  \setbox0\hbox{${\mathaccent"0362{#1}}^H$}%
  \setbox2\hbox{${\mathaccent"0362{\kern0pt#1}}^H$}%
  \ifdim\ht0=\ht2 #3\else #2\fi
  }
\newcommand*\rel@kern[1]{\kern#1\dimexpr\macc@kerna}
\newcommand*\widebar[1]{\@ifnextchar^{{\wide@bar{#1}{0}}}{\wide@bar{#1}{1}}}
\newcommand*\wide@bar[2]{\if@single{#1}{\wide@bar@{#1}{#2}{1}}{\wide@bar@{#1}{#2}{2}}}
\newcommand*\wide@bar@[3]{%
  \begingroup
  \def\mathaccent##1##2{%
    \let\mathaccent\save@mathaccent
    \if#32 \let\macc@nucleus\first@char \fi
    \setbox\z@\hbox{$\macc@style{\macc@nucleus}_{}$}%
    \setbox\tw@\hbox{$\macc@style{\macc@nucleus}{}_{}$}%
    \dimen@\wd\tw@
    \advance\dimen@-\wd\z@
    \divide\dimen@ 3
    \@tempdima\wd\tw@
    \advance\@tempdima-\scriptspace
    \divide\@tempdima 10
    \advance\dimen@-\@tempdima
    \ifdim\dimen@>\z@ \dimen@0pt\fi
    \rel@kern{0.6}\kern-\dimen@
    \if#31
      \overline{\rel@kern{-0.6}\kern\dimen@\macc@nucleus\rel@kern{0.4}\kern\dimen@}%
      \advance\dimen@0.4\dimexpr\macc@kerna
      \let\final@kern#2%
      \ifdim\dimen@<\z@ \let\final@kern1\fi
      \if\final@kern1 \kern-\dimen@\fi
    \else
      \overline{\rel@kern{-0.6}\kern\dimen@#1}%
    \fi
  }%
  \macc@depth\@ne
  \let\math@bgroup\@empty \let\math@egroup\macc@set@skewchar
  \mathsurround\z@ \frozen@everymath{\mathgroup\macc@group\relax}%
  \macc@set@skewchar\relax
  \let\mathaccentV\macc@nested@a
  \if#31
    \macc@nested@a\relax111{#1}%
  \else
    \def\gobble@till@marker##1\endmarker{}%
    \futurelet\first@char\gobble@till@marker#1\endmarker
    \ifcat\noexpand\first@char A\else
      \def\first@char{}%
    \fi
    \macc@nested@a\relax111{\first@char}%
  \fi
  \endgroup
}
\makeatother

\DeclareFontFamily{U}{dutchcal}{\skewchar\font=45}
\DeclareFontShape{U}{dutchcal}{m}{n}{<-> s*[1.2] dutchcal-r}{}
\DeclareFontShape{U}{dutchcal}{b}{n}{<-> s*[1.2] dutchcal-b}{}
\DeclareMathAlphabet{\mathdutchcal}{U}{dutchcal}{m}{n}
\SetMathAlphabet{\mathdutchcal}{bold}{U}{dutchcal}{b}{n}
\DeclareMathAlphabet{\mathdutchbcal}{U}{dutchcal}{b}{n}

\newlist{steps}{enumerate}{1}
\setlist[steps, 1]{label = {Step \arabic*:}, ref = {Step \arabic*}}

\newlist{alglist}{enumerate}{1}
\setlist[alglist, 1]{label = {\arabic*.}, ref = {\arabic*}}

\newcommand{\cmark}{\ding{51}}
\newcommand{\xmark}{\ding{55}}

\newcolumntype{L}[1]{>{\raggedright\arraybackslash}p{#1}}
\newcolumntype{C}[1]{>{\centering\arraybackslash}p{#1}}
\newcolumntype{R}[1]{>{\raggedleft\arraybackslash}p{#1}}

\definecolor{dark_green}{RGB}{102,166,30}
\definecolor{applegreen}{rgb}{0.55, 0.71, 0.0}

\definecolor{dark_red}{RGB}{217,95,2}
\definecolor{bittersweet}{rgb}{1.0, 0.44, 0.37}

\definecolor{dark_yellow}{RGB}{230,171,2}
\definecolor{bananayellow}{rgb}{1.0, 0.88, 0.21}

\mathtoolsset{showonlyrefs=true} 
\restylefloat{table}

\usepackage{xr}
\begin{document}

\title{The Terminating-Random Experiments Selector:\\ Fast High-Dimensional Variable Selection with False Discovery Rate Control}

\author{
Jasin Machkour,
Michael Muma, and
Daniel P. Palomar
\thanks{J. Machkour and M. Muma are with the Robust Data Science Group at Technische Universit\"at Darmstadt, Germany (e-mail: jasin.machkour@tu-darmstadt.de; michael.muma@tu-darmstadt.de).}
\thanks{D. P. Palomar is with the Convex Optimization Group, Hong Kong University of Science and Technology, Hong Kong (e-mail: palomar@ust.hk).}
\thanks{The work of the first author has been funded by the Deutsche Forschungsgemeinschaft (DFG, German Research Foundation) under grant number $425884435$. The work of the second author has been funded by the LOEWE initiative (Hesse, Germany) within the emergenCITY center and is supported by the ERC Starting Grant ScReeningData. The work of the third author has been funded by the Hong Kong GRF 16207820 research grant.}
}

\maketitle

\begin{abstract}
We propose the Terminating-Random Experiments (T-Rex) selector, a fast variable selection method for high-dimensional data. The T-Rex selector controls a user-defined target false discovery rate (FDR) while maximizing the number of selected variables. This is achieved by fusing the solutions of multiple early terminated random experiments. The experiments are conducted on a combination of the original predictors and multiple sets of randomly generated dummy predictors. A finite sample proof based on martingale theory for the FDR control property is provided. Numerical simulations confirm that the FDR is controlled at the target level while allowing for high power. We prove that the dummies can be sampled from any univariate probability distribution with finite expectation and variance. The computational complexity of the proposed method is linear in the number of variables. The T-Rex selector outperforms state-of-the-art methods for FDR control in numerical experiments and on a simulated genome-wide association study (GWAS), while its sequential computation time is more than two orders of magnitude lower than that of the strongest benchmark methods. The open source R package TRexSelector containing the implementation of the T-Rex selector is available on CRAN.
\end{abstract}


\section{Introduction and Motivation}
\label{sec: Introduction and Motivation}
Determining the set of active signals or variables is crucial, e.g., in detection~\cite{chung2007detection, chen2020false, chen2018sparse}, antenna array processing~\cite{tan2014direction}, distributed learning~\cite{di2012sparse}, portfolio optimization~\cite{benidis2017sparse}, and robust estimation~\cite{zoubir2012robust, zoubir2018robust, machkour2017outlier, machkour2020robust, yang2019weakly}. In this work, we focus on genome-wide association studies (GWAS)~\cite{gwasCatalog}, where only a few common genetic variations called single nucleotide polymorphisms (SNPs) among potentially millions of candidates are associated with a phenotype (e.g., disease) of interest~\cite{gwasCatalog}. To enable reproducible discoveries, it is essential that (i) the proportion of falsely selected variables among all selected variables is low while (ii) the proportion of correctly selected variables among all true active variables is high. The expected values of these quantities are referred to as the false discovery rate (FDR) and the true positive rate (TPR), respectively. Without FDR control, expensive functional genomics studies and biological laboratory experiments are wasted on researching false positives~\cite{chanock2007replicating, visscher201710, huffman2018examining,gallagher2018post}.
\begin{figure}[t]
  \centering
  		\scalebox{1}{
  			\includegraphics[width=0.98\linewidth]{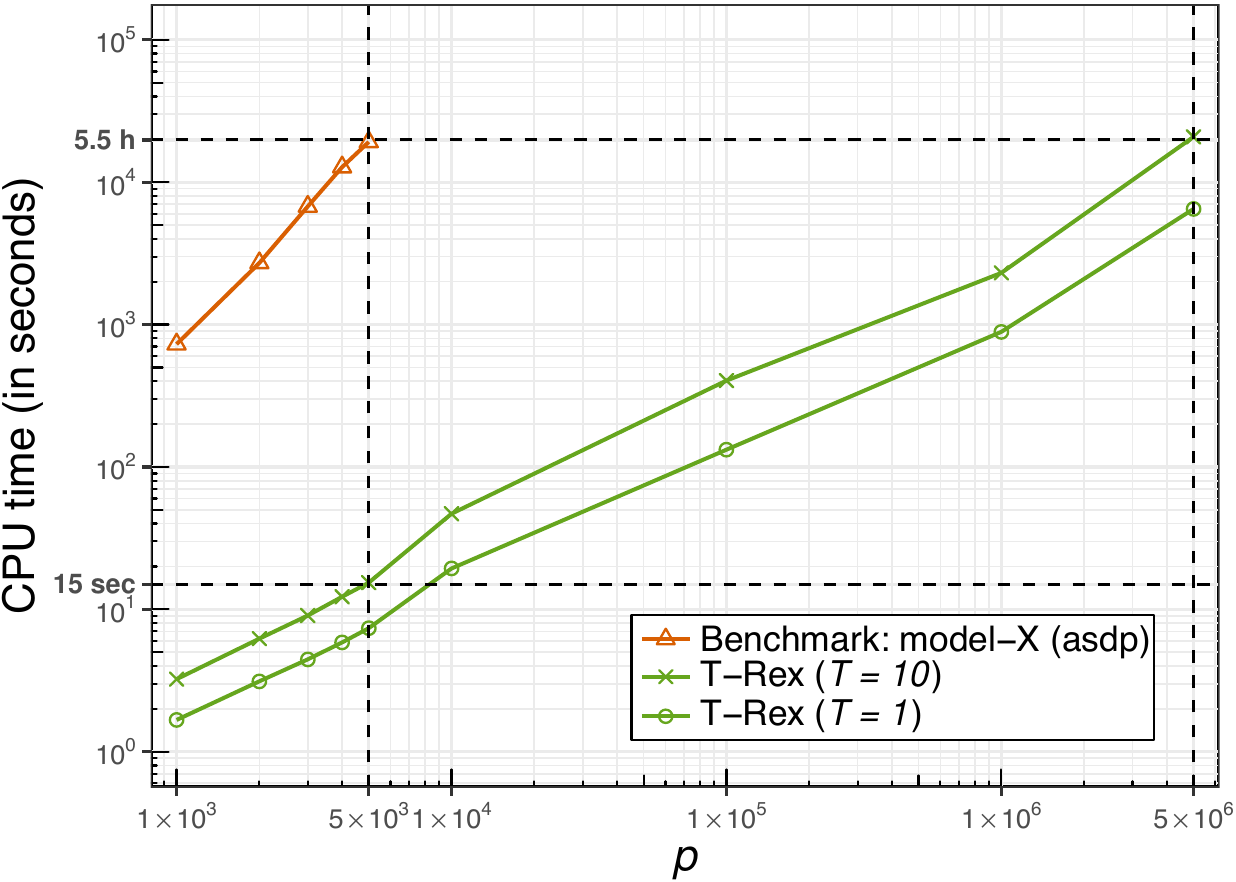}
  		}
   		\label{fig: comp_time_sweep_p_five_million}
 \caption{The sequential computation time of the \textit{T-Rex} selector is multiple orders of magnitude lower than that of the \textit{model-X} knockoff method. Note that, e.g., for $p = 5{,}000$ variables the absolute sequential computation time of the \textit{T-Rex} selector for $T = 10$ included dummies is only $15$ seconds as compared to more than $5.5$ hours for the \textit{model-X} knockoff method. Moreover, the sequential computation time of the \textit{T-Rex} selector for $5{,}000{,}000$ variables is comparable to that of the \textit{model-X} knockoff method for only $5{,}000$ variables. Note that both axes are scaled logarithmically. Setup: $n = 300$ (observations), $p_{1} = 10$ (true active variables), $L = p$ (generated dummies), $K = 20$ (random experiments), $\text{SNR} = 1$, $MC = 955$ (Monte Carlo replications) for $p \leq 5{,}000$ and $MC = 100$ for $p > 5{,}000$.}
  \label{fig: computation time sweep p five million}
\end{figure}

Establishing FDR control in high-dimensional settings is challenging and, unfortunately, established FDR-controlling methods for low-dimensional data, e.g., ~\cite{benjamini1995controlling,benjamini2001control,barber2015controlling}, do not apply to high-dimensional settings. In recent years, the \textit{model-X} knockoff method~\cite{candes2018panning} and derandomized versions thereof~\cite{ren2021derandomizing,ren2024derandomised} have been proposed. However, they are computationally demanding. In fact, creating knockoff predictors that mimic the covariance structure of the original predictors renders them infeasible for settings beyond a few thousand variables (see Figure~\ref{fig: computation time sweep p five million}). Moreover, the original derandomized knockoffs approach controls the conservative per family error rate (\textit{PFER}) and the $k$-family-wise error rate (\textit{$k$-FWER}) but does not consider the less conservative FDR metric~\cite{ren2021derandomizing}. Only the derandomized approach based on e-values controls the FDR~\cite{ren2024derandomised}. Nevertheless, the need for running the \textit{model-X} knockoff method multiple times renders both derandomized knockoffs approaches practically infeasible for large-scale high-dimensional settings.

Alternative FDR-controlling approaches that rely on conditional randomization test (\textit{CRT}) $p$-values~\cite{candes2018panning} are computationally significantly more demanding than the \textit{model-X} knockoff methods (see~\cite{candes2018panning} for a discussion), which renders them infeasible in even relatively small settings (i.e., $p \approx 1{,}000$ candidate variables).

Related lines of research on error-controlled high-dimensional variable selection are centered around stability selection methods~\cite{meinshausen2010stability, shah2013variable}, data-splitting methods~\cite{cox1975note, wasserman2009high, meinshausen2009p, barber2019knockoff}, and post-selection inference~\cite{lockhart2014significance, fithian2014optimal, lee2016exact, tibshirani2016exact}.

In this work, we propose the Terminating-Random Experiments (\textit{T-Rex}) selector, a scalable framework (see Section~\ref{subsec: The T-Rex Selector: Methodology}) that turns forward variable selection methods into FDR-controlling methods. The \textit{T-Rex} selector fuses the solutions of $K$ early terminated random experiments, in which original and dummy variables compete to be selected in a forward variable selection process. It utilizes dummies in a fundamentally different manner than existing methods (e.g.,~\cite{miller1984selection, miller2002subset, wu2007controlling}), which are not designed for FDR control. The \textit{T-Rex} calibration algorithm automatically determines its parameters, i.e., (i) the number of generated dummies $L$, (ii) the number of included dummies before terminating the random experiments $T$, and (iii) the voting level in the fusion process, such that the FDR is controlled at the target level.

Our main theoretical results are summarized as follows:
\begin{enumerate}[label=\arabic*., ref=\arabic*]
\item Using martingale theory\cite{williams1991probability}, we provide a finite sample FDR control proof (Theorem~\ref{theorem: FDR control}) that applies to low- ($p \leq n$) and high-dimensional ($p > n$) settings.
\label{main result 1}
\item We prove that, for the \textit{T-Rex} selector, the dummies can be sampled from any univariate distribution with finite mean and variance (Theorem~\ref{theorem: Dummy generation}). This is a fundamentally new result, and it does not hold for, e.g., knockoff methods~\cite{barber2015controlling, candes2018panning} that require mimicking the covariance structure of the predictors, which is computationally expensive (see Figure~\ref{fig: Ingredient 1 - dummy generation} in the supplementary materials.
\label{main result 2}
\item We also prove that the proposed calibration algorithm is optimal in the sense that it maximizes the number of selected variables while controlling the FDR at the target level (Theorem~\ref{theorem: T-Rex algorithm}).
\label{main result 3}
\end{enumerate}

The major advantages compared to existing methods are:
\begin{enumerate}[label=\arabic*., ref=\arabic*]
\item The computation time of the \textit{T-Rex} selector is multiple orders of magnitude lower compared to that of the current benchmark method (see Figure~\ref{fig: computation time sweep p five million}). Its complexity stems from the computation of $K$ terminated random experiments with expected complexity $\mathcal{O}(np)$ (see Appendix~\ref{sec: Computational Complexity} in the supplementary materials).
\label{advantage 1}
\item As inputs, the \textit{T-Rex} selector requires only the data and the target FDR level. The tuning of the sparsity parameter for \textit{Lasso}-type methods~\cite{tibshirani1996regression,efron2004least,zou2005regularization,zou2006adaptive} is no longer required when incorporating them into the \textit{T-Rex} selector framework.
\label{advantage 2}
\end{enumerate}
In summary the \textit{T-Rex} selector is, to the best of our knowledge, the first multivariate high-dimensional FDR-controlling method that scales to millions of variables in a reasonable amount of computation time (see Figure~\ref{fig: computation time sweep p five million}), which makes it a suitable method for large-scale GWAS, i.e., our major use-case. The open source R software packages \textit{TRexSelector}~\cite{machkour2022TRexSelector} and \textit{tlars}~\cite{machkour2022tlars} contain the implementation of the proposed \textit{T-Rex} selector.

Notation: Column vectors and matrices are denoted by boldface lowercase and uppercase letters, respectively. Scalars are denoted by non-boldface lowercase or uppercase letters. With the exceptions of $\mathcal{N}$ and $\varnothing$, which stand for the normal distribution and the empty set, respectively, sets are denoted by calligraphic uppercase letters, e.g., $\mathcal{A}$ with $| \mathcal{A} |$ denoting the associated cardinality. The symbols $\mathbb{E}$ and $\Var$ denote the expectation and the variance operator, respectively.

Organization: The remainder of this paper is organized as follows: Section~\ref{sec: The T-Rex Selector} introduces the methodology of the proposed \textit{T-Rex} selector. Section~\ref{sec: Main Results} presents the main theoretical results regarding the properties of the proposed method and its algorithmic details. Section~\ref{sec: Numerical Simulations} discusses the results of numerical simulations while Section~\ref{sec: Simulated Genome-Wide Association Study} evaluates the performances of the proposed \textit{T-Rex} selector and the benchmark methods on a simulated genome-wide association study (GWAS). Section~\ref{sec: Conclusion} concludes the paper. Technical proofs, numerical verifications, additional simulations, and other appendices are deferred to the supplementary materials.

\section{The T-Rex Selector}
\label{sec: The T-Rex Selector}
This section introduces the proposed \textit{T-Rex} selector. First, a mathematical formulation of the FDR and TPR is given and some forward variable selection methods that are used within the \textit{T-Rex} selector are briefly revisited. Then, the underlying methodology is described and the optimization problem of calibrating the \textit{T-Rex} selector to perform FDR control at the target level is formulated.

\subsection{FDR and TPR}
\label{sec: FDR and TPR}
The FDR and TPR are expressed mathematically as follows: Given the index set of the active variables $\mathcal{A} \subseteq \lbrace 1, \ldots, p \rbrace$, where $p$ is the number of candidate variables, and the index set of the selected active variables $\widehat{\mathcal{A}} \subseteq \lbrace 1, \ldots, p \rbrace$, the FDR and the TPR are defined by
\begin{equation}
\FDR \coloneqq \mathbb{E} \bigg [ \dfrac{| \widehat{\mathcal{A}} \backslash \mathcal{A}|}{1 \lor| \widehat{\mathcal{A}} |} \bigg ] \quad \text{and} \quad \TPR \coloneqq \mathbb{E} \bigg [ \dfrac{| \mathcal{A}  \cap \widehat{\mathcal{A}} |}{1 \lor | \mathcal{A} |} \bigg ],
\label{eq: FDR and TPR}
\end{equation}
respectively, where $|\cdot|$ denotes the cardinality operator and the symbol $\lor$ stands for the maximum operator, i.e., $a \lor b = \max \lbrace a, b \rbrace$, $a,b \in \mathbb{R}$.\footnote{Throughout this paper, the original definition of the FDR in~\cite{benjamini1995controlling} is used. Other definitions of the FDR, such as the positive FDR~\cite{storey2003positive}, exist. The interested reader is referred to both papers for discussions on different potential definitions of the FDR.} Note that by definition the FDR and TPR are zero when $| \widehat{\mathcal{A}} | = 0$ and $ | \mathcal{A} | = 0$, respectively. While the FDR and the TPR of an oracle variable selection procedure are $0$\% and $100$\%, respectively, in practice, a tradeoff must be accomplished.

\subsection{High-Dimensional Variable Selection Methods}
\label{subsec: High-Dimensional Variable Selection Methods}
The \textit{T-Rex} selector framework is versatile in the sense that it can incorporate many different forward selection algorithms. In this paper, we will focus on \textit{Lasso}-type methods~\cite{tibshirani1996regression,zou2006adaptive,zou2005regularization,tibshirani2005sparsity} and especially the \textit{LARS} algorithm~\cite{efron2004least}. Although, in general, the FDR control proof of the T-Rex selector (see Section~\ref{subsec: FDR Control}) does not assume a linear relationship between the explanatory variables and the response variable, we will introduce the linear regression model because it is required by the high-dimensional forward variable selection methods that are considered in this paper.

The linear regression model is defined by
\begin{equation}
\y = \X\bbeta + \bepsilon,
\label{eq: linear model}
\end{equation}
where $\X = [\x_{1} \,\, \x_{2} \,\, \cdots \,\, \x_{p}]$ with $\x_{j}\in \mathbb{R}^{n}$, $j = 1, \ldots, p$, is the fixed predictor matrix containing $p$ predictors and $n$ observations, $\y\in\mathbb{R}^{n}$ is the response vector, $\bbeta\in\mathbb{R}^{p}$ is the parameter vector, and $\bepsilon\sim\mathcal{N}(\boldsymbol{0},\sigma^{2}\boldsymbol{I})$, with $\boldsymbol{I}$ being the identity matrix, is an additive Gaussian noise vector with standard deviation~$\sigma$. Variables whose associated coefficients in $\bbeta$ are non-zero (zero) are called actives or active variables (nulls or null variables).

To obtain a sparse estimate $\hatbbeta$ of $\bbeta$, sparsity-inducing methods, such as the \textit{Lasso}~\cite{tibshirani1996regression} and related methods~\cite{efron2004least,zou2006adaptive,zou2005regularization,tibshirani2005sparsity}, can be used. The \textit{Lasso} solution is defined by  
\begin{equation}
\hatbbeta(\lambda) = \underset{\bbeta}{\arg\min} \, \dfrac{1}{2}\| \y - \X\bbeta \|_{2}^{2} + \lambda \| \bbeta \|_{1},
\label{eq: Lasso}
\end{equation}
where $\lambda > 0$ is a tuning parameter that controls the sparsity of $\hatbbeta$. Throughout this paper, we will use the closely related \textit{LARS} algorithm~\cite{efron2004least} as a forward variable selection method to conduct the random experiments of the \textit{T-Rex} selector. The solution path of the \textit{Lasso} over $\lambda$ is efficiently computed by applying a slightly modified \textit{LARS} algorithm.\footnote{An alternative approach to obtain the solution path of the \textit{Lasso} is to apply the pathwise coordinate descent algorithm~\cite{friedman2007pathwise}. However, this approach is not a forward variable selection method and, therefore, not applicable within the \textit{T-Rex} selector framework.} That is, instead of adding one variable at a time based on the highest correlation with the current residual, the \textit{Lasso} modification requires the removal of previously added variables when the associated coefficients change their sign~\cite{efron2004least}. However, removed variables can enter the solution path again in later steps. Since the solution paths are terminated early by the \textit{T-Rex} selector, there are only very few or even no zero crossings at all along the early terminated solution paths and, thus, in most cases, the \textit{Lasso} in~\eqref{eq: Lasso} and the \textit{LARS} algorithm produce very similar or exactly the same solution paths when they are incorporated into the \textit{T-Rex} selector.
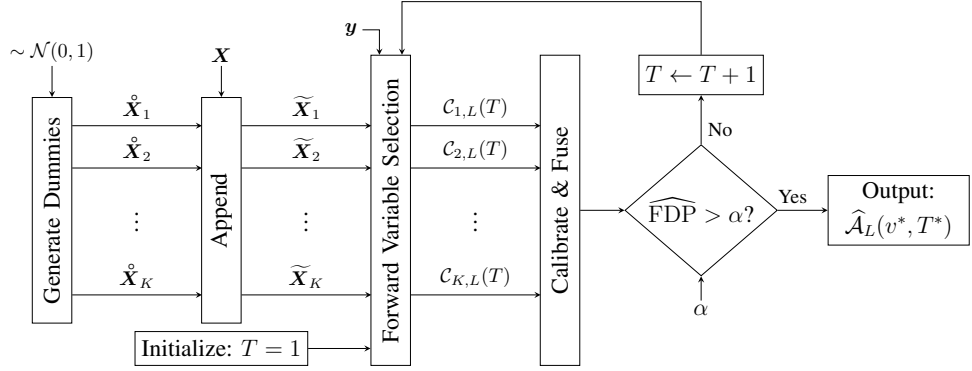
\begin{figure*}[ht]
\begin{center}
\scalebox{0.75}{
\begin{tikzpicture}[>=stealth]

  \coordinate (orig)   at (0,0);
  \coordinate (sample)   at (4,0.5);
  \coordinate (merge)   at (7,0.5);
  \coordinate (varSelect)   at (10,0.5);
  \coordinate (tFDR)   at (15.5,-1.1);
  \coordinate (fuse)   at (13,0.5);
  \coordinate (decision)   at (15.5,0.5);
  \coordinate (increment)   at (15.5,2.9);
  \coordinate (init)   at (7,-1.95);
  \coordinate (output)   at (19,0.5);
  
  \coordinate (between_scale_rank)   at (0.5,0.31);
  \coordinate (X_prime_to_tFDR_point)   at (0.5,6);
  \coordinate (X_prime_to_tFDR_point_point)   at (9,6);
  \coordinate (X_prime_to_merge_point)   at (0.5,-2.5);
  \coordinate (center_to_tFDR_point)   at (9.8,3.7);
  \coordinate (tFDR_to_fuse_point)   at (14.00,5.2);
  \coordinate (tFDR_to_sample_point)   at (4,5);
  
  \coordinate (decision_to_selection_1)   at (15.5,4.2);
  \coordinate (decision_to_selection_2)   at (10.2,4.2);

  \coordinate (Arrow_N_GenDummy)   at (4,3.06);
  \coordinate (Arrow_X_indVar)   at (7,3.06);
  \coordinate (Arrow_y_center)   at (9.5,3.7);
  \coordinate (Arrow_targetFDR_tFDR)   at (10,5.9);
  
   \coordinate (inference_Arrow)   at (15,-1.06);
   \coordinate (fuse_Arrow)   at (16.1,2.06);
  
  \coordinate (vdots1)   at (5.5,0.4);
  \coordinate (vdots2)   at (8.5,0.4);
  \coordinate (vdots3)   at (11.5,0.4);
  
  \coordinate (fuse_node)   at (15.00,0.5);
  
  \tikzstyle{decision} = [diamond, draw, 
										minimum width=2cm, minimum height=0.5cm, node distance=3cm, inner sep=0pt]

  \node[draw, minimum width=.7cm, minimum height=4cm, anchor=center , align=center] (C) at (sample) {\rotatebox{90}{\large Generate Dummies}};
  \node[draw, minimum width=.7cm, minimum height=4cm, anchor=center, align=center] (D) at (merge) {\rotatebox{90}{\large Append}};   
  \node[draw, minimum width=.7cm, minimum height=5.5cm, anchor=center, align=center] (E) at (varSelect) {\rotatebox{90}{\large Forward Variable Selection}};
  \node[draw, minimum width=.7cm, minimum height=5.5cm, anchor=center, align=center] (H) at (fuse) {\rotatebox{90}{\large Calibrate \& Fuse}};
  \node[decision, minimum width=2.7cm, minimum height=0.4cm, anchor=center, align=center] (M) at (decision) {\large $\widehat{\FDP} > \alpha$?};
  \node[draw, minimum width=2.5cm, minimum height=.7cm, anchor=center, align=center] (N) at (output) {\large Output: \\[0.3em] \large $\widehat{\mathcal{A}}_{L}(v^{*}, T^{*})$};
  \node[draw, minimum width=1.5cm, minimum height=.7cm, anchor=center, align=center] (O) at (increment) {\large $T \leftarrow T + 1$};
  \node[draw, minimum width=1.5cm, minimum height=.6cm, anchor=center, align=center] (P) at (init) {\large Initialize: $T = 1$};
  \node (J) at (vdots1) {\large $\vdots$};
  \node (K) at (vdots2) {\large $\vdots$};
  \node (L) at (vdots3) {\large $\vdots$};
  
  \draw[->] (Arrow_N_GenDummy) -- node[above, pos = 0.1]{$\sim\mathcal{N}(0, 1)$} ($(C.90)$); 
  \draw[->] (Arrow_X_indVar) -- node[above, pos = 0.1]{$\X$} ($(D.90)$); 
     
  \draw[->] ($(C.0) + (0,1.5)$) -- node[above]{$\XK_{1}$} ($(D.0) + (-0.7,1.5)$);
  \draw[->] ($(C.0) + (0,0.75)$) -- node[above]{$\XK_{2}$} ($(D.0) + (-0.7,0.75)$);
  \draw[->] ($(C.0) + (0,-1.5)$) -- node[above]{$\XK_{K}$} ($(D.0) + (-0.7,-1.5)$);
     
  \draw[->] ($(D.0) + (0,1.5)$) -- node[above]{$\XWK_{1}$} ($(E.0) + (-0.7,1.5)$);
  \draw[->] ($(D.0) + (0,0.75)$) -- node[above]{$\XWK_{2}$} ($(E.0) + (-0.7,0.75)$);
  \draw[->] ($(D.0) + (0,-1.5)$) -- node[above]{$\XWK_{K}$} ($(E.0) + (-0.7,-1.5)$);
     
  \draw[->] ($(E.0) + (0,1.5)$) -- node[above]{$\C_{1, L}(T)$} ($(H.0) + (-0.7,1.5)$);
  \draw[->] ($(E.0) + (0,0.75)$) -- node[above]{$\C_{2, L}(T)$} ($(H.0) + (-0.7,0.75)$);
  \draw[->] ($(E.0) + (0,-1.5)$) -- node[above]{$\C_{K, L}(T)$} ($(H.0) +  (-0.7,-1.5)$);
     
  \path[draw,->] ($(Arrow_y_center.0)$) -- node[left,pos=0.1] {$\y$} (center_to_tFDR_point) --  ($(E.90) + (-0.2,0)$);
 
  \coordinate (between_varSelect_fuse1)   at ($(E.0) + (2.75,1.5)$);
  \coordinate (between_varSelect_fuse2)   at ($(E.0) + (2.75,0.75)$);
  \coordinate (between_varSelect_fuse3)   at ($(E.0) + (2.75,-1.5)$);
 
  \draw[->] (tFDR) -- node[below, pos=0.1]{\large $\alpha$}(M);
  \draw[->] (H) -- (M);
  \draw[->] ($(M.0)$) -- node[above, pos=0.3]{Yes}(N);
  \draw[->] ($(M.90)$) -- node[right, pos=0.3]{No}(O);
  \path[draw,->] ($(O.90)$) -- (decision_to_selection_1) -- (decision_to_selection_2) --  ($(E.90) + (0.2,0)$);
  \draw[->] ($(P.0)$) -- ($(E.180) + (0,-2.45)$);
  
\end{tikzpicture}}
\end{center}
\caption{Simplified overview of the \textit{T-Rex} selector framework: For each random experiment $k \in \lbrace 1, \ldots, K \rbrace$, the \textit{T-Rex} selector generates a dummy matrix $\XK_{k}$ containing $L$ dummies and appends it to $\X$ to obtain the enlarged predictor matrix $\XWK_{k} = \big[ \X \,\, \XK_{k} \big]$. With $\XWK_{k}$ and the response $\y$ as inputs, a forward variable selection method is applied to obtain the candidate sets $\C_{1, L}(T), \ldots, \C_{K, L}(T)$, where $T$ is iteratively increased from one until $\widehat{\FDP}$ (i.e., an estimate of the proportion of false discoveries among all selected variables that is determined by the calibration process) exceeds the target FDR level $\alpha \in [0, 1]$. Finally, a fusion procedure determines the selected active set $\widehat{\mathcal{A}}_{L}(v^{*}, T^{*})$ for which the calibration procedure provides the optimal parameters $v^{*}$ and $T^{*}$, such that the FDR is controlled at the target level $\alpha$ while maximizing the number of selected variables.
}
\label{fig: T-Rex selector}
\end{figure*}

\subsection{The T-Rex Selector: Methodology}
\label{subsec: The T-Rex Selector: Methodology}
The general methodology underpinning the \textit{T-Rex} selector consists of several steps that are illustrated in Figure~\ref{fig: T-Rex selector}. In the following, we will introduce the framework and the notation, which will be crucial for understanding why the \textit{T-Rex} selector efficiently controls the FDR at the target level:

\begin{steps}[align=left, leftmargin=0.3cm]
\item Generate $K > 1$ dummy matrices $\XK_{k}$, $k = 1, \ldots, K$, each containing $L \geq 1$ dummy predictors that are sampled from a standard normal distribution.

\item Append each dummy matrix to the original predictor matrix $\X$, resulting in the enlarged predictor matrices
\begin{align}
\XWK_{k} \coloneqq& \big[ \X \,\, \XK_{k} \big] \\
=& \big[ \x_{1} \,\, \cdots \,\, \x_{p} \,\, \xK_{k, 1} \,\, \cdots  \,\, \xK_{k, L} \big], \quad  k = 1, \ldots, K,
\label{eq: predictor matrix with dummies}
\end{align}
where $ \xK_{k, 1}, \ldots, \xK_{k, L}$ are the dummies (see Figure~\ref{fig: predictor matrix with dummies}).

\item
Apply a forward variable selection procedure to $\big\lbrace \XWK_{k}, \y \big\rbrace$, $k = 1, \ldots, K$. For each random experiment, terminate the forward selection process after $T \geq 1$ dummy variables are included. This results in the candidate active sets $\C_{k, L}(T)$, $k = 1, \ldots, K$. After terminating the forward selection process remove all dummies from the candidate active sets.\footnote{Since we use the \textit{LARS} method throughout this paper, variables can only be included but not dropped along the solution paths. Nevertheless, the \textit{T-Rex} selector can also incorporate forward selection methods that remove some previously included variables from the candidate set along the solution path (e.g., \textit{Lasso}). For such methods, the number of currently active dummies can decrease along the solution path. However, because the solution paths are terminated after $T$ dummies are included for the first time, there is no ambiguity regarding the step in which the forward selection process ends.}
\label{T-Rex: step 3}
 
\item Iteratively increase $T$ and carry out~\ref{T-Rex: step 3} until $\widehat{\FDP}$ (i.e., a conservative estimate of the proportion of false discoveries among all selected variables) exceeds the target FDR level $\alpha \in [0, 1]$. The calibration process for determining $\widehat{\FDP}$ and the optimal values $v^{*}$ and $T^{*}$ such that the FDR is controlled at the target level $\alpha \in [0, 1]$ while maximizing the number of selected variables is derived in Section~\ref{sec: Main Results}.

\tikzset{
    ncbar angle/.initial=90,
    ncbar/.style={
        to path=(\tikztostart)
        -- ($(\tikztostart)!#1!\pgfkeysvalueof{/tikz/ncbar angle}:(\tikztotarget)$)
        -- ($(\tikztotarget)!($(\tikztostart)!#1!\pgfkeysvalueof{/tikz/ncbar angle}:(\tikztotarget)$)!\pgfkeysvalueof{/tikz/ncbar angle}:(\tikztostart)$)
        -- (\tikztotarget)
    },
    ncbar/.default=0.5cm,
}

\tikzset{square left brace/.style={ncbar=0.5cm}}
\tikzset{square right brace/.style={ncbar=-0.5cm}}

\tikzset{round left paren/.style={ncbar=0.5cm,out=120,in=-120}}
\tikzset{round right paren/.style={ncbar=0.5cm,out=60,in=-60}}

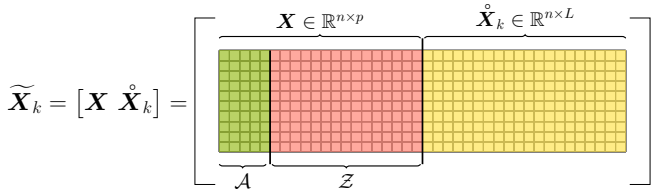
\begin{figure}[!t]
\begin{center}
\tdplotsetmaincoords{0}{0}
\scalebox{0.54}{
\begin{tikzpicture}
    [tdplot_main_coords,
        grid/.style={very thin,gray}]
\node at (-7.5,0) {\LARGE $\XWK_{k} = \big[ \X \,\, \XK_{k} \big] = $};
\draw [black] (-4.6,-2.05) to [ square left brace ] (-4.6,2.05); 
\draw [black] (5.60,-2.05) to [ square right brace ] (5.60,2.05); 
\foreach \x in {-4.5,-4.25,...,5.5}{
    \foreach \y in {-1.25,-1.00,...,1.25}
    {
        \draw[grid] (\x,-1.25) -- (\x,1.25);
        \draw[grid] (-4.5,\y) -- (5.5,\y);
    }
    }
\fill[applegreen, opacity = 0.6] (-4.5,-1.25) rectangle (-3.25,1.25);
\fill[bittersweet, opacity = 0.6] (-3.25,-1.25) rectangle (0.50,1.25);
\fill[bananayellow, opacity = 0.6] (0.50,-1.25) rectangle (5.50,1.25);

\draw[very thick, color=blue] (-3.25,-1.55) -- (-3.25,1.25);
\draw[very thick, color=blue] (0.50,-1.55) -- (0.50,1.55);

\draw [
    thick,
    decoration={
        brace,
        mirror,
        raise=0.5cm
    },
    decorate
] (-4.5,-1.05) -- (-3.35,-1.05) 
node [pos=0.5,anchor=north,yshift=-0.6cm] {\Large $\mathcal{A}$};

\draw [
    thick,
    decoration={
        brace,
        mirror,
        raise=0.5cm
    },
    decorate
] (-3.20,-1.05) -- (0.45,-1.05) 
node [pos=0.5,anchor=north,yshift=-0.6cm] {\Large $\Z$};

\draw [
    thick,
    decoration={
        brace,
        raise=0.5cm
    },
    decorate
] (-4.5,1.05) -- (0.45,1.05) 
node [pos=0.5,anchor=south,yshift=0.6cm] {\Large $\X \in \mathbb{R}^{n \times p}$};

\draw [
    thick,
    decoration={
        brace,
        raise=0.5cm
    },
    decorate
] (0.55,1.05) -- (5.50,1.05)
node [pos=0.5,anchor=south,yshift=0.6cm] {\Large $\XK_{k} \in \mathbb{R}^{n \times L}$}; 

\end{tikzpicture}}
\end{center}
\caption{The enlarged predictor matrices $\XWK_{k}$, $k = 1, \ldots, K$, replace the original predictor matrix $\X$ in each random experiment within the \textit{T-Rex} selector framework. They contain the original and the dummy predictors. The index set of the active variables and the index set of the null variables are denoted by $\A$ and $\Z$, respectively. The number of active variables and the number of null variables are denoted by $p_{1} \coloneqq | \A |$ and $p_{0} \coloneqq | \Z |$, respectively.}
\label{fig: predictor matrix with dummies}
\end{figure}
\item Fuse the candidate active sets to determine the estimate of the active set $\widehat{\mathcal{A}}_{L}(v^{*}, T^{*})$. The fusion step is based on the \textit{relative occurrence} of the original variables:
\begin{defn}[Relative occurrence]
Let $K~\in~\mathbb{N}_{+} \setminus \lbrace 1 \rbrace$ be the number of random experiments, $L~\in~\mathbb{N}_{+}$ the number of dummies, and $T~\in~\lbrace 1, \ldots, L \rbrace$ the number of included dummies after which the forward variable selection process in each random experiment is terminated. The relative occurrence of variable $j~\in~\lbrace 1, \ldots, p \rbrace$ is defined by
\begin{equation*}
\Phi_{T,L}(j) \coloneqq 
\begin{cases}
\dfrac{1}{K} \sum\limits_{k = 1}^{K} \mathbbm{1}_{k}(j, T, L), & T \geq 1 \\
0, \quad & T = 0
\end{cases},
\end{equation*}
where $\mathbbm{1}_{k}(j, T, L)$ is the indicator function for which
\begin{equation*}
\mathbbm{1}_{k}(j, T, L) =
\begin{cases}
1, \quad & j \in \C_{k, L}(T) \\
0, \quad & \text{otherwise}
\end{cases}.
\end{equation*}
\label{definition: relative occurrence}
\end{defn}
All variables whose relative occurrences at $T = T^{*}$ exceed the voting level $v^{*} \in [0.5, 1)$ are selected and the estimator of the active set is defined by
\begin{equation}
\widehat{\mathcal{A}}_{L}(v^{*}, T^{*}) \coloneqq \lbrace j : \Phi_{T^{*}, L}(j) > v^{*} \rbrace.
\label{eq: selected active set}
\end{equation}
\end{steps}
The details of how the calibration process determines $T^{*}$ and $v^{*}$ such that, for any choice of $L$, the \textit{T-Rex} selector controls the FDR at the target level while maximizing the number of selected variables are deferred to Section~\ref{subsec: The T-Rex Selector: Calibration Algorithm}. Moreover, an extension to the calibration process to jointly determine $T^{*}$, $v^{*}$, and $L$ is also proposed in Section~\ref{subsec: The T-Rex Selector: Calibration Algorithm}. The number of random experiments $K$ is not subject to optimization. However, choosing $K \geq 20$ provides excellent empirical results and we never observed notable improvements for $K \geq 100$.\footnote{Instead of fixing the number of random experiments, it could be increased until the relative occurrences $\Phi_{T,L}(j)$, $j = 1, \ldots, p$, converge. However, since a significant reduction of computation time is achieved by executing the independent random experiments in parallel on multicore computers or high-performance clusters, fixing $K$ to a multiple of the number of available CPUs is preferable.}

An example that helps in developing an intuition for the three main ingredients of the \textit{T-Rex} selector, which are (i) sampling dummies from a univariate distribution, (ii) early terminating the solution paths of the random experiments, and (iii) fusing the candidate sets based on relative occurrences, is deferred to Appendix~\ref{sec: Main Ingredients of the T-Rex Selector} in the supplementary materials.

\subsection{Problem Statement}
\label{subsec: Problem Statement}
An optimization problem formalizing the task of selecting as many true positives as possible while controlling the FDR at the target level is formulated. We start with some remarks on notation followed by definitions of the FDR and the TPR, which particularize the generic definitions in~\eqref{eq: FDR and TPR} for the \textit{T-Rex} selector.
For better readability, from now on, the arguments $T$ and $L$ of the estimator of the active set are dropped, i.e., $\widehat{\A}(v) \coloneqq \widehat{\A}_{L}(v, T)$, except when referring specifically to the set in~\eqref{eq: selected active set} for which the values $v^{*}$ and $T^{*}$ result from the calibration that will be discussed in Section~\ref{sec: Main Results}. Note that the term ``included candidates'' refers to the variables that were picked (and not dropped) along the solution path of each random experiment while the term ``selected variables'' refers to the variables whose relative occurrences exceed the voting level $v \in [0.5, 1)$.
\begin{defn}[$V_{T,L}(v)$, $S_{T,L}(v)$ and $R_{T,L}(v)$]
The number of selected null variables $V_{T,L}(v)$, the number of selected active variables $S_{T,L}(v)$, and the number of selected variables $R_{T,L}(v)$ are defined, respectively, by
\begin{align}
V_{T,L}(v) &\coloneqq \big| \widehat{\mathcal{A}}^{\, 0}(v) \big| \coloneqq  \big| \lbrace \text{null } j : \Phi_{T, L}(j) > v \rbrace  \big|, \\
S_{T,L}(v) &\coloneqq \big| \widehat{\mathcal{A}}^{\, 1}(v) \big| \coloneqq  \big| \lbrace \text{active } j : \Phi_{T, L}(j) > v \rbrace  \big|, \text{ and}\\
R_{T,L}(v) &\coloneqq V_{T,L}(v) + S_{T,L}(v) = \big| \widehat{\mathcal{A}}(v) \big|.
\end{align}
\label{definition: Empirical processes}
\end{defn}
The FDR and TPR expressions in~\eqref{eq: FDR and TPR} are rewritten using Definition~\ref{definition: Empirical processes} as follows:
\begin{defn}[FDP and FDR]
The false discovery proportion (FDP) is defined by
\begin{equation*}
\FDP(v, T, L) \coloneqq \dfrac{V_{T, L}(v)}{R_{T, L}(v) \lor 1}
\end{equation*}
and the FDR is defined by
\begin{equation*}
\FDR(v, T, L) \coloneqq \mathbb{E} \big[ \FDP(v, T, L) \big],
\end{equation*}
where the expectation is taken with respect to the noise in~\eqref{eq: linear model}.
\label{definition: FDP and FDR}
\end{defn}
\begin{defn}[TPP and TPR]
The true positive proportion (TPP) is defined by 
\begin{equation*}
\TPP(v, T, L) \coloneqq \dfrac{S_{T, L}(v)}{p_{1} \lor 1}
\end{equation*}
and the TPR is defined by
\begin{equation*}
\TPR(v, T, L) \coloneqq \mathbb{E} \big[ \TPP(v, T, L) \big],
\end{equation*}
where the expectation is taken with respect to the noise in~\eqref{eq: linear model}.
\label{definition: TPP and TPR}
\end{defn}
\begin{rmk}
Note that if $R_{T, L}(v)$ is equal to zero, then $V_{T, L}(v)$ is zero as well. In this case, the denominator in the expression for the FDP is set to one and, thus, the FDP becomes zero. This is a reasonable solution to the ``$0/0$'' case, because when no variables are selected there exist no false discoveries. Similarly, when there exist no true active variables among the candidates, i.e. $p_{1} = S_{T, L}(v) = 0$, the TPP equals zero.
\end{rmk}
A major result of this work is to determine $T^{*}$ and $v^{*}$, such that, for any fixed $L \in \mathbb{N}_{+}$, the \textit{T-Rex} selector maximizes $\TPR(v, T, L)$ while provably controlling $\FDR(v, T, L)$ at any given target level $\alpha \in [0, 1]$. In practice, this amounts to finding the solution of the optimization problem
\begin{equation}
\max_{v, T} \, \TPP(v, T, L) \quad \text{s.t.} \quad \widehat{\FDP}(v, T, L) \leq \alpha,
\label{eq: T-Rex optimization problem 1}
\end{equation}
which is equivalent to
\begin{equation}
\max_{v, T} \, S_{T, L}(v) \quad \text{s.t.} \quad \widehat{\FDP}(v, T, L) \leq \alpha
\label{eq: T-Rex optimization problem 2}
\end{equation}
because $p_{1}$ is a constant. Note that $\widehat{\FDP}(v, T, L)$ is a conservative estimator of $\FDP(v, T, L)$, i.e., it holds that $\FDR(v, T, L) = \mathbb{E} \big[ \FDP(v, T, L) \big] \leq \mathbb{E} \big[ \widehat{\FDP}(v, T, L) \big] =  \widehat{\FDR}(v, T, L)$. The details of the conservative FDP estimator are discussed in Section~\ref{sec: Main Results}. Since we cannot observe $S_{T, L}(v)$, it is replaced by $R_{T, L}(v)$. This results in the final optimization problem:
\begin{equation}
\max_{v, T} \, R_{T, L}(v) \quad \text{s.t.} \quad \widehat{\FDP}(v, T, L) \leq \alpha.
\label{eq: T-Rex relaxed optimization problem}
\end{equation}
In words: \textit{The T-Rex selector maximizes the number of selected variables while controlling a conservative estimator of the FDP at the target level $\alpha$.}

In Section~\ref{sec: Main Results}, it is shown that the \textit{T-Rex} selector efficiently solves~\eqref{eq: T-Rex relaxed optimization problem} and that any solution of~\eqref{eq: T-Rex relaxed optimization problem} is a feasible solution of~\eqref{eq: T-Rex optimization problem 1} and~\eqref{eq: T-Rex optimization problem 2}.

\section{Main Results}
\label{sec: Main Results}
This section contains our main results about the proposed \textit{T-Rex} selector, which concern: FDR-control (Theorem~\ref{theorem: FDR control}), dummy generation (Theorem~\ref{theorem: Dummy generation}), and the optimal calibration algorithm (Theorem~\ref{theorem: T-Rex algorithm}). We use martingale theory~\cite{williams1991probability} to prove the FDR control property of the \textit{T-Rex} selector. The developed FDR control theory relies on standard assumptions that are extensively verified especially for GWAS, i.e., the main use-case of this paper (see Appendices~\ref{sec: General Assumptions}, \ref{sec: Exemplary Numerical Verification of Assumptions}, and~\ref{sec: Setup, Preprocessing, and Additional Results: Simulated Genome-Wide Association Study} in the supplementary materials). Additionally, the computational complexity of the \textit{T-Rex} selector, which stems from the computation of $K$ terminated random experiments with expected complexity $\mathcal{O}(np)$, is derived in Appendix~\ref{sec: Computational Complexity} in the supplementary materials.
\subsection{FDR Control}
\label{subsec: FDR Control}
In Definition~\ref{definition: relative occurrence}, the relative occurrence $\Phi_{T, L}(j)$ of the $j$th candidate variable has been introduced. It can be decomposed into the changes in relative occurrence, i.e., 
\begin{equation}
\Phi_{T, L}(j) = \sum\limits_{t = 1}^{T} \Delta\Phi_{t, L}(j), \quad j = 1, \ldots, p,
\label{eq: reformulation of relative occurrence}
\end{equation}
where $\Delta\Phi_{t, L}(j) \coloneqq \Phi_{t, L}(j) - \Phi_{t - 1, L}(j)$ is the change in relative occurrence from step~$t - 1$ to $t$ for variable $j$.\footnote{When using a forward selection method within the \textit{T-Rex} selector framework that does not drop variables along the solution path (e.g. \textit{LARS}), all $\Phi_{t, L}(j)$'s are non-decreasing in $t$ and, therefore, $\Delta\Phi_{t, L}(j) \geq 0$ for all $j$. In contrast, when using forward selection methods that might drop variables along the solution path (e.g. \textit{Lasso}), the $\Phi_{t, L}(j)$'s might decrease in $t$ and, therefore, the $\Delta\Phi_{t, L}(j)$'s can be negative. Nevertheless, the relative occurrence $\Phi_{T, L}(j)$ is non-negative for all $j$ and any forward selection method.} Since the active and the null variables are interspersed in the solution paths of the random experiments, some null variables might appear earlier on the solution paths than some active variables.\footnote{Many researchers have observed that active and null variables are interspersed in solution paths obtained from sparsity-inducing methods, such as the \textit{LARS} algorithm or the \textit{Lasso}~\cite{su2017false, barber2015controlling}.} Therefore, it is unavoidable that the $\Delta\Phi_{t, L}(j)$'s of the null variables are inflated along the solution paths of the random experiments. Moreover, we observe interspersion not only for active and null variables but also for dummies, which is expected since dummies can be interpreted as flagged null variables.

The above considerations motivate the definition of the \textit{deflated relative occurrence} to harness the information about the fraction of included dummies in each step along the solution paths in order to deflate the $\Delta\Phi_{t, L}(j)$'s of the null variables and, thus, account for the interspersion effect.
\begin{defn}[Deflated relative occurrence]
The deflated relative occurrence of variable $j$ is defined by
\begin{align}
&\Phi^{\prime}_{T, L}(j) \coloneqq \\
&\sum\limits_{t = 1}^{T} \bigg( 1 - \dfrac{p - \sum_{q = 1}^{p} \Phi_{t, L}(q)}{L - (t - 1)} \dfrac{1}{\sum_{q \in \widehat{\A}(0.5)} \Delta\Phi_{t, L}(q)} \bigg) \Delta\Phi_{t, L}(j),
\label{eq: path-adapted relative occurrence}
\end{align}
$j = 1, \ldots, p$.
\label{definition: path-adapted relative occurrence}
\end{defn}
In words: \textit{The deflated relative occurrence is the sum over the deflated $\Delta \Phi_{t, L}(j)$'s from step $t = 1$ until step $t = T$}. As detailed and intuitively explained in Appendix~\ref{sec: The Deflated Relative Occurrence} in the supplementary materials, the $\Delta\Phi_{t, L}(j)$'s are multiplied by a deflation factor that takes into account the ratio between the fraction of selected dummies and the fraction of selected candidate variables in each step $t \in \lbrace 1, \ldots, T \rbrace$.

Using the deflated relative occurrences, the estimator of $V_{T, L}(v)$, i.e., the number of selected null variables (see Definition~\ref{definition: Empirical processes}), and the corresponding FDP estimator are defined as follows:
\begin{defn}[FDP estimator]
The estimator of $V_{T, L}(v)$ is defined by
\begin{equation}
\widehat{V}_{T, L}(v) \coloneqq \sum\limits_{j \in \widehat{\A}(v)} \big(1 - \Phi^{\prime}_{T, L}(j) \big)
\end{equation}
and the corresponding estimator of $\FDP(v, T, L)$ is defined by
\begin{equation}
\widehat{\FDP}(v, T, L) = \dfrac{\widehat{V}_{T, L}(v)}{R_{T, L}(v) \lor 1}
\label{eq: FDP hat}
\end{equation}
with 
\begin{equation}
\widehat{\FDR}(v, T, L) \coloneqq \mathbb{E} \big[ \widehat{\FDP}(v, T, L) \big]
\end{equation}
being its expected value.
\label{definition: FDP estimator}
\end{defn}
The main idea behind FDR control for the \textit{T-Rex} selector is that controlling $\widehat{\FDP}(v, T, L)$ at the target level $\alpha \in [0, 1]$ guarantees that $\FDR(v, T, L)$ is controlled at the target level as well. To achieve this, we define $v \in [0.5, 1)$ as the voting level at which $\widehat{\FDP}(v, T, L)$ is controlled at the target level. Note that $v$ has to be at least $50\%$ to ensure that all selected variables occur in at least more than the majority of the candidate sets within the \textit{T-Rex} selector.
\begin{defn}[Voting level]
Let $T \in \lbrace 1, \ldots, L \rbrace$ and $L \in \mathbb{N}_{+}$ be fixed. Then, the voting level is defined by
\begin{equation}
v \coloneqq \inf \lbrace \nu \in [0.5, 1) : \widehat{\FDP}(\nu, T, L) \leq \alpha \rbrace
\label{eq: Voting level}
\end{equation}
with the convention that $v = 1$ if the infimum does not exist.\footnote{The voting level can be interpreted as a stopping time. The term `stopping time' stems from martingale theory~\cite{williams1991probability}. In the proof of Lemma~\ref{lemma: martingale} in Appendix~\ref{sec: Proofs} in the supplementary materials, it is shown that indeed $v$ is a stopping time with respect to some still to be defined filtration of a still to be defined stochastic process. Note that the convention of setting $v = 1$ if the infimum does not exist ensures that no variables are selected when there exists no triple $(T, L, v)$ that satisfies Equation~\eqref{eq: Voting level}.}
\label{definition: Voting level}
\end{defn}
\begin{rmk}
Recall that the aim that is stated in the optimization problem in~\eqref{eq: T-Rex relaxed optimization problem} is to select as many variables as possible while controlling $\widehat{\FDP}(v, T, L)$ at the target level. For fixed $T$ and $L$, this is achieved by the smallest voting level that satisfies the constraint on $\widehat{\FDP}(v, T, L)$. We can easily see that for any fixed $T$ and $L$, the voting level in~\eqref{eq: Voting level} solves the optimization problem in~\eqref{eq: T-Rex relaxed optimization problem}. The reason is that for any two voting levels $v_{1}, \, v_{2} \in [0.5, 1)$ with $v_{2} \geq v_{1}$ satisfying the $\widehat{\FDP}$-constraint in~\eqref{eq: Voting level}, it holds that $R_{T, L}(v_{1}) \geq R_{T, L}(v_{2})$.
\label{remark: Voting level}
\end{rmk}
\begin{rmk}
If $v$, $T$, and $L$ satisfy Equation~\eqref{eq: Voting level}, then the FDP from Definition~\ref{definition: FDP and FDR} can be upper-bounded as follows:
\begingroup
\allowdisplaybreaks
\begin{align}
\FDP(v, T, L) 
& = \dfrac{V_{T, L}(v)}{R_{T, L}(v) \lor 1} = \widehat{\FDP}(v, T, L) \cdot \dfrac{V_{T, L}(v)}{\widehat{V}_{T, L}(v)}
\label{eq: upper-bounded FDP-3}
\\
& \leq \alpha \cdot \dfrac{V_{T, L}(v)}{\widehat{V}_{T, L}(v)} \leq \alpha \cdot \dfrac{V_{T, L}(v)}{\widehat{V}_{T, L}^{\prime}(v)},
\label{eq: upper-bounded FDP-5}
\end{align}
\endgroup
where $\widehat{V}_{T, L}^{\prime}(v)$, which is supposed to be greater than zero, is defined by
\begin{equation}
\widehat{V}_{T, L}^{\prime}(v) \coloneqq \widehat{V}_{T, L}(v) - \sum\limits_{j \in \widehat{\A}(v)} \big( 1 - \Phi_{T, L}(j) \big).
\label{eq: upper-bounded FDP-6}
\end{equation}
\label{remark: upper-bounded FDP}
\end{rmk}
Before the FDR control theorem is formulated, we introduce a lemma that contains the backbone of our FDR control theorem, which is rooted in martingale theory~\cite{williams1991probability}:
\begin{customlemma}{5}
Define $\mathcal{V} \coloneqq \lbrace \Phi_{T, L}(j) : \Phi_{T, L}(j) > 0.5, \, j = 1, \ldots, p \rbrace$ and
\begin{equation}
H_{T, L}(v) \coloneqq \dfrac{V_{T, L}(v)}{\widehat{V}_{T, L}^{\prime}(v)}.
\end{equation}
Let $\mathcal{F}_{v} \coloneqq \sigma \big( \lbrace V_{T, L}(u) \rbrace_{u \geq v}, \lbrace \widehat{V}_{T, L}^{\prime}(u) \rbrace_{u \geq v} \big)$ be a backward-filtration with respect to $v$. Then, for all tuples $(T, L) \in \lbrace 1, \ldots, L \rbrace \times \mathbb{N}_{+}$, $\lbrace H_{T, L}(v) \rbrace_{v \in \mathcal{V}}$ is a backward-running super-martingale with respect to $\mathcal{F}_{v}$. That is,
\begin{equation}
\mathbb{E} \big[ H_{T, L} \big( v - \epsilon_{T, L}^{*}(v) \big) \bigm| \mathcal{F}_{v} \big] \geq H_{T, L}(v),
\end{equation}
where
\begin{equation}
\epsilon_{T, L}^{*}(v) \coloneqq \inf \lbrace \epsilon \in (0, v) : R_{T, L}(v - \epsilon) - R_{T, L}(v) = 1 \rbrace
\label{eq: epsilon star in martingale lemma}
\end{equation}
with $v \in [0.5, 1)$ and the convention that $\epsilon_{T, L}^{*}(v) = 0$ if the infimum does not exist.
\label{lemma: martingale}
\end{customlemma}
\begin{proof}
The proof is deferred to Appendix~\ref{sec: Proofs} in the supplementary materials.
\label{eq: proof lemma martingale}
\end{proof}

\begin{thm}[FDR control]
Suppose that $\widehat{V}_{T, L}^{\prime}(v) > 0$. Then, for all triples $(T, L, v) \in \lbrace 1, \ldots, L \rbrace \times \mathbb{N}_{+} \times [0.5, 1)$ that satisfy Equation~\eqref{eq: Voting level} and as $K \to \infty$, the \textit{T-Rex} selector controls the FDR at any fixed target level $\alpha \in [0, 1]$, i.e.,
\begin{equation}
\FDR(v, T, L) = \mathbb{E} \big[ \FDP(v, T, L) \big] \leq \alpha.
\label{eq: FDR control}
\end{equation}
\label{theorem: FDR control}
\end{thm}
\begin{proof}
With Lemma~\ref{lemma: martingale} and since the stopping time in~\eqref{eq: Voting level} is adapted to the filtration, i.e., it is $\mathcal{F}_{v}$-measurable, the optional stopping theorem can be applied to upper bound $\mathbb{E} \big[ H_{T, L}(v) \big]$. This yields, as $K \rightarrow \infty$,
\begingroup
\allowdisplaybreaks
\begin{align}
\mathbb{E} \big[ H_{T, L}(v) \big] &\leq \mathbb{E} \big[ H_{T, L}(0.5) \big] = \dfrac{1}{\widehat{V}_{T, L}^{\prime}(0.5)} \cdot \mathbb{E} \big[ V_{T, L}(0.5) \big]
\label{eq: proof-theorem-1.19}
\\[0.5em]
&\leq \dfrac{1}{\widehat{V}_{T, L}^{\prime}(0.5)} \cdot \dfrac{T}{L + 1} \cdot p_{0}
\label{eq: proof-theorem-1.20}
\\[0.5em]
&= \dfrac{1}{\dfrac{T}{L + 1} \cdot p_{0}} \cdot \dfrac{T}{L + 1} \cdot p_{0} = 1.
\label{eq: proof-theorem-1.21}
\end{align}
\endgroup
The first inequality is a consequence of the optional stopping theorem and Lemma~\ref{lemma: martingale} and the equation in the first line follows from $\widehat{V}_{T, L}^{\prime}(0.5)$ being deterministic as $K \rightarrow \infty$. The second line follows from $\mathbb{E}[\NHG(p_{0} + L, p_{0}, T)] = T \cdot p_{0} / (L + 1)$ and $V_{T, L}(v) \overset{d}{\leq} \NHG(p_{0} + L, p_{0}, T)$, $v \in [0.5, 1)$, i.e. $V_{T, L}(v)$ is stochastically dominated by the negative hypergeometric distribution (NHG) with $p_{0} + L$ total elements, $p_{0}$ success elements, and $T$ failures after which a random experiment is terminated (for more details, see Appendix~\ref{sec: General Assumptions} in the supplementary materials). The third line holds since
\begingroup
\allowdisplaybreaks
\begin{align}
\widehat{V}_{T, L}^{\prime}(0.5) &= \sum\limits_{t = 1}^{T} \dfrac{p_{0} - \sum_{q \in \Z} \Phi_{t, L}(q)}{L - (t - 1)} = \sum\limits_{t = 1}^{T} \dfrac{p_{0} - \frac{t}{L + 1} \cdot p_{0}}{L - (t - 1)}
\label{eq: proof-theorem-2.23}
\\[0.5em]
&= \dfrac{p_{0}}{L + 1} \cdot \sum\limits_{t = 1}^{T} \dfrac{L - t + 1}{L - t + 1} = \dfrac{T}{L + 1} \cdot p_{0},
\label{eq: proof-theorem-2.25}
\end{align}
\endgroup
where the second equation follows from Lemma~\ref{lemma: 6} in Appendix~\ref{sec: Proofs} in the supplementary materials. Finally, it follows
\begin{align}
\FDR(v, T, L) &= \mathbb{E} \big[ \FDP(v, T, L) \big] \leq \alpha \cdot \mathbb{E} \big[ H_{T, L}(v) \big] \leq \alpha,
\label{eq: proof-theorem-2.28}
\end{align}
i.e., FDR control at the target level $\alpha$ is achieved.
\end{proof}
\subsection{Dummy Generation}
\label{subsec: Dummy Generation}
As shown in Figure~\ref{fig: T-Rex selector}, the \textit{T-Rex} selector generates $L$ i.i.d. dummies for each random experiment by sampling each element of the dummy vectors from the standard normal distribution, i.e.,
\begin{equation}
\xK_{l} = [ \xk_{1, l} \, \cdots \, \xk_{n, l} ]^{\top}, \text{ where } \xk_{i, l} \sim \mathcal{N}(0, 1),
\label{eq: dummy vectors}
\end{equation}
$i = 1, \ldots, n$, $l = 1, \ldots, L$. This raises the question whether dummies can be sampled from other distributions, as well, to serve as flagged null variables. From an asymptotic point of view, i.e., $n \to \infty$, and if some mild conditions are satisfied, the perhaps at first glance surprising answer to this question is that \textit{dummies can be sampled from any univariate probability distribution with finite expectation and variance in order to serve as flagged null variables within the \textit{T-Rex} selector}. 

We will prove the above statement for any forward selection procedure that uses sample correlations of the predictors with the response or with the current residuals in each forward selection step to determine which variable is included next. Thus, the statement is true, e.g., for the \textit{LARS} algorithm, \textit{Lasso}, \textit{adaptive Lasso}, and \textit{elastic net}.

Recall that null variables and dummies are not related to the response. For null variables this holds by definition and for dummies this holds because dummies are generated without using any information about the response.\footnote{Note that the knockoff generation processes of the \textit{fixed-X} and the \textit{model-X} knockoff method, i.e., the benchmark methods, are fundamentally different from our approach that uses dummies. Although these methods also do not use any information about the response to generate the knockoffs, unlike the proposed \textit{T-Rex} selector, they must incorporate the covariance structure of the predictor matrix, which leads to a large computation time, especially for high dimensions (see Appendix~\ref{sec: Main Ingredients of the T-Rex Selector} in the supplementary materials and Figure~\ref{fig: computation time sweep p five million}).} Moreover, the sample correlations of the dummies with the response are random. Thus, the higher the number of generated dummies, the higher the probability of including a dummy instead of a null or even a true active variable in the next step of a random experiment. These considerations suggest that only the number of dummies within the enlarged predictor matrices is relevant for the behavior of the forward selection process in each random experiment. That is, for $n \to \infty$, the distribution from which the dummies are sampled has no influence on the distribution of the correlation variables
\begin{equation}
\GKvar_{l, m, k} \coloneqq \sum\limits_{i = 1}^{n} \gamma_{i, m, k} \cdot \XKvar_{i, l, k},
\label{eq: correlation variables}
\end{equation}
$l \in \Dum_{m, k}$, $m \geq 1$, $k = 1, \ldots, K$, where $\gamma_{i, m, k}$ is the $i$th element of $\boldsymbol{\gamma}_{m, k} \coloneqq \y - \X \hatbbeta_{m, k}$ (i.e., the residual vector in the $m$th forward selection step of the $k$th random experiment) with $\hatbbeta_{m, k}$ and $\Dum_{m, k}$ being the estimator of the parameter vector and the index set of the non-included dummies in the $m$th forward selection step of the $k$th random experiment, respectively. Note that $\boldsymbol{\gamma}_{1, k} = \y$ for all $k$, since $\hatbbeta_{1, k} = \boldsymbol{0}$ for all $k$, i.e., the residual vector in the first step of the forward selection process is simply the response vector $\y$. The random variable $\XKvar_{i, l, k}$ represents the $i$th element of the $l$th dummy within the $k$th random experiment. Summarizing, $\GKvar_{l, m, k}$ can be interpreted as the weighted sum of the i.i.d. random variables $\XKvar_{1, l, k}, \ldots, \XKvar_{n, l, k}$ with fixed weights $\gamma_{1, m, k}, \ldots, \gamma_{n, m, k}$. With these preliminaries in place, the second main theorem is formulated as follows:

\begin{thm}[Dummy generation]
Let $\XKvar_{i, l, k}$, $i = 1, \ldots, n$, $l \in \Dum_{m, k}$, $m \geq 1$, $k = 1, \ldots, K$, be standardized i.i.d. dummy random variables (i.e., $\mathbb{E} \big[ \XKvar_{i, l, k} \big] = 0$ and $\Var \big[ \XKvar_{i, l, k} \big] = 1$ for all $i, l, m, k$) following any probability distribution with finite expectation and variance. Define
\begin{equation}
D_{n, l, m, k} \coloneqq \dfrac{1}{\Gamma_{n, m, k}} \cdot \GKvar_{l, m, k},
\label{eq: weighted correlation variables}
\end{equation}
where $\Gamma_{n, m, k}^{2} \coloneqq \sum_{i = 1}^{n} \gamma_{i, m, k}^{2}$ with $\Gamma_{n, m, k} > 0$ for all $n, m, k$ and with fixed $\gamma_{i, m, k} \in \mathbb{R}$ for all $i, m, k$. Suppose that 
\begin{equation}
\lim_{n \to \infty} \dfrac{\gamma_{i, m, k}}{\Gamma_{n, m, k}} = 0, \quad i = 1, \ldots, n,
\label{eq: asymptotic condition dummy generation}
\end{equation}
for all $m, k$. Then, as $n \to \infty$,
\begin{equation}
D_{n, l, m, k} \overset{d}{\to} D, \quad D \sim \mathcal{N}(0, 1),
\label{eq: convergence of dummy distribution}
\end{equation}
for all $l, m, k$.
\label{theorem: Dummy generation}
\end{thm}

\begin{proof}[Proof sketch]
The Lindeberg-Feller central limit theorem is applicable because $\XKvar_{i, l, k}$, $i = 1, \ldots, n$, $l \in \Dum_{m, k}$, $m \geq 1$, $k = 1, \ldots, K$, are i.i.d. random variables and it holds that $\mathbb{E} \big[ D_{n, l, m, k} \big] = 0$ and $\Var \big[ D_{n, l, m, k} \big] = 1$. Moreover, since $\QKvar_{i, l, m, k} \coloneqq \gamma_{i, m, k} \cdot \XKvar_{i, l, k} \, / \, \Gamma_{n, m, k}$ satisfies the Lindeberg condition for all $l, m, k$, the theorem follows.
\label{proof: [Sketch] Dummy generation}
\end{proof}
The details of the proof and illustrative examples with non-Gaussian dummies are deferred to Appendix~\ref{sec: Proofs} and Appendix~\ref{sec: Illustration of Theorem 2 (Dummy Generation)}, respectively, in the supplementary materials.
\begin{rmk}
Note that sampling dummies from any univariate probability distribution with finite expectation and variance to serve as flagged null variables is only reasonable in combination with multiple random experiments as conducted by the proposed \textit{T-Rex} selector. We emphasize that Theorem~\ref{theorem: Dummy generation} is not applicable to knockoff generation procedures of, e.g., \textit{fixed-X} and \textit{model-X} knockoffs.
\label{remark: dummy genertion 2}
\end{rmk}
\subsection{The T-Rex Selector: Optimal Calibration Algorithm}
\label{subsec: The T-Rex Selector: Calibration Algorithm}
This section describes the proposed \textit{T-Rex} calibration algorithm, which efficiently solves the optimization problem in~\eqref{eq: T-Rex relaxed optimization problem} and provides feasible solutions for~\eqref{eq: T-Rex optimization problem 1} and~\eqref{eq: T-Rex optimization problem 2}. The pseudocode of the \textit{T-Rex} calibration method is provided in Algorithm~\ref{algorithm: T-Rex}.
\begin{algorithm}[h]
\caption{\textit{T-Rex} Calibration.}
\begin{alglist}
\item \textbf{Input}: $\alpha \in [0, 1]$, $K$, $L$, $\X$, $\y$.
\label{algorithm: T-Rex Step 1}
\item \textbf{Set} $T = 1$, $\Delta v = \dfrac{1}{K}$, $\widehat{\FDP}(v = 1 - \Delta v, T, L) = 0$.
\label{algorithm: T-Rex Step 2}
\item \textbf{While} $\widehat{\FDP}(v = 1 - \Delta v, T, L) \leq \alpha$ and $T \leq L$ \textbf{do}:
\begin{enumerate}[label*=\arabic*.]
\item[3.1.] \textbf{For} $v = 0.5, 0.5 + \Delta v, 0.5 + 2 \cdot \Delta v, \ldots, 1 - \Delta v$ \textbf{do}:
\begin{enumerate}
\item[i.] \textbf{Compute} $\widehat{\FDP}(v, T, L)$ as in~\eqref{eq: FDP hat}.
\item[ii.] \textbf{If} $\widehat{\FDP}(v, T, L) \leq \alpha$
\begin{enumerate}
\item[] \textbf{Compute} $\widehat{\A}_{L}(v, T)$ as in~\eqref{eq: selected active set}.
\end{enumerate}
\textbf{Else}
\begin{enumerate}
\item[] \textbf{Set} $\widehat{\A}_{L}(v, T) = \varnothing$.
\end{enumerate}
\end{enumerate}
\item[3.2.] \textbf{Set} $T \leftarrow T + 1$.
\end{enumerate}
\label{algorithm: T-Rex Step 3}
\item \textbf{Solve}
\begin{alignat*}{2}
& \max_{v^{\prime}, T^{\prime}} &{}& \, \big| \widehat{\A}_{L}(v^{\prime}, T^{\prime}) \big| \\
& \text{s.t.} &{}& T^{\prime} \in \lbrace 1, \ldots, T - 1 \rbrace \\
&{}&{}& v^{\prime} \in \lbrace  0.5, 0.5 + \Delta v, 0.5 + 2 \cdot \Delta v, \ldots, 1 - \Delta v \rbrace
\end{alignat*}
and let $(v^{*}, T^{*})$ be a solution.
\label{algorithm: T-Rex Step 4}
\item \textbf{Output}: $(v^{*}, T^{*})$ and $\widehat{\A}_{L}(v^{*}, T^{*})$.
\label{algorithm: T-Rex Step 5}
\end{alglist}
\label{algorithm: T-Rex}
\end{algorithm}
The algorithm flow is as follows: First, the number of dummies $L$ and the number of random experiments $K$ are set (usually $L = p$ and $K = 20$).\footnote{As already mentioned in Section~\ref{subsec: The T-Rex Selector: Methodology}, $K$ is not subject to optimization. In practice, choosing $K = 20$ already provides excellent results (see Section~\ref{sec: Numerical Simulations}) and only incremental improvements are achieved with larger values of $K$.} Then, setting $v = 1 - \Delta v$ and starting at $T = 1$, the number of included dummies is iteratively increased until reaching the value of $T$ for which the FDP estimate at a voting level of $v = 1 - \Delta v$ exceeds the target level for the first time. In each iteration, before the target level is exceeded, $\widehat{\A}_{L}(v, T)$ is computed as in~\eqref{eq: selected active set} on a grid for $v$, while for values of $v$ for which $\widehat{\FDP}(v, T, L)$ exceeds the target level $\widehat{\A}_{L}(v, T)$ is equal to the empty set. Picking the $v^{\prime}$ and $T^{\prime}$ that maximize the number of selected variables yields the final solution.\footnote{In case of multiple solutions, we recommend to choose the solution with the largest $v$ because such a solution provides the variables that were selected most frequently. Nevertheless, all solutions to the calibration problem that are computed using Algorithm~\ref{algorithm: T-Rex} provide FDR control while maximizing the number of selected variables.}

The reason for exiting the loop in Step~\ref{algorithm: T-Rex Step 3} when the FDP estimate at a voting level of $1 - \Delta v$ exceeds the target level for the first time is based on two key observations from our still to be presented simulation results (see Figure~\ref{fig: 3d-plots (FDP, FDP_hat and TPP)}):
\begin{enumerate}[label=\arabic*., ref=\arabic*]
\item For any fixed $T$ and $L$ the average value of $\widehat{\FDP}(v, T, L)$ decreases as $v$ increases.
\item For any fixed $v$ and $L$ the average value of $\widehat{\FDP}(v, T, L)$ increases as $T$ increases.
\end{enumerate}

\begin{rmk}
To foster the intuition behind these observations, we note that Equation~\eqref{eq: FDP hat} can be written as follows:
\begin{equation}
\widehat{\FDP}(v, T, L) =  \dfrac{\widehat{V}_{T, L}(v)}{\big( V_{T, L}(v) + S_{T, L}(v) \big) \lor 1}.
\label{eq: reformulate FDP_hat}
\end{equation}
Taking Definition~\ref{definition: Empirical processes}, Definition~\ref{definition: FDP estimator}, and the reformulation of Equation~\eqref{eq: FDP hat} into account, we see that the observations suggest that we can expect the rather conservative estimate $\widehat{V}_{T, L}(v)$ of $V_{T, L}(v)$ in the numerator to decrease faster than the total number of selected variables $V_{T, L}(v) + S_{T, L}(v)$ in the denominator when increasing the voting level $v$. This is something that can be expected since, in general, assuming a variable selection method that performs better than random selection, active variables are expected to have higher relative occurrences than null variables and, therefore, remain selected even for large values of the voting level~$v$. A similar reasoning can be applied to intuitively understand the monotonical increase of $\widehat{\FDP}(v, T, L)$ with respect to $T$.
\end{rmk}

With these preliminaries in place, the third main theorem of this paper can be formulated:
\begin{thm}[Optimality of Algorithm~\ref{algorithm: T-Rex}]
Let $(v^{*}, T^{*})$ be a solution determined by Algorithm~\ref{algorithm: T-Rex} and suppose that, ceteris paribus, $\widehat{\FDP}(v, T, L)$ is monotonically decreasing in $v$ and monotonically increasing in $T$. Then, $(v^{*}, T^{*})$ is an optimal solution of~\eqref{eq: T-Rex relaxed optimization problem} and a feasible solution of~\eqref{eq: T-Rex optimization problem 1} and~\eqref{eq: T-Rex optimization problem 2}.
\label{theorem: T-Rex algorithm}
\end{thm}

\begin{proof}[Proof sketch]
Since the objective functions of the optimization problems in Step~\ref{algorithm: T-Rex Step 4} of Algorithm~\ref{algorithm: T-Rex} and in~\eqref{eq: T-Rex relaxed optimization problem} are equivalent, i.e., $\big| \widehat{\A}_{L}(v, T) \big| = R_{T, L}(v)$, it only needs to be shown that the feasible set in Step~\ref{algorithm: T-Rex Step 4} of the algorithm contains the feasible set of~\eqref{eq: T-Rex relaxed optimization problem}. Since the conditions of the optimization problems in~\eqref{eq: T-Rex optimization problem 1},~\eqref{eq: T-Rex optimization problem 2}, and~\eqref{eq: T-Rex relaxed optimization problem} are equivalent, this also proves that $(v^{*}, T^{*})$ is a feasible solution of~\eqref{eq: T-Rex optimization problem 1} and~\eqref{eq: T-Rex optimization problem 2}.
\label{proof: [Sketch] T-Rex algorithm}
\end{proof}
The details of the proof are deferred to Appendix~\ref{sec: Proofs} in the supplementary materials.

\subsection{Extension to the Calibration Algorithm}
\label{subsec: Extension to The Calibration Algorithm}
In Theorem~\ref{theorem: FDR control}, we have also established that the \textit{T-Rex} selector controls the FDR at the target level for any choice of the number of dummies $L$. However, the choice of $L$ has an influence on how tightly the FDR is controlled at the target level (see Figure~\ref{fig: 3d-plots (FDP, FDP_hat and TPP)}). Since controlling the FDR more tightly usually increases the TPR (i.e., power), it is desirable to choose the parameters of the \textit{T-Rex} selector accordingly. We will see in the simulations in Section~\ref{sec: Numerical Simulations} that with increasing $L$, the FDR can be more tightly controlled at low target levels. In order to harness the positive effects that come with larger values of $L$ while limiting the increased memory requirement for high values of $L$, we propose an extended version of the calibration algorithm that jointly determines $v$, $T$, and $L$ such that the FDR is more tightly controlled at the target FDR level while not running out of memory.\footnote{The reader might raise the question whether also the computation time increases with increasing $L$. There is no definite answer to this question. On the one hand, for very large values of $L$ the computation time might increase. On the other hand, with increasing $L$ the solution paths of the experiments are terminated earlier because the probability of selecting dummies grows with increasing $L$. Thus, increasing $L$ might increase or decrease the computation time depending on whether the first or the second effect dominates.} The major difference to Algorithm~\ref{algorithm: T-Rex} is that the number of dummies $L$ is iteratively increased until the estimate of the FDP falls below the target FDR level $\alpha$. The pseudocode of the extended \textit{T-Rex} calibration algorithm is provided in Algorithm~\ref{algorithm: Extended T-Rex}.\footnote{The R package \textit{TRexSelector}~\cite{machkour2022TRexSelector} contains the implementation of the extended calibration algorithm in Algorithm~\ref{algorithm: Extended T-Rex}.}
\begin{algorithm}[h]
\caption{Extended \textit{T-Rex} Calibration.}
\begin{alglist}
\item \textbf{Input}: $\alpha \in [0, 1]$, $K$, $\X$, $\y$, $\tilde{v}$, $L_{\max}$, $T_{\max}$.
\label{algorithm: Extended T-Rex Step 1}
\item \textbf{Set} $L = p$, $T = 1$.
\label{algorithm: Extended T-Rex Step 2}
\item \textbf{While} $\widehat{\FDP}(v = \tilde{v}, T, L) > \alpha$ and $L \leq L_{\max}$ \textbf{do}:
\label{algorithm: Extended T-Rex Step 3}
\begin{enumerate}
\item[] \textbf{Set} $L \leftarrow L + p$.
\end{enumerate}
\item \textbf{Set} $\Delta v = \dfrac{1}{K}$, $\widehat{\FDP}(v = 1 - \Delta v, T, L) = 0$.
\label{algorithm: Extended T-Rex Step 4}
\item \textbf{While} $\widehat{\FDP}(v = 1 - \Delta v, T, L) \leq \alpha$ and $T \leq T_{\max}$ \textbf{do}:
\begin{enumerate}[label*=\arabic*.]
\item[5.1.] \textbf{For} $v = 0.5, 0.5 + \Delta v, 0.5 + 2 \cdot \Delta v, \ldots, 1 - \Delta v$ \textbf{do}:
\begin{enumerate}
\item[i.] \textbf{Compute} $\widehat{\FDP}(v, T, L)$ as in~\eqref{eq: FDP hat}.
\item[ii.] \textbf{If} $\widehat{\FDP}(v, T, L) \leq \alpha$
\begin{enumerate}
\item[] \textbf{Compute} $\widehat{\A}_{L}(v, T)$ as in~\eqref{eq: selected active set}.
\end{enumerate}
\textbf{Else}
\begin{enumerate}
\item[] \textbf{Set} $\widehat{\A}_{L}(v, T) = \varnothing$.
\end{enumerate}
\end{enumerate}
\item[5.2.] \textbf{Set} $T \leftarrow T + 1$.
\end{enumerate}
\label{algorithm: Extended T-Rex Step 5}
\item \textbf{Solve}
\begin{alignat*}{2}
& \max_{v^{\prime}, T^{\prime}} &{}& \, \big| \widehat{\A}_{L}(v^{\prime}, T^{\prime}) \big| \\
& \text{s.t.} &{}& T^{\prime} \in \lbrace 1, \ldots, T - 1 \rbrace \\
&{}&{}& v^{\prime} \in \lbrace  0.5, 0.5 + \Delta v, 0.5 + 2 \cdot \Delta v, \ldots, 1 - \Delta v \rbrace
\end{alignat*}
and let $(v^{*}, T^{*})$ be a solution.
\label{algorithm: Extended T-Rex Step 6}
\item \textbf{Output}: $(v^{*}, T^{*})$ and $\widehat{\A}_{L}(v^{*}, T^{*})$.
\label{algorithm: Extended T-Rex Step 7}
\end{alglist}
\label{algorithm: Extended T-Rex}
\end{algorithm}

Note that the extension to Algorithm~\ref{algorithm: T-Rex} lies in Step~\ref{algorithm: Extended T-Rex Step 2} and Step~\ref{algorithm: Extended T-Rex Step 3}. Additionally, and in contrast to Algorithm~\ref{algorithm: T-Rex}, the input to the algorithm is extended by a reference voting level $\tilde{v} \in [0.5, 1)$ and the maximum values of $L$ and $T$, namely $L_{\max}$ and $T_{\max}$. The algorithm flow is as follows: First $L$ and $T$ are set as follows: $L = p$ and $T = 1$. Then, starting at $L = p$ the number of dummies $L$ is iteratively increased in steps of $p$ until the estimate of the FDP at the voting level $\tilde{v}$ falls below the target FDR level $\alpha$ or $L$ exceeds $L_{\max}$. The rest of the algorithm is as in Algorithm~\ref{algorithm: T-Rex} except that the loop in Step~\ref{algorithm: Extended T-Rex Step 5} is exited when $T$ exceeds $T_{\max}$. 

What remains to be discussed are the choices of the hyperparameters $\tilde{v}$, $L_{\max}$, and $T_{\max}$. Throughout this paper, we have set $\tilde{v} = 0.75$, $L_{\max} = 10p$, and $T_{\max} = \lceil n / 2 \rceil$, where $\lceil n / 2 \rceil$ denotes the smallest integer that is equal to or larger than $n / 2$. An explanation and a discussion of these choices are deferred to Appendix~\ref{sec: The Deflated Relative Occurrence} in the supplementary materials.

\section{Numerical Simulations}
\label{sec: Numerical Simulations}
In this section, the performances of the proposed \textit{T-Rex} selector and the benchmark methods are compared in a simulation study. The benchmark methods in low-dimensional settings (i.e., $p \leq n$) are the well-known Benjamini-Hochberg (\textit{BH}) method~\cite{benjamini1995controlling}, the Benjamini-Yekutieli (\textit{BY}) method~\cite{benjamini2001control}, and the \textit{fixed-X} knockoff methods~\cite{barber2015controlling}, while the \textit{model-X} knockoff methods~\cite{candes2018panning} are the benchmarks in high-dimensional settings (i.e., $p > n$). Knockoff methods come in two variations called ``knockoff'' and ``knockoff+''. Only the ``knockoff+'' version is an FDR controlling method. For a detailed explanation and discussion of the benchmark methods, the reader is referred to Appendix~\ref{sec: Benchmark Methods for FDR Control} in the supplementary materials.

\begin{figure*}[!htbp]
  \centering
  \hspace{-1.5em}
  \subfloat[Average FDP and $\widehat{\FDP}$ ($L = p$).]{
  		\scalebox{0.94}{
  			\includegraphics[width=0.33\linewidth]{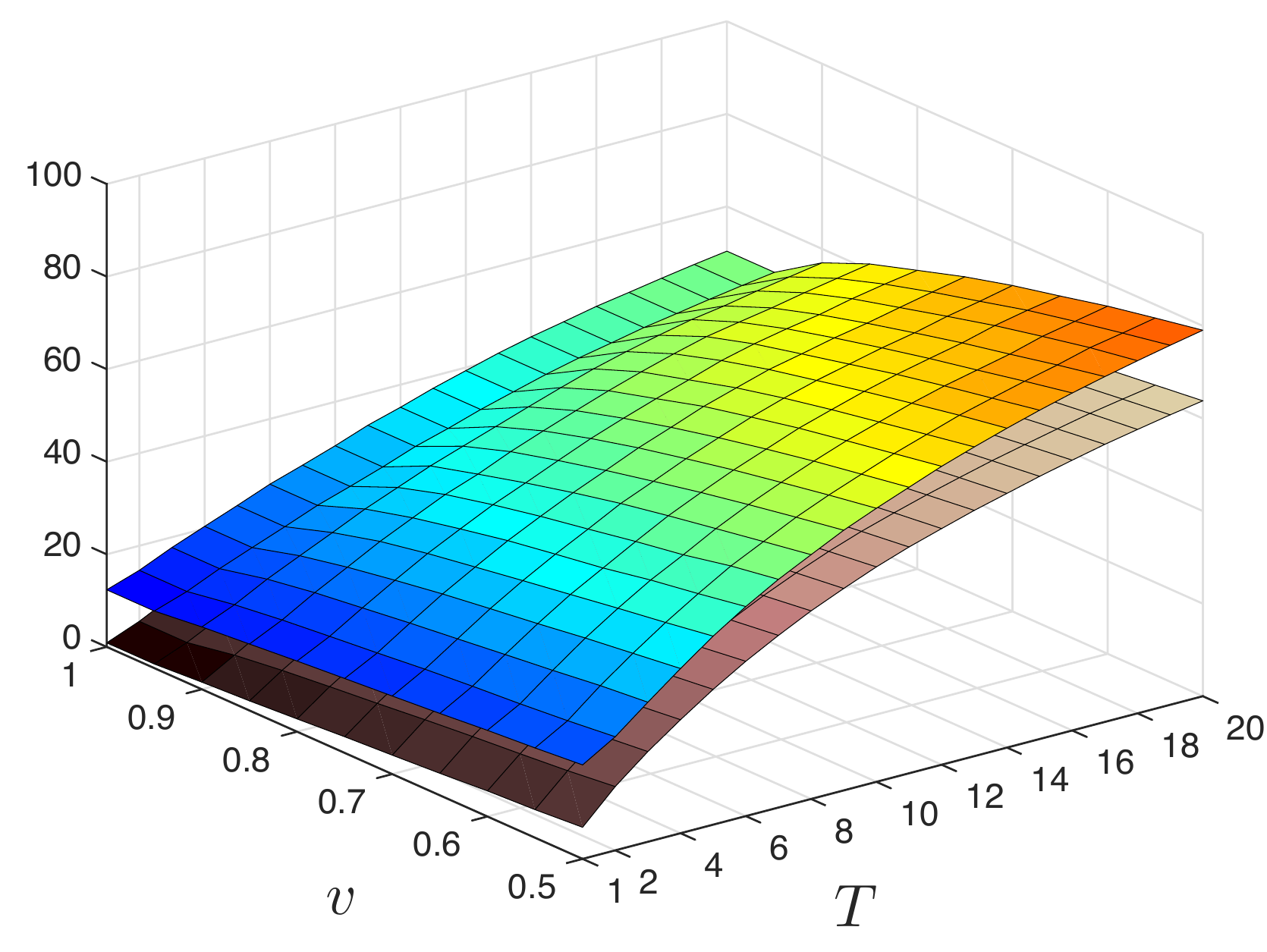}
  		}
   		\label{fig: FDP_L_p}
   }
   \hspace*{-1.6em}
  \subfloat[Average FDP and $\widehat{\FDP}$ ($L = 3p$).]{
  		\scalebox{0.94}{
  			\includegraphics[width=0.33\linewidth]{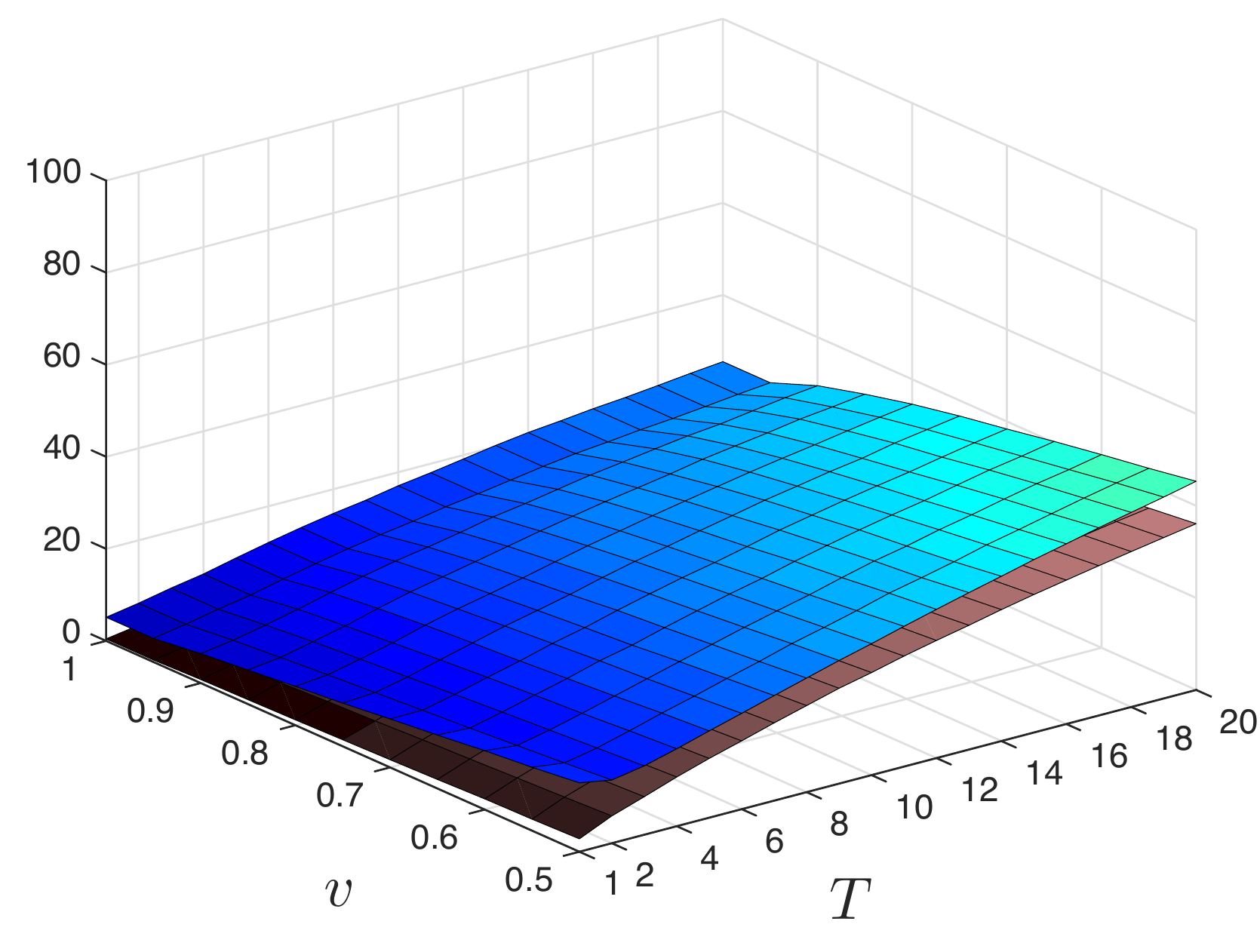}
  		}
   		\label{fig: FDP_L_3p}
   }
   \hspace*{-1.6em}
  \subfloat[Average FDP and $\widehat{\FDP}$ ($L = 5p$).]{
  		\scalebox{0.94}{
  			\includegraphics[width=0.4\linewidth]{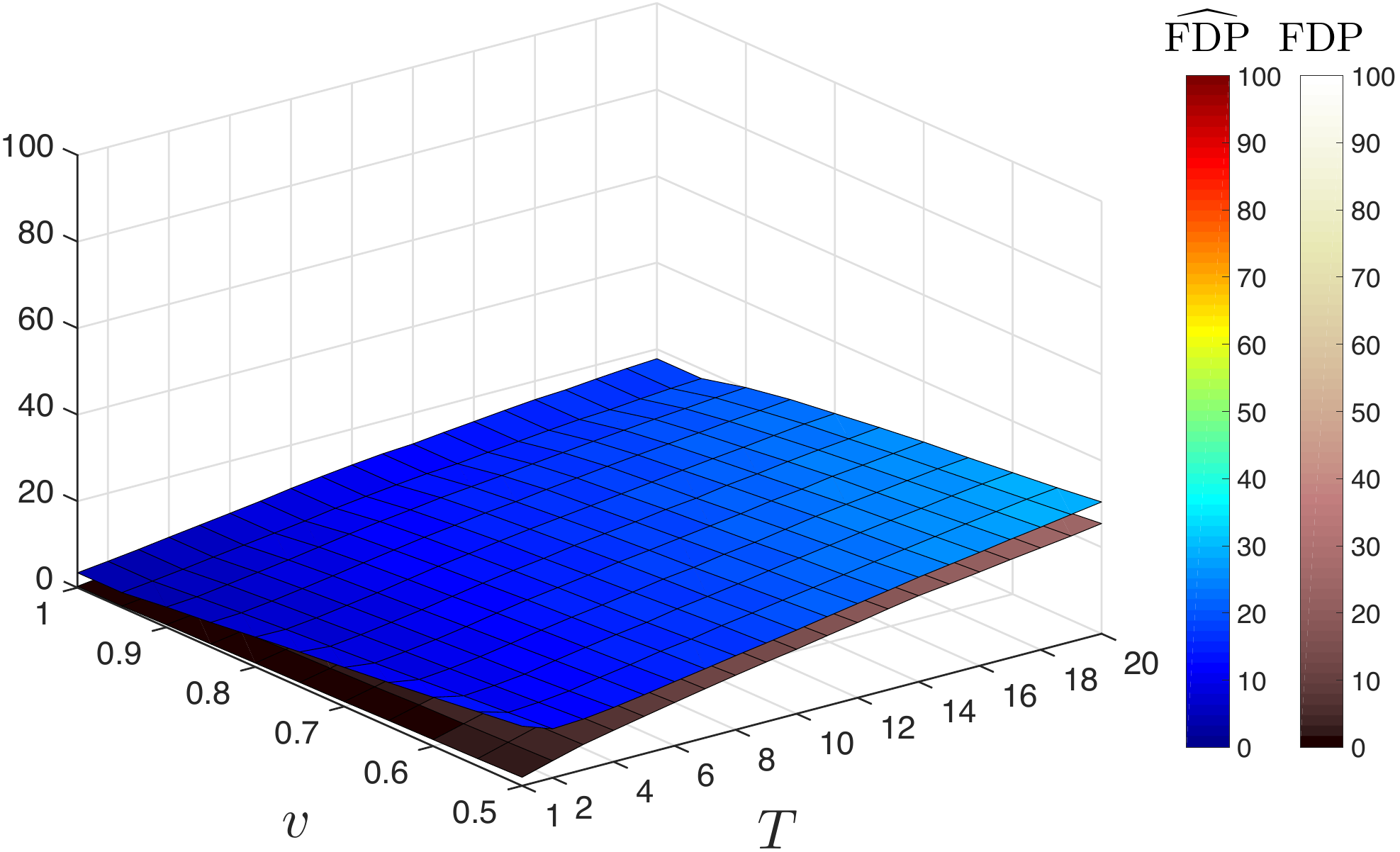}
  		}
   		\label{fig: FDP_L_5p}
   }
   \\
   \hspace{-1.5em}
  \subfloat[Average TPP ($L = p$).]{
  		\scalebox{0.95}{
  			\includegraphics[width=0.33\linewidth]{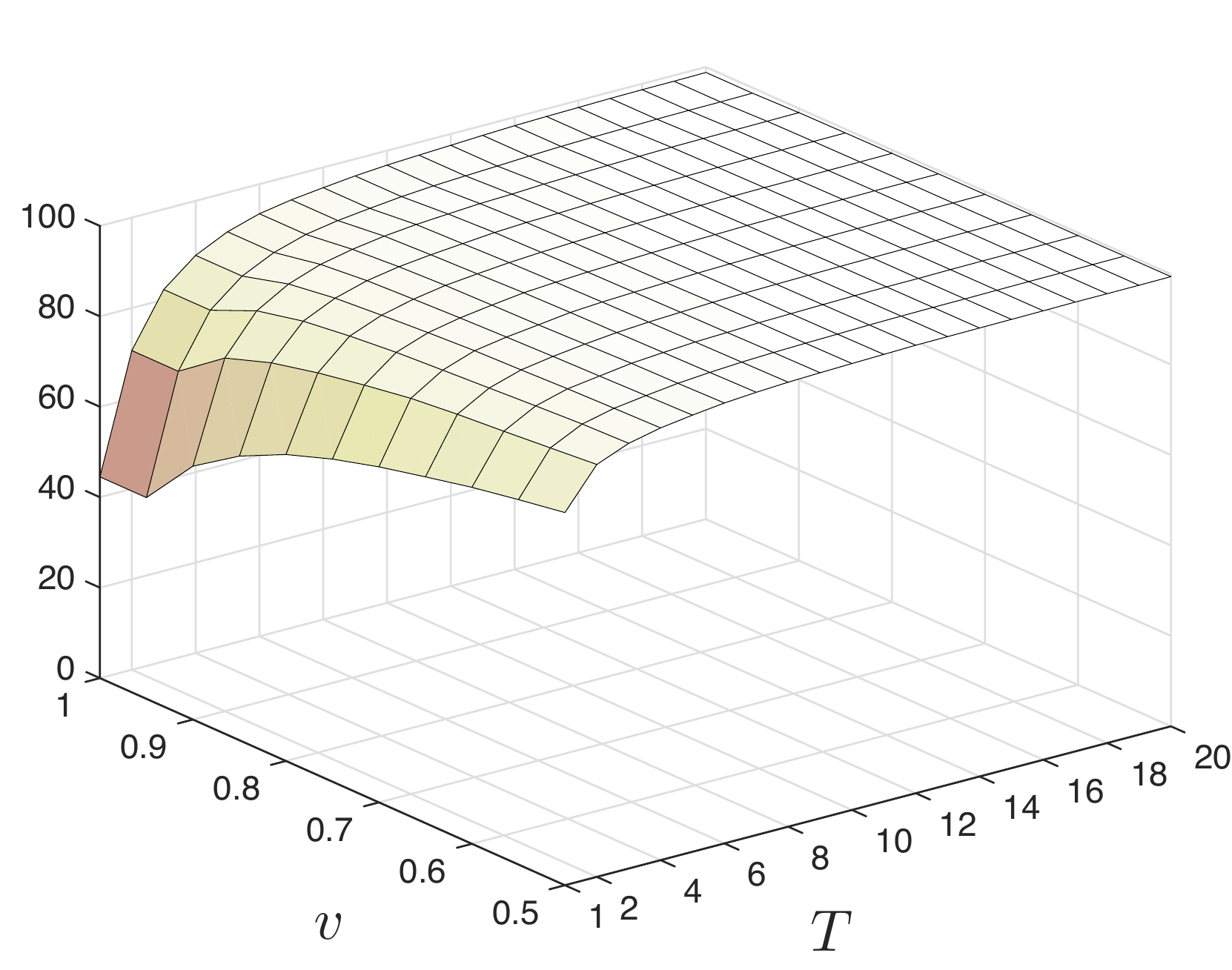}
  		}
   		\label{fig: FDP_L_p}
   }
   \hspace*{-1.5em}
  \subfloat[Average TPP ($L = 3p$).]{
  		\scalebox{0.95}{
  			\includegraphics[width=0.33\linewidth]{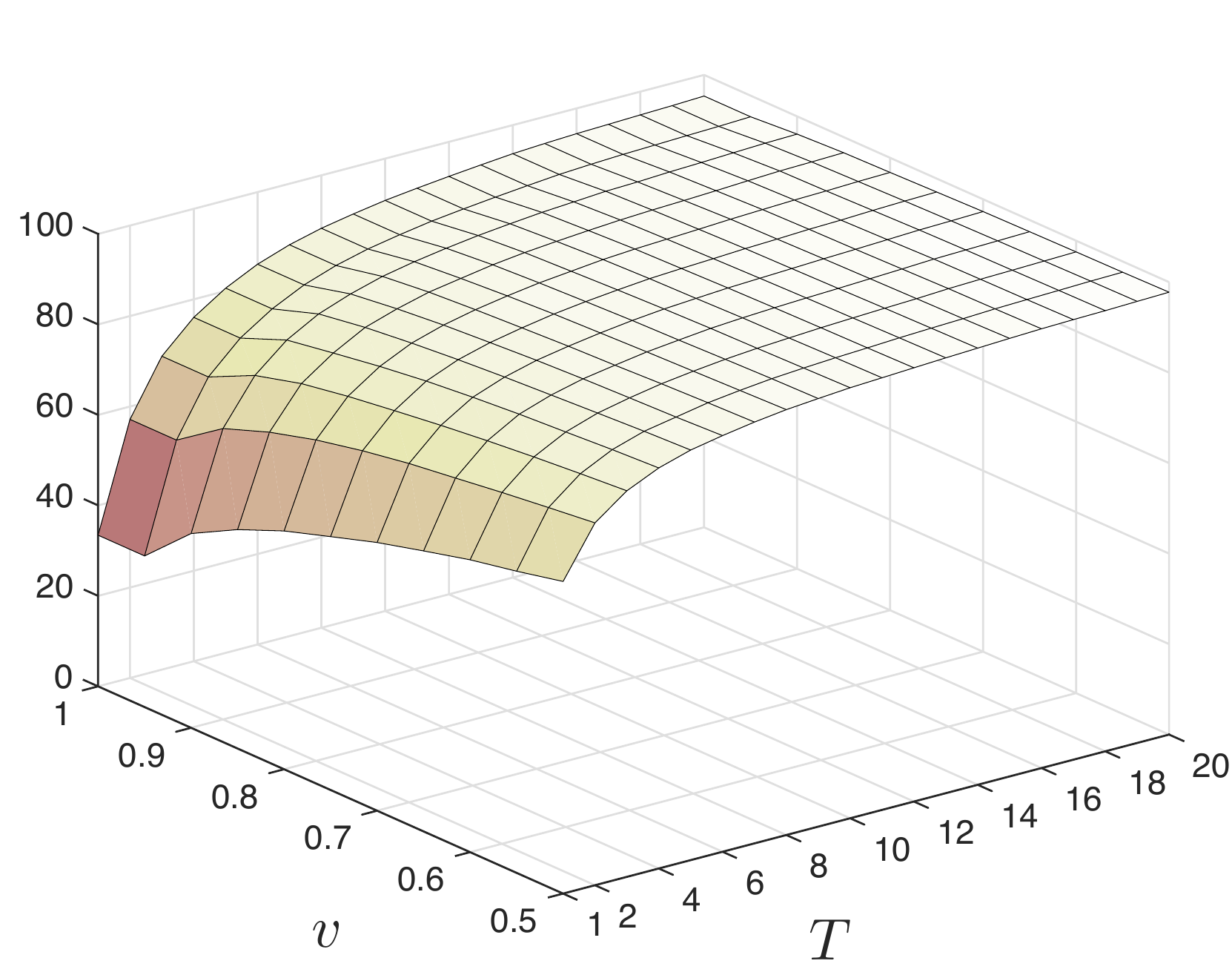}
  		}
   		\label{fig: TPP_L_3p}
   }
   \hspace*{-1.5em}
  \subfloat[Average TPP ($L = 5p$).]{
  		\scalebox{0.95}{
  			\includegraphics[width=0.355\linewidth]{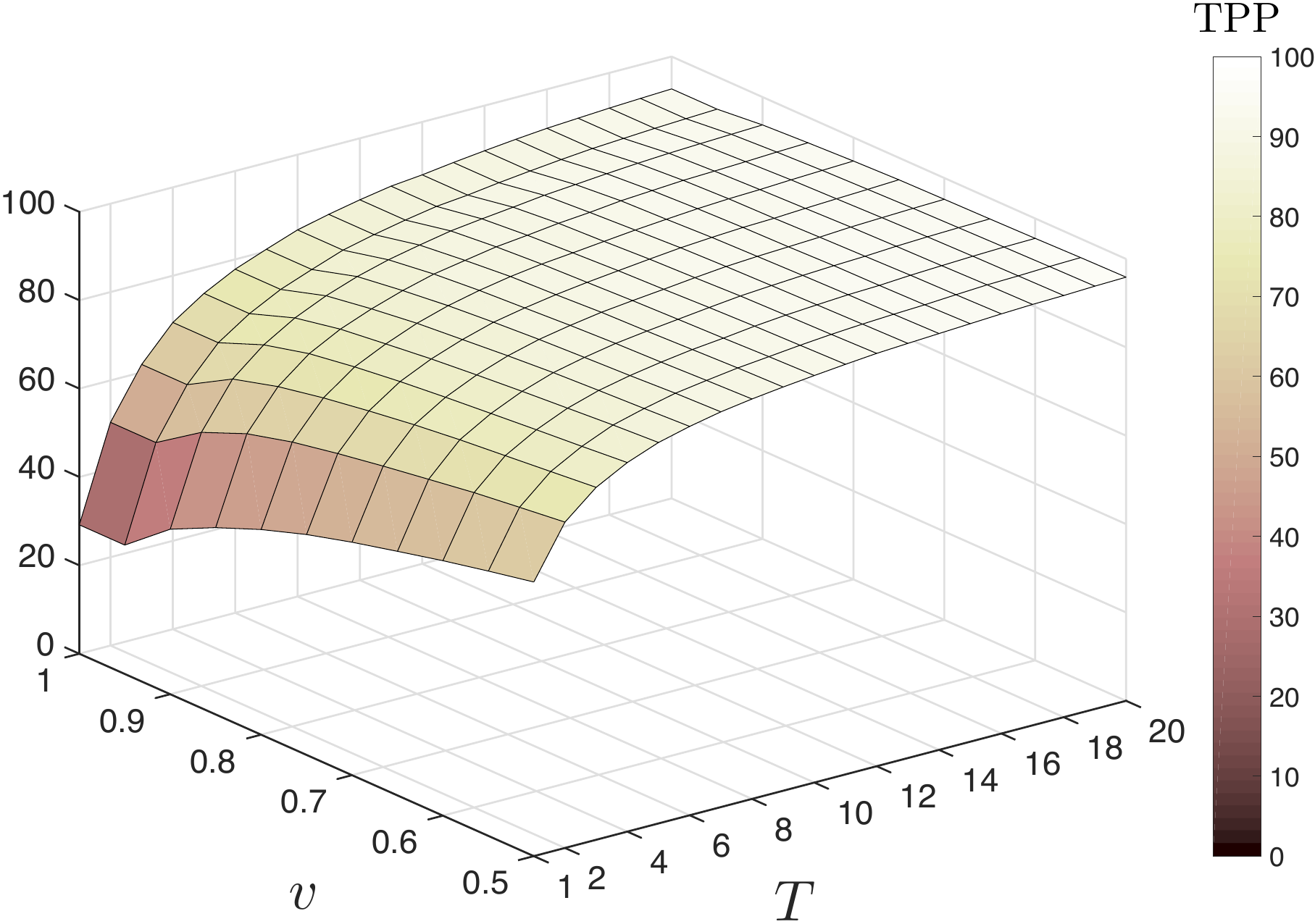}
  		}
   		\label{fig: TPP_L_5p}
   }
  \caption{The \textit{T-Rex} selector controls the FDR for all values of $v$ and $T$ while achieving a high power, even at low values of $T$. Note that the FDR control is tighter for large values of $L$. This observation led to the development of Algorithm~\ref{algorithm: Extended T-Rex}. Moreover, we observe that the conditions in Theorem~\ref{theorem: T-Rex algorithm} hold on average (i.e., ceteris paribus, $\widehat{\FDP}(v, T, L)$ is monotonically decreasing in $v$ and monotonically increasing in $T$). Setup: $n = 300$, $p = 1{,}000$, $p_{1} = 10$, $K = 20$, $\text{SNR} = 1$, $MC = 955$.}
  \label{fig: 3d-plots (FDP, FDP_hat and TPP)}
\end{figure*}
%
\begin{figure*}[!htbp]
  \centering
	\hspace*{-0.8em}
  \subfloat[]{
  		\scalebox{1}{
  			\includegraphics[width=0.24\linewidth]{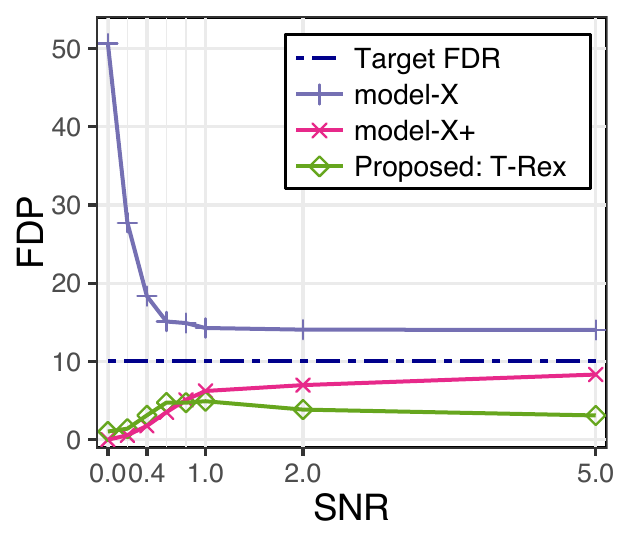}
  		}
   		\label{fig: FDP_vs_SNR_p_1000_Optimal_T_L}
   }
	\hspace*{-1.4em}
  \subfloat[]{
  		\scalebox{1}{
  			\includegraphics[width=0.245\linewidth]{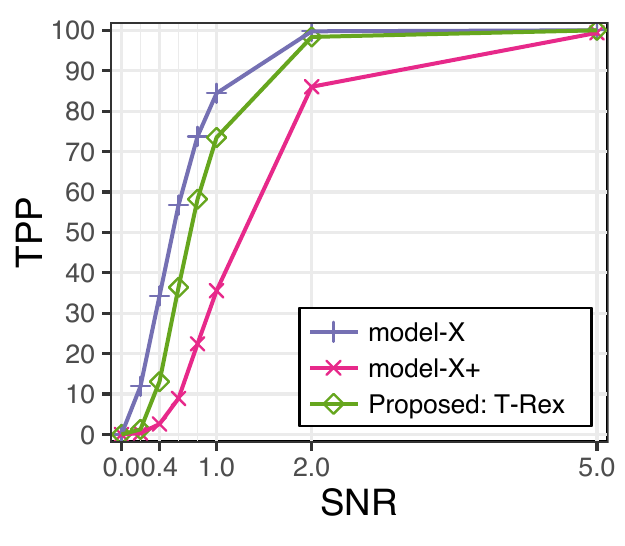}
  		}
   		\label{fig: TPP_vs_SNR_p_1000_Optimal_T_L}
   }
	\hspace*{-1.4em}
    \subfloat[]{
  		\scalebox{1}{
  			\includegraphics[width=0.24\linewidth]{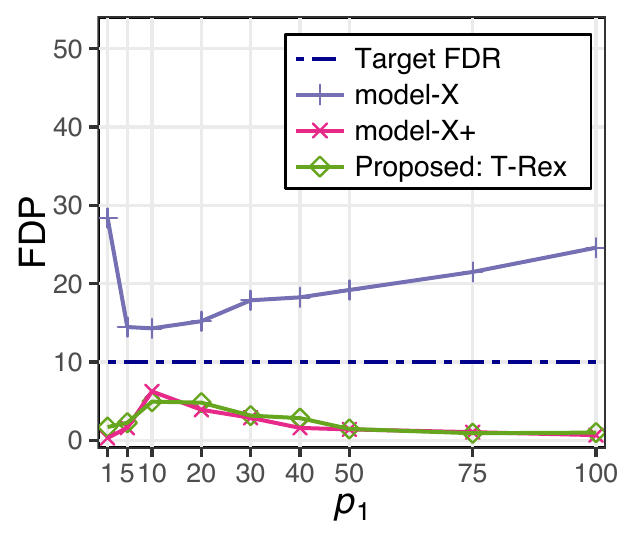}
  		}
   		\label{fig: FDP_vs_p1_p_1000_Optimal_T_L}
   }
	\hspace*{-1.4em}
  \subfloat[]{
  		\scalebox{1}{
  			\includegraphics[width=0.244\linewidth]{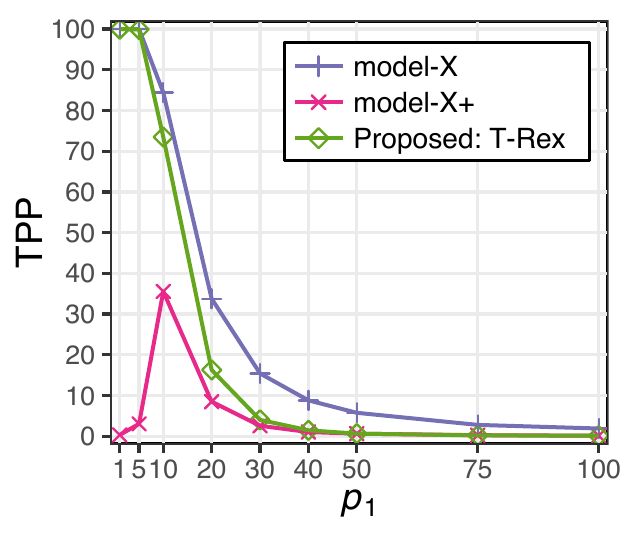}
  		}
   		\label{fig: TPP_vs_p1_p_1000_Optimal_T_L}
   }
  \caption{\textbf{General}: The \textit{model-X} knockoff method fails to control the FDR. Among the FDR-controlling methods, the \textit{T-Rex} selector outperforms the \textit{model-X} knockoff+ method in terms of power. \textbf{Details}: \textbf{(a)}~The \textit{T-Rex} selector and the \textit{model-X} knockoff+ method control the FDR at a target level of $10\%$ for the whole range of SNR values while the \textit{model-X} knockoff method fails to control the FDR and performs poorly at low SNR values. Setup: $n = 300$, $p = 1{,}000$, $p_{1} = 10$, $T_{\max} = \lceil n/2 \rceil$, $L_{\max} = 10p$, $K = 20$, $MC = 955$. \textbf{(b)}~As expected, the TPR (i.e., power) increases with respect to the SNR. It is remarkable that even though the FDR of the \textit{T-Rex} selector lies below that of the \textit{model-X} knockoff+ method for SNR values larger than $0.6$, its power exceeds that of its strongest FDR-controlling competitor. The high power of the \textit{model-X} knockoff method cannot be interpreted as an advantage, because the method does not control the FDR. Setup: Same as in Figure~(a). \textbf{(c)}~As in Figure~(a), only the \textit{T-Rex} selector and the \textit{model-X} knockoff+ method control the FDR at a target level of $10\%$, whereas the \textit{model-X} knockoff method always exceeds the target level. Setup: $n = 300$, $p = 1{,}000$, $T_{\max} = \lceil n/2 \rceil$, $L_{\max} = 10p$, $K = 20$, $\text{SNR} = 1$, $MC = 955$. \textbf{(d)}~Among the FDR-controlling methods, the \textit{T-Rex} selector has by far the highest power for sparse settings. The power of the \textit{model-X} knockoff method exceeds that of the FDR-controlling methods, but this cannot be interpreted as an advantage of the method since it exceeds the target FDR level. Note that for an increasing number of active variables the power drops for all methods since apparently the number of data points $n = 300$ does not suffice in the simulated settings with a low sparsity level, i.e., settings with many active variables. Setup: Same as in Figure~(c).}
  \label{fig: sweep p1 and sweep snr plots p = 1000}
\end{figure*}
%
\begin{figure}[!htbp]
  \centering
	\hspace*{-0.8em}
  \subfloat[]{
  		\scalebox{1}{
  			\includegraphics[width=0.49\linewidth]{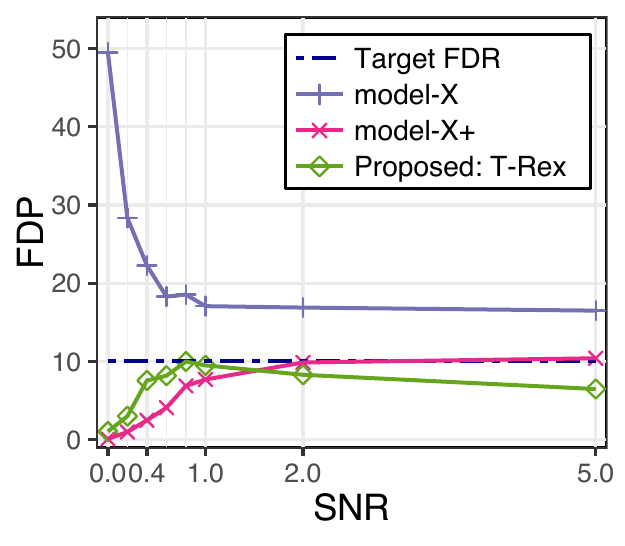}
  		}
   		\label{fig: FDP_vs_SNR_p_1000_Optimal_T_L_AR_1}
   }
	\hspace*{-1.4em}
  \subfloat[]{
  		\scalebox{1}{
  			\includegraphics[width=0.49\linewidth]{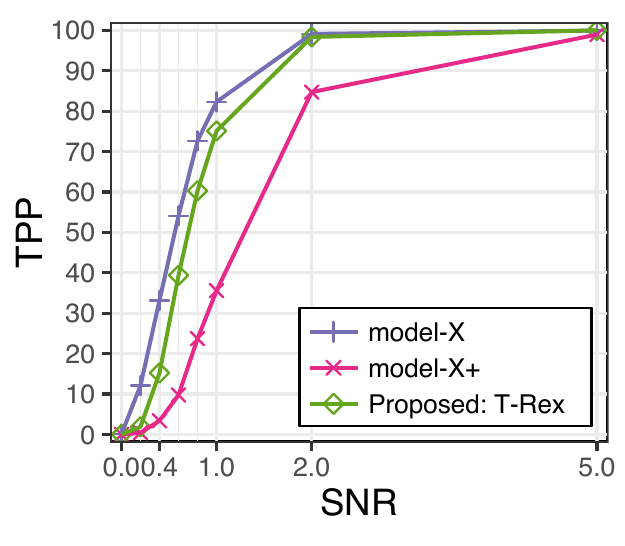}
  		}
   		\label{fig: TPP_vs_SNR_p_1000_Optimal_T_L_AR_1}
   }
  \caption{\textbf{Average FDP and TPP in the case of dependent predictors}: The \textit{T-Rex} selector controls the FDR, has the highest power among the FDR-controlling methods, and reaches the almost highest possible TPR level at an SNR of $2$ while the \textit{model-X} knockoff+ method requires an SNR of $5$ to reach the same TPR level.  The \textit{model-X} knockoff+ method also controls the FDR except for an SNR of $5$, where it slightly exceeds the target FDR, and the \textit{model-X} knockoff method does not control the FDR. The predictors were sampled from an autoregressive model of order one (AR($1$)) with Gaussian noise and an autocorrelation coefficient $\rho = 0.5$. Setup: $n = 300$, $p = 1{,}000$, $p_{1} = 10$, $T_{\max} = \lceil n/2 \rceil$, $L_{\max} = 10p$, $K = 20$, $MC = 955$.}
  \label{fig: sweep snr plots p = 1000 _Optimal_T_L_AR_1}
\end{figure}
%

\subsection{Setup and Results}
\label{subsec: Setup and Results}
We generate a sparse high-dimensional setting\footnote{Additional simulation results that allow for a performance comparison of the proposed \textit{T-Rex} selector to the \textit{BH} method, the \textit{BY} method, and the \textit{fixed-X} knockoff methods in a low-dimensional setting are deferred to Appendix~\ref{sec: Additional Simulation Results} in the supplementary materials.} with $n$ observations, $p$ predictors, and a response given by the linear model in~\eqref{eq: linear model}. Further, $\beta_{j} = 1$ for $p_{1}$ randomly selected $j$'s while $\beta_{j} = 0$ for the others. The predictors are (i) sampled independently from the standard normal distribution (Figures~\ref{fig: 3d-plots (FDP, FDP_hat and TPP)} and~\ref{fig: sweep p1 and sweep snr plots p = 1000}) and (ii) sampled from an autoregressive model of order one with autocorrelation coefficient $\rho = 0.5$ (Figure~\ref{fig: sweep snr plots p = 1000 _Optimal_T_L_AR_1}). The standard deviation of the noise~$\sigma$ is chosen such that the signal-to-noise ratio (SNR), which is given by $\Var[ \X\bbeta] \, / \, \sigma^{2}$, is equal to the desired value. In Appendices~\ref{sec: Illustration of Theorem 2 (Dummy Generation)} and~\ref{sec: Robustness of The T-Rex Selector} of the supplementary materials, we show results for non-Gaussian predictors and heavy-tailed noise settings. The specific values of the above described simulation setting and the parameters of the \textit{T-Rex} selector, i.e., the values of $n$, $p$, $p_{1}$, $\text{SNR}$, $K$, $L$, $T$, $v$, are specified in the figure captions. The results are averaged over $MC = 955$ Monte Carlo replications.\footnote{The reason for running $955$ Monte Carlo replications is that the simulations were conducted on the Lichtenberg High-Performance Computer of the Technische Universität Darmstadt, which consists of multiple nodes of $96$ CPUs each. In order to run computationally efficient simulations, our computation jobs are designed to request $2$ nodes and run $5$ cycles on each CPU while one CPU acts as the master, i.e., $(2 \cdot 96 - 1) \cdot 5 = 955$.}

First, in order to assess the FDR control performance and the achieved power of the \textit{T-Rex} selector, respectively, the average FDP, $\widehat{\FDP}$, and TPP are computed over a two-dimensional grid for $v$ and $T$ for different values of $L$. Then, leaving all other parameters in this setup fixed, we compare the performance of the proposed \textit{T-Rex} selector in combination with the proposed extended calibration algorithm in Algorithm~\ref{algorithm: Extended T-Rex} with the benchmark methods for different values of $p_{1}$ and the SNR at a target FDR level of $10\%$.

The reported average FDP, $\widehat{\FDP}$, and TPP (all averaged over $955$ Monte Carlo replications) in Figures~\ref{fig: 3d-plots (FDP, FDP_hat and TPP)},~\ref{fig: sweep p1 and sweep snr plots p = 1000}, and~\ref{fig: sweep snr plots p = 1000 _Optimal_T_L_AR_1} are estimates of the FDR, $\widehat{\FDR}$, and TPR, respectively. For this reason, the results are discussed in terms of the FDR, $\widehat{\FDR}$, and TPR in the captions of the figures, while the axes labels emphasize that the average FDP, $\widehat{\FDP}$, and TPP are plotted.

The simulation results confirm that the proposed \textit{T-Rex} selector possesses the FDR control property. Moreover, the simulation results show that the \textit{T-Rex} selector outperforms the benchmark methods and that its computation time is multiple orders of magnitude lower than that of its competitors (see Figure~\ref{fig: computation time sweep p five million} in Section~\ref{sec: Introduction and Motivation} and Table~\ref{table: results simulated GWAS}). The detailed descriptions and discussions of the simulation results are given in the captions of Figures~\ref{fig: 3d-plots (FDP, FDP_hat and TPP)},~\ref{fig: sweep p1 and sweep snr plots p = 1000}, and~\ref{fig: sweep snr plots p = 1000 _Optimal_T_L_AR_1}. Furthermore, Appendix~\ref{sec: Robustness of The T-Rex Selector} in the supplementary materials discusses in more detail the robustness of the T-Rex selector in the presence of non-Gaussian heavy-tailed noise.
\begin{table*}[!htbp]
\centering
\begin{tabular}{@{}L{\dimexpr0.18\linewidth-2\tabcolsep\relax}@{}C{\dimexpr0.12\linewidth-2\tabcolsep\relax}@{}C{\dimexpr0.15\linewidth-2\tabcolsep\relax}@{}C{\dimexpr0.15\linewidth-2\tabcolsep\relax}@{}C{\dimexpr0.26\linewidth-2\tabcolsep\relax}@{}R{\dimexpr0.26\linewidth-2\tabcolsep\relax}}
\bfseries Methods
& \bfseries FDR control?
& \bfseries Average FDP (in~$\%$)
& \bfseries Average TPP (in~$\%$)
& \bfseries Average sequential com- putation time (hh:mm:ss)
& \bfseries Average relative sequential computation time \\
\midrule
		\textbf{Proposed: \textit{T-Rex}} & \cmark & $\boldsymbol{6.45}$ & $\boldsymbol{38.50}$ & $\boldsymbol{00$:$04$:$05}$ & $\boldsymbol{1}$ \\[0.0em]
		\textit{model-X+} & \cmark & $0.00$ & $0.00$ & $12$:$32$:$47$ & $183.71$ \\[0.0em]
		\textit{model-X} & \xmark & $13.07$ & $41.40$ & $12$:$32$:$47$ & $183.71$ \\[0.0em]
		\textit{BY} & \xmark & $94.00$ & $0.00$ & $00$:$00$:$00$ & $0.00$ \\[0.0em]
		\textit{BH} & \xmark & $99.00$ & $0.00$ & $00$:$00$:$00$ & $0.00$ \\
\end{tabular}
\caption{The proposed \textit{T-Rex} selector is the only method whose average FDP lies below the target FDR level of $10\%$ while achieving a non-zero power. The only competitor that provably possesses the FDR control property, namely the \textit{model-X} knockoff+ method, has an average FDP of $0\%$ but also an average TPP of $0\%$, i.e., it has no power. The \textit{model-X} knockoff method exceeds the target FDR level. The computationally cheap procedure of plugging the marginal $p$-values into the \textit{BH} method or the \textit{BY} method, which has been a standard procedure in GWAS, fails in this high-dimensional setting. The sequential computation time of the proposed \textit{T-Rex} selector in combination with the extended calibration algorithm in Algorithm~\ref{algorithm: Extended T-Rex} is roughly $4$ minutes as compared to more than $12.5$ hours for the \textit{model-X} methods. That is, the \textit{T-Rex} selector is $183$ times faster than its strongest competitors. Note that this is only a comparison of the sequential computation times. Since the random experiments of the proposed \textit{T-Rex} selector are independent and, therefore, can be run in parallel on multicore computers, an additional substantial speedup can be achieved.}
\label{table: results simulated GWAS}
\end{table*}

\section{Simulated Genome-Wide Association Study}
\label{sec: Simulated Genome-Wide Association Study}
The \textit{T-Rex} selector and the benchmark methods are applied to conduct a high-dimensional simulated case-control GWAS. The size of the GWAS was chosen, such that it was still practically feasible to compute the computationally intensive benchmark methods. The goal is to detect the single nucleotide polymorphisms (SNPs) that are associated with a disease of interest (i.e., active variables), while keeping the number of selected SNPs that are not associated with that disease (i.e., null variables) low.

\subsection{Setup}
\label{subsec: GWAS Setup}
The genotypes of $700$ cases and $300$ controls are simulated based on haplotypes from phase $3$ of the  International HapMap project~\cite{international2010integrating} using the software HAPGEN2~\cite{su2011hapgen2}. We simulated $10$ randomly selected disease loci on the first $20{,}000$ SNPs of chromosome $15$ (contains $42{,}351$ SNPs in total) with randomly selected risk alleles (either $0$ or $1$ with $\mathbb{P}(\text{``$0$''}) = \mathbb{P}(\text{``$1$''}) = 0.5$) and with the heterozygote risks and the homozygote risks being sampled from the uniform distribution on the intervals $[1.5, 2]$ and $[2.5, 3]$, respectively. Since we are conducting a case-control study, the control and case phenotypes are $0$ and $1$, respectively. Note that the SNPs and the phenotype represent the candidate variables and the response, respectively, while the disease loci represent the indices of the active variables. Thus, we have $p_{1} = 10$ active variables and $p_{0} = 19{,}990$ null variables. The number of observations is $n = 1{,}000$ ($700$ cases and $300$ controls). The results are averaged over $100$ data sets satisfying the above specifications. The detailed description of the setup and the preprocessing of the data is deferred to Appendix~\ref{sec: Setup, Preprocessing, and Additional Results: Simulated Genome-Wide Association Study} in the supplementary materials.

\subsection{Results}
\label{subsec: GWAS Results}
In this HAPGEN2-based GWAS benchmarking, the \textit{T-Rex} selector demonstrates its real-life applicability, as it  is the only FDR-controlling method with a positive TPR, and its sequential computation time is $4$ minutes (vs. more than $12$ hours for the knockoff methods). The results (i.e., FDR, TPR, and sequential computation time) and a discussion thereof are given in Table~\ref{table: results simulated GWAS}, while additional results are deferred to Appendix~\ref{sec: Setup, Preprocessing, and Additional Results: Simulated Genome-Wide Association Study} in the supplementary materials.

\section{Conclusion}
\label{sec: Conclusion}
The \textit{T-Rex} selector, a new fast FDR-controlling variable selection framework for high-dimensional settings, was proposed and benchmarked against existing methods in numerical simulations and a simulated GWAS. The \textit{T-Rex} selector is, to the best of our knowledge, the first multivariate high-dimensional FDR-controlling method that scales to millions of variables in a reasonable amount of computation time. Since the \textit{T-Rex} random experiments can be computed in parallel, multicore computers allow for additional substantial savings in computation time. These properties make the \textit{T-Rex} selector a suitable method especially for large-scale GWAS.

For other use-cases (e.g., high-dimensional survival analysis, sparse financial index tracking), where strong dependencies among the variables (e.g., gene expression levels, stock returns) exist and common SNP pruning or other preprocessing techniques are not applicable, the dependency-aware \textit{T-Rex} (\textit{T-Rex+DA}) selector has been proposed~\cite{machkour2024dependency}. The \textit{T-Rex+DA} selector performs high-dimensional FDR-controlled variable selection in the presence of strong dependencies at the cost of a reduced power compared to the \textit{T-Rex} selector.

Moreover, the \textit{T-Rex} selector has already proven to be a useful and versatile framework for screening large-scale genomics biobanks~\cite{machkour2023ScreenTRex}, efficient computation in big data applications~\cite{scheidt2023FDRControlLaptop}, grouped variable selection~\cite{machkour2022TRexGVS,machkour2023InformedEN}, Gaussian graphical models~\cite{koka2024false}, sparse principal component analysis~\cite{machkour2024TRexPCA}, sparse financial index tracking~\cite{machkour2024TRexIndexTracking}, and survival analysis~\cite{machkour2024dependency}.

In order to ensure the reproducibility of the presented results and enhance the usability of the proposed \textit{T-Rex} selector, the actively maintained open-source R software package TRexSelector has been made available on CRAN~\cite{machkour2022TRexSelector}.

As a next step, we shall conduct multiple reproducibility studies applying the \textit{T-Rex} selector on large-scale genotype and phenotype data from the UK Biobank~\cite{sudlow2015uk} in order to reproduce some of the reported results in the GWAS catalog~\cite{gwasCatalog}. Our aim is to confirm past discoveries, discover new genetic associations, and flag potentially false reported genetic associations. We plan to publish our results as a curated catalog of reproducible genetic associations and hope that this endeavor helps scientists to focus their efforts in revealing the causal mechanisms behind the genetic associations on the most promising and reproducible genetic associations.

\section*{Acknowledgments}
We thank Michael Fauss, Taulant Koka, and Fabian Scheidt for many discussions and helpful feedback. We also thank Simon Tien for his help in developing the R software packages \textit{TRexSelector} and \textit{tlars}. Extensive computations on the Lichtenberg High-Performance Computer of the Technische Universität Darmstadt were conducted for this research.

\bibliographystyle{IEEEtran_newLineInBibURL}
\IEEEtriggeratref{45}
\bibliography{bibliography}

\begin{thebibliography}{10}
\providecommand{\url}[1]{#1}
\csname url@samestyle\endcsname
\providecommand{\newblock}{\relax}
\providecommand{\bibinfo}[2]{#2}
\providecommand{\BIBentrySTDinterwordspacing}{\spaceskip=0pt\relax}
\providecommand{\BIBentryALTinterwordstretchfactor}{4}
\providecommand{\BIBentryALTinterwordspacing}{\spaceskip=\fontdimen2\font plus
\BIBentryALTinterwordstretchfactor\fontdimen3\font minus
  \fontdimen4\font\relax}
\providecommand{\BIBforeignlanguage}[2]{{%
\expandafter\ifx\csname l@#1\endcsname\relax
\typeout{** WARNING: IEEEtran.bst: No hyphenation pattern has been}%
\typeout{** loaded for the language `#1'. Using the pattern for}%
\typeout{** the default language instead.}%
\else
\language=\csname l@#1\endcsname
\fi
#2}}
\providecommand{\BIBdecl}{\relax}
\BIBdecl

\bibitem{chung2007detection}
P.-J. Chung, J.~F. Bohme, C.~F. Mecklenbrauker, and A.~O. Hero, ``Detection of
  the number of signals using the benjamini-hochberg procedure,'' \emph{IEEE
  Trans. Signal Process.}, vol.~55, no.~6, pp. 2497--2508, 2007.

\bibitem{chen2020false}
J.~Chen, W.~Zhang, and H.~V. Poor, ``A false discovery rate oriented approach
  to parallel sequential change detection problems,'' \emph{IEEE Trans. Signal
  Process.}, vol.~68, pp. 1823--1836, 2020.

\bibitem{chen2018sparse}
Z.~Chen, F.~Sohrabi, and W.~Yu, ``Sparse activity detection for massive
  connectivity,'' \emph{IEEE Trans. Signal Process.}, vol.~66, no.~7, pp.
  1890--1904, 2018.

\bibitem{tan2014direction}
Z.~Tan, Y.~C. Eldar, and A.~Nehorai, ``Direction of arrival estimation using
  co-prime arrays: A super resolution viewpoint,'' \emph{IEEE Trans. Signal
  Process.}, vol.~62, no.~21, pp. 5565--5576, 2014.

\bibitem{di2012sparse}
P.~Di~Lorenzo and A.~H. Sayed, ``Sparse distributed learning based on diffusion
  adaptation,'' \emph{IEEE Trans. Signal Process.}, vol.~61, no.~6, pp.
  1419--1433, 2012.

\bibitem{benidis2017sparse}
K.~Benidis, Y.~Feng, and D.~P. Palomar, ``Sparse portfolios for
  high-dimensional financial index tracking,'' \emph{IEEE Trans. Signal
  Process.}, vol.~66, no.~1, pp. 155--170, 2017.

\bibitem{zoubir2012robust}
A.~M. Zoubir, V.~Koivunen, Y.~Chakhchoukh, and M.~Muma, ``Robust estimation in
  signal processing: A tutorial-style treatment of fundamental concepts,''
  \emph{IEEE Signal Process. Mag.}, vol.~29, no.~4, pp. 61--80, 2012.

\bibitem{zoubir2018robust}
A.~M. Zoubir, V.~Koivunen, E.~Ollila, and M.~Muma, \emph{Robust statistics for
  signal processing}.\hskip 1em plus 0.5em minus 0.4em\relax Cambridge Univ.
  Press, 2018.

\bibitem{machkour2017outlier}
J.~Machkour, B.~Alt, M.~Muma, and A.~M. Zoubir, ``The
  outlier-corrected-data-adaptive lasso: A new robust estimator for the
  independent contamination model,'' in \emph{2017 25th Eur. Signal Process.
  Conf. (EUSIPCO)}, 2017, pp. 1649--1653.

\bibitem{machkour2020robust}
J.~Machkour, M.~Muma, B.~Alt, and A.~M. Zoubir, ``A robust adaptive lasso
  estimator for the independent contamination model,'' \emph{Signal Process.},
  vol. 174, p. 107608, 2020.

\bibitem{yang2019weakly}
C.~Yang, X.~Shen, H.~Ma, B.~Chen, Y.~Gu, and H.~C. So, ``Weakly convex
  regularized robust sparse recovery methods with theoretical guarantees,''
  \emph{IEEE Trans. Signal Process.}, vol.~67, no.~19, pp. 5046--5061, 2019.

\bibitem{gwasCatalog}
A.~Buniello, J.~A.~L. MacArthur, M.~Cerezo, L.~W. Harris, J.~Hayhurst,
  C.~Malangone, A.~McMahon, J.~Morales, E.~Mountjoy, E.~Sollis \emph{et~al.},
  ``{The {NHGRI-EBI GWAS} Catalog of published genome-wide association studies,
  targeted arrays and summary statistics 2019},'' \emph{Nucleic Acids Res.},
  vol.~47, no.~D1, pp. D1005--D1012, 2019.

\bibitem{chanock2007replicating}
S.~J. Chanock, T.~Manolio, M.~Boehnke, E.~Boerwinkle, D.~J. Hunter, G.~Thomas,
  J.~N. Hirschhorn, G.~Abecasis, D.~Altshuler, J.~E. Bailey-Wilson
  \emph{et~al.}, ``Replicating genotype--phenotype associations,''
  \emph{Nature}, vol. 447, no. 7145, pp. 655--660, 2007.

\bibitem{visscher201710}
P.~M. Visscher, N.~R. Wray, Q.~Zhang, P.~Sklar, M.~I. McCarthy, M.~A. Brown,
  and J.~Yang, ``10 years of {GWAS} discovery: Biology, function, and
  translation,'' \emph{Am. J. Hum. Genet.}, vol. 101, no.~1, pp. 5--22, 2017.

\bibitem{huffman2018examining}
J.~E. Huffman, ``Examining the current standards for genetic discovery and
  replication in the era of mega-biobanks,'' \emph{Nat. Commun.}, vol.~9,
  no.~1, pp. 1--4, 2018.

\bibitem{gallagher2018post}
M.~D. Gallagher and A.~S. Chen-Plotkin, ``The post-{GWAS} era: From association
  to function,'' \emph{Am. J. Hum. Genet.}, vol. 102, no.~5, pp. 717--730,
  2018.

\bibitem{benjamini1995controlling}
Y.~Benjamini and Y.~Hochberg, ``Controlling the false discovery rate: a
  practical and powerful approach to multiple testing,'' \emph{J. R. Stat. Soc.
  Ser. B. Stat. Methodol.}, vol.~57, no.~1, pp. 289--300, 1995.

\bibitem{benjamini2001control}
Y.~Benjamini and D.~Yekutieli, ``The control of the false discovery rate in
  multiple testing under dependency,'' \emph{Ann. Statist.}, vol.~29, no.~4,
  pp. 1165--1188, 2001.

\bibitem{barber2015controlling}
R.~F. Barber and E.~J. Cand{\`e}s, ``Controlling the false discovery rate via
  knockoffs,'' \emph{Ann. Statist.}, vol.~43, no.~5, pp. 2055--2085, 2015.

\bibitem{candes2018panning}
E.~J. Cand{\`e}s, Y.~Fan, L.~Janson, and J.~Lv, ``Panning for gold:
  ‘model-{X}’ knockoffs for high dimensional controlled variable
  selection,'' \emph{J. R. Stat. Soc. Ser. B. Stat. Methodol.}, vol.~80, no.~3,
  pp. 551--577, 2018.

\bibitem{ren2021derandomizing}
Z.~Ren, Y.~Wei, and E.~Cand{\`e}s, ``Derandomizing knockoffs,'' \emph{J. Amer.
  Statist. Assoc.}, pp. 1--11, 2021.

\bibitem{ren2024derandomised}
Z.~Ren and R.~F. Barber, ``Derandomised knockoffs: leveraging e-values for
  false discovery rate control,'' \emph{J. R. Stat. Soc. Ser. B. Stat.
  Methodol.}, vol.~86, no.~1, pp. 122--154, 2024.

\bibitem{meinshausen2010stability}
N.~Meinshausen and P.~B{\"u}hlmann, ``Stability selection,'' \emph{J. R. Stat.
  Soc. Ser. B. Stat. Methodol.}, vol.~72, no.~4, pp. 417--473, 2010.

\bibitem{shah2013variable}
R.~D. Shah and R.~J. Samworth, ``Variable selection with error control: another
  look at stability selection,'' \emph{J. R. Stat. Soc. Ser. B. Stat.
  Methodol.}, vol.~75, no.~1, pp. 55--80, 2013.

\bibitem{cox1975note}
D.~R. Cox, ``A note on data-splitting for the evaluation of significance
  levels,'' \emph{Biometrika}, vol.~62, no.~2, pp. 441--444, 1975.

\bibitem{wasserman2009high}
L.~Wasserman and K.~Roeder, ``High dimensional variable selection,'' \emph{Ann.
  Statist.}, vol.~37, no.~5A, pp. 2178--2201, 2009.

\bibitem{meinshausen2009p}
N.~Meinshausen, L.~Meier, and P.~B{\"u}hlmann, ``P-values for high-dimensional
  regression,'' \emph{J. Amer. Statist. Assoc.}, vol. 104, no. 488, pp.
  1671--1681, 2009.

\bibitem{barber2019knockoff}
R.~F. Barber and E.~J. Cand{\`e}s, ``A knockoff filter for high-dimensional
  selective inference,'' \emph{Ann. Statist.}, vol.~47, no.~5, pp. 2504--2537,
  2019.

\bibitem{lockhart2014significance}
R.~Lockhart, J.~Taylor, R.~J. Tibshirani, and R.~Tibshirani, ``A significance
  test for the lasso,'' \emph{Ann. Statist.}, vol.~42, no.~2, pp. 413--468,
  2014.

\bibitem{fithian2014optimal}
W.~Fithian, D.~Sun, and J.~Taylor, ``Optimal inference after model selection,''
  \emph{arXiv preprint, arXiv:1410.2597}, 2014.

\bibitem{lee2016exact}
J.~D. Lee, D.~L. Sun, Y.~Sun, and J.~E. Taylor, ``Exact post-selection
  inference, with application to the lasso,'' \emph{Ann. Statist.}, vol.~44,
  no.~3, pp. 907--927, 2016.

\bibitem{tibshirani2016exact}
R.~J. Tibshirani, J.~Taylor, R.~Lockhart, and R.~Tibshirani, ``Exact
  post-selection inference for sequential regression procedures,'' \emph{J.
  Amer. Statist. Assoc.}, vol. 111, no. 514, pp. 600--620, 2016.

\bibitem{miller1984selection}
A.~J. Miller, ``Selection of subsets of regression variables,'' \emph{J. R.
  Stat. Soc. Ser. A. Gen.}, vol. 147, no.~3, pp. 389--410, 1984.

\bibitem{miller2002subset}
------, \emph{Subset selection in regression}.\hskip 1em plus 0.5em minus
  0.4em\relax CRC Press, 2002.

\bibitem{wu2007controlling}
Y.~Wu, D.~D. Boos, and L.~A. Stefanski, ``Controlling variable selection by the
  addition of pseudovariables,'' \emph{J. Amer. Statist. Assoc.}, vol. 102, no.
  477, pp. 235--243, 2007.

\bibitem{williams1991probability}
D.~Williams, \emph{Probability with martingales}.\hskip 1em plus 0.5em minus
  0.4em\relax Cambridge Univ. Press, 1991.

\bibitem{tibshirani1996regression}
R.~Tibshirani, ``Regression shrinkage and selection via the lasso,'' \emph{J.
  R. Stat. Soc. Ser. B. Stat. Methodol.}, vol.~58, no.~1, pp. 267--288, 1996.

\bibitem{efron2004least}
B.~Efron, T.~Hastie, I.~Johnstone, and R.~Tibshirani, ``Least angle
  regression,'' \emph{Ann. Statist.}, vol.~32, no.~2, pp. 407--499, 2004.

\bibitem{zou2005regularization}
H.~Zou and T.~Hastie, ``Regularization and variable selection via the elastic
  net,'' \emph{J. R. Stat. Soc. Ser. B. Stat. Methodol.}, vol.~67, no.~2, pp.
  301--320, 2005.

\bibitem{zou2006adaptive}
H.~Zou, ``The adaptive lasso and its oracle properties,'' \emph{J. Amer.
  Statist. Assoc.}, vol. 101, no. 476, pp. 1418--1429, 2006.

\bibitem{machkour2022TRexSelector}
\BIBentryALTinterwordspacing
J.~Machkour, S.~Tien, D.~P. Palomar, and M.~Muma, \emph{TRexSelector: T-Rex
  Selector: {H}igh-Dimensional Variable Selection \& FDR Control}, 2024, {R}
  package version 1.0.0. [Online].\par Available:
  \url{https://CRAN.R-project.org/package=TRexSelector}
\BIBentrySTDinterwordspacing

\bibitem{machkour2022tlars}
\BIBentryALTinterwordspacing
------, \emph{{tlars: The T-LARS Algorithm: {E}arly-Terminated Forward Variable
  Selection}}, 2024, {R} package version 1.0.1. [Online].\par Available:
  \url{https://CRAN.R-project.org/package=tlars}
\BIBentrySTDinterwordspacing

\bibitem{storey2003positive}
J.~D. Storey, ``The positive false discovery rate: a bayesian interpretation
  and the q-value,'' \emph{Ann. Statist.}, vol.~31, no.~6, pp. 2013--2035,
  2003.

\bibitem{tibshirani2005sparsity}
R.~Tibshirani, M.~Saunders, S.~Rosset, J.~Zhu, and K.~Knight, ``Sparsity and
  smoothness via the fused lasso,'' \emph{J. R. Stat. Soc. Ser. B. Stat.
  Methodol.}, vol.~67, no.~1, pp. 91--108, 2005.

\bibitem{friedman2007pathwise}
J.~Friedman, T.~Hastie, H.~H{\"o}fling, and R.~Tibshirani, ``Pathwise
  coordinate optimization,'' \emph{Ann. Appl. Stat.}, vol.~1, no.~2, pp.
  302--332, 2007.

\bibitem{su2017false}
W.~Su, M.~Bogdan, and E.~J. Cand{\`e}s, ``False discoveries occur early on the
  lasso path,'' \emph{Ann. Statist.}, vol.~45, no.~5, pp. 2133--2150, 2017.

\bibitem{international2010integrating}
{The International HapMap 3 Consortium}, ``Integrating common and rare genetic
  variation in diverse human populations,'' \emph{Nature}, vol. 467, no. 7311,
  pp. 52--58, 2010.

\bibitem{su2011hapgen2}
Z.~Su, J.~Marchini, and P.~Donnelly, ``{HAPGEN2}: simulation of multiple
  disease {SNP}s,'' \emph{Bioinformatics}, vol.~27, no.~16, pp. 2304--2305,
  2011.

\bibitem{machkour2024dependency}
J.~Machkour, M.~Muma, and D.~P. Palomar, ``High-dimensional false discovery
  rate control for dependent variables,'' \emph{arXiv preprint
  arXiv:2401.15796}, 2024.

\bibitem{machkour2023ScreenTRex}
------, ``False discovery rate control for fast screening of large-scale
  genomics biobanks,'' in \emph{Proc. 22nd IEEE Statist. Signal Process.
  Workshop (SSP)}, 2023, pp. 666--670.

\bibitem{scheidt2023FDRControlLaptop}
F.~Scheidt, J.~Machkour, and M.~Muma, ``Solving {FDR}-controlled sparse
  regression problems with five million variables on a laptop,'' in \emph{Proc.
  IEEE 9th Int. Workshop Comput. Adv. Multi-Sensor Adapt. Process. (CAMSAP)},
  2023, pp. 116--120.

\bibitem{machkour2022TRexGVS}
J.~Machkour, M.~Muma, and D.~P. Palomar, ``False discovery rate control for
  grouped variable selection in high-dimensional linear models using the
  {T-Knock} filter,'' in \emph{30th Eur. Signal Process. Conf. (EUSIPCO)},
  2022, pp. 892--896.

\bibitem{machkour2023InformedEN}
------, ``The informed elastic net for fast grouped variable selection and
  {FDR} control in genomics research,'' in \emph{Proc. IEEE 9th Int. Workshop
  Comput. Adv. Multi-Sensor Adapt. Process. (CAMSAP)}, 2023, pp. 466--470.

\bibitem{koka2024false}
T.~Koka, J.~Machkour, and M.~Muma, ``False discovery rate control for
  {G}aussian graphical models via neighborhood screening,'' \emph{arXiv
  preprint arXiv:2401.09979}, 2024.

\bibitem{machkour2024TRexPCA}
J.~Machkour, A.~Breloy, M.~Muma, D.~P. Palomar, and F.~Pascal, ``Sparse {PCA}
  with false discovery rate controlled variable selection,'' \emph{arXiv
  preprint arXiv:2401.08375}, 2024.

\bibitem{machkour2024TRexIndexTracking}
J.~Machkour, D.~P. Palomar, and M.~Muma, ``{FDR}-controlled portfolio
  optimization for sparse financial index tracking,'' \emph{arXiv preprint
  arXiv:2401.15139}, 2024.

\bibitem{sudlow2015uk}
C.~Sudlow, J.~Gallacher, N.~Allen, V.~Beral, P.~Burton, J.~Danesh, P.~Downey,
  P.~Elliott, J.~Green, M.~Landray \emph{et~al.}, ``U{K} {B}iobank: An open
  access resource for identifying the causes of a wide range of complex
  diseases of middle and old age,'' \emph{PLOS Med.}, vol.~12, no.~3, p.
  e1001779, 2015.

\end{thebibliography}


\begin{thebibliography}{10}
\providecommand{\url}[1]{#1}
\csname url@samestyle\endcsname
\providecommand{\newblock}{\relax}
\providecommand{\bibinfo}[2]{#2}
\providecommand{\BIBentrySTDinterwordspacing}{\spaceskip=0pt\relax}
\providecommand{\BIBentryALTinterwordstretchfactor}{4}
\providecommand{\BIBentryALTinterwordspacing}{\spaceskip=\fontdimen2\font plus
\BIBentryALTinterwordstretchfactor\fontdimen3\font minus
  \fontdimen4\font\relax}
\providecommand{\BIBforeignlanguage}[2]{{%
\expandafter\ifx\csname l@#1\endcsname\relax
\typeout{** WARNING: IEEEtran.bst: No hyphenation pattern has been}%
\typeout{** loaded for the language `#1'. Using the pattern for}%
\typeout{** the default language instead.}%
\else
\language=\csname l@#1\endcsname
\fi
#2}}
\providecommand{\BIBdecl}{\relax}
\BIBdecl

\bibitem{efron2004least}
B.~Efron, T.~Hastie, I.~Johnstone, and R.~Tibshirani, ``Least angle
  regression,'' \emph{Ann. Statist.}, vol.~32, no.~2, pp. 407--499, 2004.

\bibitem{tibshirani1996regression}
R.~Tibshirani, ``Regression shrinkage and selection via the lasso,'' \emph{J.
  R. Stat. Soc. Ser. B. Stat. Methodol.}, vol.~58, no.~1, pp. 267--288, 1996.

\bibitem{zou2006adaptive}
H.~Zou, ``The adaptive lasso and its oracle properties,'' \emph{J. Amer.
  Statist. Assoc.}, vol. 101, no. 476, pp. 1418--1429, 2006.

\bibitem{zou2005regularization}
H.~Zou and T.~Hastie, ``Regularization and variable selection via the elastic
  net,'' \emph{J. R. Stat. Soc. Ser. B. Stat. Methodol.}, vol.~67, no.~2, pp.
  301--320, 2005.

\bibitem{Friedman2010glmnet}
J.~Friedman, T.~Hastie, and R.~Tibshirani, ``Regularization paths for
  generalized linear models via coordinate descent,'' \emph{J. Stat. Softw.},
  vol.~33, no.~1, pp. 1--22, 2010.

\bibitem{machkour2022tlars}
\BIBentryALTinterwordspacing
J.~Machkour, S.~Tien, D.~P. Palomar, and M.~Muma, \emph{{tlars: The T-LARS
  Algorithm: {E}arly-Terminated Forward Variable Selection}}, 2024, {R} package
  version 1.0.1. [Online].\par Available:
  \url{https://CRAN.R-project.org/package=tlars}
\BIBentrySTDinterwordspacing

\bibitem{candes2018panning}
E.~J. Cand{\`e}s, Y.~Fan, L.~Janson, and J.~Lv, ``Panning for gold:
  ‘model-{X}’ knockoffs for high dimensional controlled variable
  selection,'' \emph{J. R. Stat. Soc. Ser. B. Stat. Methodol.}, vol.~80, no.~3,
  pp. 551--577, 2018.

\bibitem{su2011hapgen2}
Z.~Su, J.~Marchini, and P.~Donnelly, ``{HAPGEN2}: simulation of multiple
  disease {SNP}s,'' \emph{Bioinformatics}, vol.~27, no.~16, pp. 2304--2305,
  2011.

\bibitem{barber2015controlling}
R.~F. Barber and E.~J. Cand{\`e}s, ``Controlling the false discovery rate via
  knockoffs,'' \emph{Ann. Statist.}, vol.~43, no.~5, pp. 2055--2085, 2015.

\bibitem{benjamini1995controlling}
Y.~Benjamini and Y.~Hochberg, ``Controlling the false discovery rate: a
  practical and powerful approach to multiple testing,'' \emph{J. R. Stat. Soc.
  Ser. B. Stat. Methodol.}, vol.~57, no.~1, pp. 289--300, 1995.

\bibitem{storey2004strong}
J.~D. Storey, J.~E. Taylor, and D.~Siegmund, ``Strong control, conservative
  point estimation and simultaneous conservative consistency of false discovery
  rates: a unified approach,'' \emph{J. R. Stat. Soc. Ser. B. Stat. Methodol.},
  vol.~66, no.~1, pp. 187--205, 2004.

\bibitem{gavrilov2009adaptive}
Y.~Gavrilov, Y.~Benjamini, and S.~K. Sarkar, ``An adaptive step-down procedure
  with proven {FDR} control under independence,'' \emph{Ann. Statist.},
  vol.~37, no.~2, pp. 619 -- 629, 2009.

\bibitem{sackrowitz1999p}
H.~Sackrowitz and E.~Samuel-Cahn, ``P values as random variables—expected {P}
  values,'' \emph{Am. Stat.}, vol.~53, no.~4, pp. 326--331, 1999.

\bibitem{benjamini2001control}
Y.~Benjamini and D.~Yekutieli, ``The control of the false discovery rate in
  multiple testing under dependency,'' \emph{Ann. Statist.}, vol.~29, no.~4,
  pp. 1165--1188, 2001.

\bibitem{reich2001linkage}
D.~E. Reich, M.~Cargill, S.~Bolk, J.~Ireland, P.~C. Sabeti, D.~J. Richter,
  T.~Lavery, R.~Kouyoumjian, S.~F. Farhadian, R.~Ward \emph{et~al.}, ``Linkage
  disequilibrium in the human genome,'' \emph{Nature}, vol. 411, no. 6834, pp.
  199--204, 2001.

\bibitem{sesia2019gene}
M.~Sesia, C.~Sabatti, and E.~J. Cand{\`e}s, ``Gene hunting with hidden markov
  model knockoffs,'' \emph{Biometrika}, vol. 106, no.~1, pp. 1--18, 2019.

\bibitem{machkour2022TRexSelector}
\BIBentryALTinterwordspacing
J.~Machkour, S.~Tien, D.~P. Palomar, and M.~Muma, \emph{TRexSelector: T-Rex
  Selector: {H}igh-Dimensional Variable Selection \& FDR Control}, 2024, {R}
  package version 1.0.0. [Online].\par Available:
  \url{https://CRAN.R-project.org/package=TRexSelector}
\BIBentrySTDinterwordspacing

\bibitem{international2010integrating}
{The International HapMap 3 Consortium}, ``Integrating common and rare genetic
  variation in diverse human populations,'' \emph{Nature}, vol. 467, no. 7311,
  pp. 52--58, 2010.

\end{thebibliography}

\vfill

\typeout{get arXiv to do 4 passes: Label(s) may have changed. Rerun}

\end{document}


\title{Supplement to\\ ``The Terminating-Random Experiments Selector: Fast High-Dimensional Variable Selection with False Discovery Rate Control''}

\author{
Jasin Machkour,
Michael Muma, and
Daniel P. Palomar
\thanks{J. Machkour and M. Muma are with the Robust Data Science Group at Technische Universit\"at Darmstadt, Germany (e-mail: jasin.machkour@tu-darmstadt.de; michael.muma@tu-darmstadt.de).}
\thanks{D. P. Palomar is with the Convex Optimization Group, Hong Kong University of Science and Technology, Hong Kong (e-mail: palomar@ust.hk).}
}

\maketitle

\begin{abstract}
This supplement is organized as follows: Appendix~\ref{sec: Proofs} presents some technical lemmas and the detailed proofs of Theorems~\ref{theorem: Dummy generation},~\ref{theorem: T-Rex algorithm}, and Corollary~\ref{corollary: 1}. In Appendix~\ref{sec: Main Ingredients of the T-Rex Selector}, the three main ingredients of the T-Rex selector are discussed and exemplified. Appendix~\ref{sec: The Deflated Relative Occurrence} provides an intuitive explanation of the deflated relative occurrence from Definition~\ref{definition: path-adapted relative occurrence}. Appendix~\ref{sec: Hyperparameter Choices for the Extended Calibration Algorithm} discusses the hyperparameter choices for the extended calibration algorithm in Algorithm~\ref{algorithm: Extended T-Rex}. In Appendix~\ref{sec: Computational Complexity}, the computational complexity of the T-Rex selector is derived. Appendices~\ref{sec: General Assumptions} and~\ref{sec: Exemplary Numerical Verification of Assumptions}, respectively, discuss and numerically verify the assumptions used by the state-of-the-art benchmark methods and the proposed approach. In Appendix~\ref{sec: Benchmark Methods for FDR Control}, some relevant details of the benchmark methods are discussed. In Appendix~\ref{sec: Additional Simulation Results}, additional simulation results for a low-dimensional setting are presented and discussed. Appendix~\ref{sec: Setup, Preprocessing, and Additional Results: Simulated Genome-Wide Association Study} provides details on the setup and the preprocessing of the analyzed data and additional results of the simulated genome-wide association study, while Appendix~\ref{sec: Illustration of Theorem 2 (Dummy Generation)} illustrates Theorem~\ref{theorem: Dummy generation}. Appendix~\ref{sec: Robustness of The T-Rex Selector} discusses the robustness of the T-Rex selector in the presence of non-Gaussian noise.
\end{abstract}

\begin{IEEEkeywords}
T-Rex selector, false discovery rate (FDR) control, high-dimensional variable selection, martingale theory, genome-wide association studies (GWAS).
\end{IEEEkeywords}

\appendices
\section{Proofs}
\label{sec: Proofs}
In this appendix, we introduce and prove some technical corollaries and lemmas. Then, the detailed proofs of Theorem~\ref{theorem: Dummy generation} (Dummy generation), Corollary~\ref{corollary: 1}, and Theorem~\ref{theorem: T-Rex algorithm} (Optimality of Algorithm~\ref{algorithm: T-Rex}) are presented. The results follow from standard assumptions in FDR control theory (for numerical verifications, see Appendices~\ref{sec: General Assumptions},~\ref{sec: Exemplary Numerical Verification of Assumptions}, and~\ref{sec: Setup, Preprocessing, and Additional Results: Simulated Genome-Wide Association Study}).  Throughout these supplementary materials, and especially in this section, all equation labels that do not start with the letter of the appendix they appear in (e.g., A.1, A.2, H.1) refer to equations from the main paper. Table~\ref{table: Notation overview} provides an overview of frequently used expressions.
\begin{table*}[!htbp]
\centering
\begin{tabular}{@{}L{\dimexpr0.31\linewidth-2\tabcolsep\relax}@{}L{\dimexpr0.67\linewidth-2\tabcolsep\relax}}
\bfseries Expression
& \bfseries Meaning\\
\midrule
		$K~\in~\mathbb{N}_{+}\backslash \lbrace 1 \rbrace$ & Number of random experiments. \\[0.2em]
  		$L~\in~\mathbb{N}_{+}$ & Number of dummies. \\[0.2em]
   		$T~\in~\lbrace 1, \ldots, L \rbrace$ & Number of included dummies after which the forward variable selection process in each random experiment is terminated. \\[0.2em]
   		$T^{*}$ & Optimal value of $T$ as determined by the calibration process. \\[0.2em]
   		$v~\in~[0.5, 1)$ & Voting level. \\[0.2em]
   		$v^{*}$ & Optimal value of $v$ as determined by the calibration process. \\[0.2em]
        $\alpha \in [0, 1]$ & Target FDR level. \\[0.2em]
        $\Z \coloneqq \big\lbrace \text{null } j : j \in \lbrace 1, \ldots, p \rbrace \big\rbrace$ & Index set of null variables. \\[0.2em]
        $\A \coloneqq \big\lbrace \text{active } j : j \in \lbrace 1, \ldots, p \rbrace \big\rbrace$ & Index set of active variables. \\[0.2em]
        $p_{0} \coloneqq | \Z |$ & Number of null variables. \\[0.2em]
        $p_{1} \coloneqq | \A |$ & Number of (true) active variables. \\[0.2em]
        $p = p_{0} + p_{1}$ & Total number of variables. \\[0.2em]
        $n$ & Number of data points. \\[0.2em]
        $\widehat{\A}(v) \coloneqq \widehat{\A}_{L}(v, T)$ & Estimator of the active set, i.e., index set of the selected variables. \\[0.2em]
        $\widehat{\A}^{\, 0}(v) \coloneqq \lbrace \text{null } j : \Phi_{T,L}(j) > v \rbrace$ & Index set of the selected null variables. \\[0.2em]
        $\widehat{\A}^{\, 1}(v) \coloneqq \lbrace \text{active } j : \Phi_{T,L}(j) > v \rbrace$ &  Index set of the selected active variables. \\[0.2em]
        $\C_{k, L}(T)$ & Candidate set of the $k$th random experiment, i.e., index set of the included variables in the $k$th random experiment.\\
\end{tabular}
\caption{Overview of frequently used expressions.}
\label{table: Notation overview}
\end{table*}

\subsection{Preliminaries: Technical Corollaries and Lemmas}
\label{subsec: Preliminaries: Technical Corollaries and Lemmas}
\begin{cor}
Let $\Z_{m, k}$ and $\Dum_{m, k}$ be the index sets of the non-included null and dummy variables in the $m$th \textit{LARS}\footnote{Note that Corollary~\ref{corollary: 1} and subsequent results apply to all forward selection methods that select one (and do not drop any) variable in each forward selection step based on the maximum absolute sample correlations between the predictors and the response or the current residual. Thus, the results hold for the \textit{LARS} algorithm~\cite{efron2004least} and approximately hold for the \textit{Lasso}~\cite{tibshirani1996regression}, \textit{adaptive Lasso}~\cite{zou2006adaptive}, \textit{elastic net}~\cite{zou2005regularization}, and many other related methods.} forward selection step of the $k$th random experiment, respectively. Then, for all $j \in \Z_{m, k} \cup \Dum_{m, k}$, the probability of including $X_{j}$ in the $m$th step of the $k$th random experiment (RE) is equal, i.e., for all $j \in \Z_{m, k} \cup \Dum_{m, k}$ it holds that
\begin{align}
    \mathbb{P}(&\text{``}X_{j} \text{ included in $m$th step of $k$th RE''} \mid j \in \Z_{m, k} \cup \Dum_{m, k} )
    \\[-1.5em]
    &= \dfrac{1}{| \Z_{m, k} \cup \Dum_{m, k} |}.
    \label{eq: equal probability picking null or dummy}
\end{align}
\label{corollary: 1}
\end{cor}

\begin{proof}
For ease of readability, the proof is deferred to Appendix~\ref{subsec: Proof of Corollary 1}.
\label{eq: proof corollary 1}
\end{proof}

\begin{cor}
The numbers of included null variables at step $t$ of all random experiments are i.i.d. random variables following the negative hypergeometric distribution, i.e., as $n \rightarrow \infty$,
\begin{equation}
\sum_{j \in \Z} \mathbbm{1}_{k}(j, t, L) \sim \NHG(p_{0} + L, p_{0}, t),
\label{eq: corollary 1}
\end{equation}
$t = 1, \ldots,T$, $k = 1, \ldots, K$, where $\Z$ is the index set of the null variables.
\label{corollary: 2}
\end{cor}

\begin{proof}
Let $t$ be the number of included dummies after which a random experiment is terminated. There exists a \textit{LARS} step $m$ at which $t$ dummies are included. From Corollary~\ref{corollary: 1}, we know that the probability of including a null variable and the probability of including a dummy variable are equal in each step of any random experiment. Therefore, it follows from Corollary~\ref{corollary: 1} that the number of included null variables in any random experiment can be described by a process that randomly picks null and dummy variables one at a time, without replacement, and with equal probability from $\Z_{m, k} \cup \Dum_{m, k}$ until the process is terminated after $t$ dummies are included. Since the included active variables in that process do not count towards the number of included null variables, the total number of variables in the process is $p_{0}$ instead of $p$. The described process exactly follows the definition of the negative hypergeometric distribution, i.e., $\NHG(p_{0} + L, p_{0}, t)$ with $p_{0} + L$ total elements, $p_{0}$ success elements, and $t$ failures after which a random experiment is terminated.
\label{proof: Corollary 2}
\end{proof}

As a consequence of A-\ref{assumption: 1} and A-\ref{assumption: 2} (see Appendix~\ref{sec: General Assumptions}), the number of selected null variables (i.e., $V_{T, L}(v)$) conditioned on the number of null variables exceeding the minimum voting level of $50\%$ (i.e., $V_{T, L}(0.5)$) is binomially distributed with $\mathbb{P} \big( \Phi_{T, L}(j_{0}) > v \big)$ being the selection probability of variable $j_{0} \in \widehat{\A}^{\, 0}(0.5)$. Thus, we obtain the following hierarchical model:
\begin{customcor}{3}
The number of selected null variables $V_{T, L}(v)$ follows the hierarchical model
\begin{align}
V_{T, L}(v) &\bigm| V_{T, L}(0.5) \\
&\sim \Bin \Big( V_{T, L}(0.5), \, \mathbb{P} \big( \Phi_{T, L}(j_{0}) > v \big) \Big), \\[0.5em]
V_{T, L}(0.5) &\overset{d}{\leq} \NHG(p_{0} + L, p_{0}, T),
\label{eq: corollary 2}
\end{align}
where $\mathbb{P} \big( \Phi_{T, L}(j_{0}) > v \big) > 0$ for all $j_{0} \in \widehat{\A}^{\, 0}(0.5)$ and for any $v \in [0.5, 1)$.
\label{corollary: 3}
\end{customcor}
\begin{customlemma}{1}
Let $v$ be any real number in $[0.5, 1)$ and $K \to \infty$. Then, for any $j_{0} \in \widehat{\A}^{\, 0}(0.5)$, the following equation is satisfied:
\begin{equation*}
\mathbb{E} \big[ V_{T, L}(v) \big] = \mathbb{P} \big( \Phi_{T, L}(j_{0}) > v \big) \cdot \mathbb{E} \big[ V_{T, L}(0.5) \big].
\label{eq: lemma-4}
\end{equation*}
\label{lemma: 4}
\end{customlemma}

\begin{proof}
Using the tower property of the expectation, we can rewrite the expectation of $V_{T, L}(v)$ as follows:
\begin{align}
\mathbb{E} \big[ V_{T, L}(v) \big] &= \mathbb{E} \Big[ \mathbb{E} \big[ V_{T, L}(v) \bigm| V_{T, L}(0.5) \big] \Big]
\label{eq: proof-lemma-4.1}
\\[0.5em]
&= \mathbb{E} \big[ V_{T, L}(0.5) \cdot \mathbb{P} \big( \Phi_{T, L}(j_{0}) > v \big) \big]
\label{eq: proof-lemma-4.2}
\\[0.5em]
&= \mathbb{P} \big( \Phi_{T, L}(j_{0}) > v \big) \cdot \mathbb{E} \big[ V_{T, L}(0.5) \big].
\label{eq: proof-lemma-4.3}
\end{align}
The second equation follows from 
\begin{align}
V_{T, L}(v) &\bigm| V_{T, L}(0.5) \\
&\sim \Bin \Big( V_{T, L}(0.5), \, \mathbb{P} \big( \Phi_{T, L}(j_{0}) > v \big) \Big)
\label{eq: proof-lemma-4.4}
\end{align}
in Corollary~\ref{corollary: 3} and the third equation holds because $\Phi_{T, L}(j_{0})$, $j_{0} \in \widehat{\A}^{\, 0}(0.5)$, are i.i.d. random variables and, therefore, the selection probability $\mathbb{P} \big( \Phi_{T, L}(j_{0}) >v \big)$ for any fixed $v$ is the same constant for all $j_{0}$.
\end{proof}

\begin{customlemma}{2}
Let $v$ be any real number in $[0.5, 1)$ and $K \to \infty$. Define
\begin{equation}
\widehat{V}_{T, L}^{\prime}(v) \coloneqq \widehat{V}_{T, L}(v) - \sum\limits_{j \in \widehat{\A}(v)} \big( 1 - \Phi_{T, L}(j) \big).
\label{eq: lemma-5.1}
\end{equation}
Then, for any $j_{0} \in \widehat{\A}^{\, 0}(0.5)$, the following equation is satisfied:
\begin{equation}
\mathbb{E} \big[ \widehat{V}_{T, L}^{\prime}(v) \big] = \mathbb{P} \big( \Phi_{T, L}(j_{0}) > v \big) \cdot \widehat{V}_{T, L}^{\prime}(0.5).
\label{eq: lemma-5.3}
\end{equation}
\label{lemma: 5}
\end{customlemma}

\begin{proof}
Taking the expectation of $\widehat{V}_{T, L}^{\prime}(v)$ yields
\begingroup
\allowdisplaybreaks
\begin{align}
&\mathbb{E} \big[ \widehat{V}_{T, L}^{\prime}(v) \big]
\\[0.5em]
&= \mathbb{E} \Bigg[ \sum\limits_{t = 1}^{T} \dfrac{p - \sum_{q = 1}^{p} \Phi_{t, L}(q)}{L - (t - 1)} \cdot \dfrac{\sum_{j \in \widehat{\A}(v)} \Delta \Phi_{t,L}(j)}{\sum_{q \in  \widehat{\A}(0.5)} \Delta \Phi_{t,L}(q)} \Bigg]
\label{eq: proof-lemma-5.2}
\\[0.5em]
&= \sum\limits_{t = 1}^{T} \dfrac{p_{0} - \sum_{q \in \Z} \Phi_{t, L}(q)}{L - (t - 1)} 
\\*
& \qquad\qquad \cdot \mathbb{E} \Bigg[ \dfrac{\sum_{j \in  \widehat{\A}^{\, 0}(v)} \Delta \Phi_{t,L}(j)}{\sum_{q \in  \widehat{\A}^{\, 0}(0.5)} \Delta \Phi_{t,L}(q)} \Bigg],
\label{eq: proof-lemma-5.3}
\end{align}
\endgroup
where the first and the second equation follow from Definitions~\ref{definition: path-adapted relative occurrence},~\ref{definition: FDP estimator}, and A-\ref{assumption: 3} (see Appendix~\ref{sec: General Assumptions}), respectively. Note that $\sum_{q \in \Z} \Phi_{t, L}(q) = \frac{1}{K} \sum_{k = 1}^{K}\sum_{q \in \Z} \mathbbm{1}_{k}(q, t, L)$ is the average number of included null variables when stopping after $t$ dummies have been included. Since $K \to \infty$, the law of large numbers allows replacing the average by its expectation. That is, $\sum_{q \in \Z} \Phi_{t, L}(q) = \mathbb{E} \big[ \sum_{q \in \Z} \mathbbm{1}_{k}(q, t, L) \big]$. Therefore, $\sum_{q \in \Z} \Phi_{t, L}(q)$ is deterministic and can be written outside the expectation.

Using the tower property, we can rewrite the expectation in~\eqref{eq: proof-lemma-5.3} as follows:
\begin{align}
&\mathbb{E} \Bigg[ \dfrac{\sum_{j \in \widehat{\A}^{\, 0}(v)} \Delta \Phi_{t,L}(j)}{\sum_{q \in \widehat{\A}^{\, 0}(0.5)} \Delta \Phi_{t,L}(q)} \Bigg]
\\[0.5em]
&= \mathbb{E} \Bigg[ \mathbb{E} \Bigg[ \dfrac{\sum_{j \in \widehat{\A}^{\, 0}(v)} \Delta \Phi_{t,L}(j)}{\sum_{q \in \widehat{\A}^{\, 0}(0.5)} \Delta \Phi_{t,L}(q)} \Biggm| \big| \widehat{\A}^{\, 0}(v) \big|, \big| \widehat{\A}^{\, 0}(0) \big| \Bigg] \Bigg]
\label{eq: proof-lemma-5.4}
\\[0.5em]
&= \mathbb{E} \Bigg[ \dfrac{\big| \widehat{\A}^{\, 0}(v) \big|}{\big| \widehat{\A}^{\, 0}(0.5) \big|} \Bigg]
\label{eq: proof-lemma-5.5}
\end{align}
The last equation follows from $\Delta \Phi_{t, L}(j_{0})$, $j_{0} \in \widehat{\A}^{\, 0}(0.5)$, being i.i.d. random variables and the well known fact that $\mathbb{E} [Q_{M} \, / \, Q_{N}] = M \, / \, N$, where $Q_{B} = \sum_{b = 1}^{B} Z_{b}$ with $Z_{1}, \ldots, Z_{B}$, $B \in \lbrace M, N \rbrace$, being non-zero i.i.d. random variables and $M \leq N$.

Noting that $| \widehat{\A}^{\, 0}(v)| = V_{T, L}(v)$ and applying the tower property again, we can rewrite the expectation in~\eqref{eq: proof-lemma-5.5} as follows:
\begin{align}
\mathbb{E} \Bigg[ &\dfrac{\big| \widehat{\A}^{\, 0}(v) \big|}{\big| \widehat{\A}^{\, 0}(0.5) \big|} \Bigg] 
\\[0.5em]
&= \mathbb{E} \Bigg[ \dfrac{V_{T, L}(v)}{V_{T, L}(0.5)} \Bigg]
\label{eq: proof-lemma-5.6}
\\[0.5em]
&= \mathbb{E} \Bigg[ \mathbb{E} \Bigg[ \dfrac{V_{T, L}(v)}{V_{T, L}(0.5)} \Biggm| V_{T, L}(0.5) \Bigg] \Bigg]
\label{eq: proof-lemma-5.7}
\\[0.5em]
&= \mathbb{E} \Bigg[ \dfrac{1}{V_{T, L}(0.5)} \cdot \mathbb{E} \big[ V_{T, L}(v) \bigm| V_{T, L}(0.5) \big] \Bigg]
\label{eq: proof-lemma-5.8}
\\[0.5em]
&= \mathbb{E} \Bigg[ \dfrac{1}{V_{T, L}(0.5)} \cdot V_{T, L}(0.5) \cdot \mathbb{P} \big( \Phi_{T, L}(j_{0}) > v \big) \Bigg]
\label{eq: proof-lemma-5.9}
\\[0.5em]
&= \mathbb{P} \big( \Phi_{T, L}(j_{0}) > v \big).
\label{eq: proof-lemma-5.10}
\end{align}
The last three equations follow from the same arguments as in the proof of Lemma~\ref{lemma: 4}. Thus,
\begin{align}
\mathbb{E} \big[ \widehat{V}_{T, L}^{\prime}(v) \big] &= \mathbb{P} \big( \Phi_{T, L}(j_{0}) > v \big) \cdot \sum\limits_{t = 1}^{T} \dfrac{p_{0} - \sum_{q \in \Z} \Phi_{t, L}(q)}{L - (t - 1)}
\label{eq: proof-lemma-5.11}
\\[0.5em]
&= \mathbb{P} \big( \Phi_{T, L}(j_{0}) > v \big) \cdot \widehat{V}_{T, L}^{\prime}(0.5).
\label{eq: proof-lemma-5.12}
\end{align}
\end{proof}
\begin{customlemma}{3}
Let $K \to \infty $. Then,
\begin{equation*}
\mathbb{E}  \Bigg[ \sum\limits_{q \in \Z} \Phi_{t, L}(q) \Bigg] = \dfrac{t}{L + 1} \cdot p_{0}.
\end{equation*}
\label{lemma: 6}
\end{customlemma}
\begin{proof}
Using Definition~\ref{definition: relative occurrence}, we obtain
\begin{equation}
\sum\limits_{q \in \Z} \Phi_{t, L}(q) = \dfrac{1}{K} \sum\limits_{k = 1}^{K} \sum\limits_{q \in \Z} \mathbbm{1}_{k}(q, t, L).
\label{eq: proof-proposition-1.18}
\end{equation}
Then, taking the expectation and noting that 
\begin{equation}
\sum_{q \in \Z} \mathbbm{1}_{k}(q, t, L) \sim \NHG(p_{0} + L, p_{0}, t), \quad k = 1, \ldots, K,
\label{eq: proof-proposition-1.19}
\end{equation}
i.e., the number of included null variables in the $K$ random experiments are i.i.d. random variables following the negative hypergeometric distribution as stated in Corollary~\ref{corollary: 2}, yields
\begin{align}
\mathbb{E}  \Bigg[ \sum\limits_{q \in\Z} \Phi_{t, L}(q) \Bigg]
&= \dfrac{1}{K} \sum\limits_{k = 1}^{K} \mathbb{E}  \Bigg[ \sum\limits_{q \in \Z} \mathbbm{1}_{k}(q, t, L) \Bigg]
\label{eq: proof-proposition-1.20}
\\[0.5em]
&= \dfrac{1}{K} \cdot K \cdot \dfrac{t}{L + 1} \cdot p_{0}
\label{eq: proof-proposition-1.21}
\\[0.5em]
&= \dfrac{t}{L + 1} \cdot p_{0}. \qedhere
\label{eq: proof-proposition-1.22}
\end{align}
\end{proof}
\begin{customlemma}{4}
Let $v$ be any real number in $[0.5, 1)$. Define 
\begin{equation}
\epsilon_{T, L}^{*}(v) \coloneqq \inf \lbrace \epsilon \in (0, v) : R_{T, L}(v - \epsilon) - R_{T, L}(v) = 1 \rbrace
\label{eq: epsilon star}
\end{equation}
with the convention that $\epsilon_{T, L}^{*}(v) = 0$ if the infimum does not exist. Suppose that $V_{T, L} \big( v - \epsilon_{T, L}^{*}(v) \big) = V_{T, L}(v) + 1$, $\mathbb{E} \big[ V_{T, L}(v) \big] > 0$, and $\mathbb{E} \big[ \widehat{V}_{T, L}^{\prime}(v) \big] > 0$. Then, for all $j_{0} \in \widehat{\A}^{\, 0}(0.5)$ it holds that
\begin{align} 
\text{(i)} \quad \mathbb{E} \big[ V_{T, L} &\big( v - \epsilon_{T, L}^{*}(v) \big) \bigm| V_{T, L}(v) \big] 
\\[0.5em]
&= V_{T, L}(v) \cdot \dfrac{\mathbb{P}\big( \Phi_{T, L}(j_{0}) > v - \epsilon_{T, L}^{*}(v) \big)}{\mathbb{P} \big( \Phi_{T, L}(j_{0}) > v \big)}
\end{align}
and
\begin{align}
\text{(ii)} \quad \mathbb{E} \big[ \widehat{V}_{T, L}^{\prime} &\big( v - \epsilon_{T, L}^{*}(v) \big) \bigm| \widehat{V}_{T, L}^{\prime}(v) \big] 
\\[0.5em]
&=  \widehat{V}_{T, L}^{\prime}(v) \cdot \dfrac{\mathbb{P}\big( \Phi_{T, L}(j_{0}) > v - \epsilon_{T, L}^{*}(v) \big)}{\mathbb{P} \big( \Phi_{T, L}(j_{0}) > v \big)}.
\end{align}
\label{lemma: 7}
\end{customlemma}

\begin{proof}
(i) Let $\delta \geq 1$ be a constant that satisfies the equation $V_{T, L} \big( v - \epsilon_{T, L}^{*}(v) \big) = \delta \cdot V_{T, L}(v)$. Then,
\begin{align}
\mathbb{E} \big[ V_{T, L} \big( v - \epsilon_{T, L}^{*}(v) \big) \bigm| V_{T, L}(v) \big] 
&= \mathbb{E} \big[ \delta \cdot V_{T, L}(v) \bigm| V_{T, L}(v) \big]
\\
&=  \delta \cdot V_{T, L}(v).
\label{eq: proof-lemma-7.1}
\end{align}
We rewrite $\delta \cdot V_{T, L}(v)$ as follows:
\begin{align}
\delta \cdot V_{T, L}(v) &= V_{T, L}(v) \cdot \dfrac{\delta \cdot \mathbb{E} \big[ V_{T, L}(v) \big]}{\mathbb{E} \big[ V_{T, L}(v) \big]}
\label{eq: proof-lemma-7.2}
\\[0.5em]
&= V_{T, L}(v) \cdot \dfrac{\mathbb{E} \big[ V_{T, L}(v - \epsilon_{T, L}^{*}(v)) \big]}{\mathbb{E} \big[ V_{T, L}(v) \big]}
\label{eq: proof-lemma-7.3}
\\[0.5em]
&= V_{T, L}(v) \cdot \dfrac{\mathbb{P} \big( \Phi_{T, L}(j_{0}) > v - \epsilon_{T, L}^{*}(v) \big)}{\mathbb{P} \big( \Phi_{T, L}(j_{0}) > v \big)}.
\label{eq: proof-lemma-7.4}
\end{align}
The last line follows from Lemma~\ref{lemma: 4}. Comparing $\delta \cdot V_{T, L}(v)$ and the last line, we see that
\begin{equation}
\delta = \mathbb{P} \big( \Phi_{T, L}(j_{0}) > v - \epsilon_{T, L}^{*}(v) \big) \, / \, \mathbb{P} \big( \Phi_{T, L}(j_{0}) > v \big)
\label{eq: proof-lemma-7.5}
\end{equation} 
and the first part of the lemma follows.
\\\\
(ii) The proof is analogous to the proof of (i). The only difference is that Lemma~\ref{lemma: 5} instead of Lemma~\ref{lemma: 4} needs to be used for rewriting the expression $\delta \cdot \widehat{V}_{T, L}^{\prime}(v)$.
\end{proof}

\subsection{Proof of Lemma~\ref{lemma: martingale} (Martingale)}
\label{subsec: Lemma 5 (Martingale)}
\begin{proof}
If there exists a variable with an index, say, $j^{*}$ that is not selected at the voting level $v$ but at the level $v - \epsilon_{T, L}^{*}(v)$ and it is a null variable, then we have
\begin{equation}
V_{T, L}(v - \epsilon_{T, L}^{*}(v)) = V_{T, L}(v) + 1.
\label{eq: proof-theorem-1.1}
\end{equation}
However, if $j^{*}$ is an active variable or if the infimum in~\eqref{eq: epsilon star} does not exist, that is, no additional variable is selected at the voting level $v - \epsilon_{T, L}^{*}(v)$ when compared to the level $v$, then we obtain
\begin{equation}
V_{T, L}(v - \epsilon_{T, L}^{*}(v)) = V_{T, L}(v).
\label{eq: proof-theorem-1.2}
\end{equation}
Summarizing both results, we have
\begin{equation}
V_{T, L}(v - \epsilon_{T, L}^{*}(v)) = 
\begin{cases}
V_{T, L}(v) + 1, \quad j^{*} \in \Z 
\\[1em]
V_{T, L}(v), \quad
\textstyle
\begin{array}{l}
j^{*} \in \A \\ 
\text{or } \epsilon_{T, L}^{*}(v) = 0
\end{array}
\end{cases}.
\label{eq: proof-theorem-1.3}
\end{equation}
Thus, using the definition of $H_{T, L}(v)$ within Lemma~\ref{lemma: martingale} in the main paper, we obtain
\begingroup
\allowdisplaybreaks
\begin{align}
&\mathbb{E} \big[ H_{T, L} \big( v - \epsilon_{T, L}^{*}(v) \big) \bigm| \mathcal{F}_{v} \big]
\\[1em]
&= \mathbb{E} \Bigg[ \dfrac{V_{T, L} \big( v - \epsilon_{T, L}^{*}(v) \big)}{\widehat{V}_{T, L}^{\prime} \big( v - \epsilon_{T, L}^{*}(v) \big)} \Biggm| V_{T, L}(v), \widehat{V}_{T, L}^{\prime}(v) \Bigg]
\label{eq: proof-theorem-1.4}
\\[1em]
&= 
\begin{cases}
\mathbb{E} \Bigg[ \dfrac{V_{T, L}(v) + 1}{\widehat{V}_{T, L}^{\prime} \big( v - \epsilon_{T, L}^{*}(v) \big)} \Biggm| V_{T, L}(v), \widehat{V}_{T, L}^{\prime}(v) \Bigg], \quad j^{*} \in \Z
\\[2em]
\mathbb{E} \Bigg[ \dfrac{V_{T, L}(v)}{\widehat{V}_{T, L}^{\prime}(v)} \Biggm| V_{T, L}(v), \widehat{V}_{T, L}^{\prime}(v) \Bigg],
\quad
\textstyle
\begin{array}{l}
j^{*} \in \A \\ 
\text{or } \epsilon_{T, L}^{*}(v) = 0
\end{array}
\end{cases}
\label{eq: proof-theorem-1.5}
\\[1em]
&= 
\begin{cases}
\dfrac{\mathbb{E} \Bigg[ \dfrac{1}{\widehat{V}_{T, L}^{\prime} \big( v - \epsilon_{T, L}^{*}(v) \big)} \Biggm| V_{T, L}(v), \widehat{V}_{T, L}^{\prime}(v) \Bigg]}{\big( V_{T, L}(v) + 1 \big)^{-1}}, \, j^{*} \in \Z 
\\[2em]
\dfrac{V_{T, L}(v)}{\widehat{V}_{T, L}^{\prime}(v)},
\,
\textstyle
\begin{array}{l}
j^{*} \in \A \\ 
\text{or } \epsilon_{T, L}^{*}(v) = 0
\end{array}
\end{cases}.
\label{eq: proof-theorem-1.6}
\end{align}
\endgroup
Using Lemma~\ref{lemma: 7}, we can rewrite the denominator within the first case of Equation~\eqref{eq: proof-theorem-1.6} as follows:
\begin{align}
V_{T, L}&(v) + 1
\\[0.5em]
&= \mathbb{E} \big[ V_{T, L} \big( v - \epsilon_{T, L}^{*}(v) \big) \bigm| V_{T, L}(v) \big]
\label{eq: proof-theorem-1.11}
\\[0.5em]
&= V_{T, L}(v) \cdot \dfrac{\mathbb{P} \big( \Phi_{T, L}(j_{0}) > v - \epsilon_{T, L}^{*}(v) \big)}{\mathbb{P} \big( \Phi_{T, L}(j_{0}) > v \big)}.
\label{eq: proof-theorem-1.13}
\end{align}
Next, we rewrite the numerator within the first case of Equation~\eqref{eq: proof-theorem-1.6} as follows:
\begin{align}
\mathbb{E} \Bigg[ &\dfrac{1}{\widehat{V}_{T, L}^{\prime} \big( v - \epsilon_{T, L}^{*}(v) \big)} \Biggm| V_{T, L}(v), \widehat{V}_{T, L}^{\prime}(v) \Bigg]
\\[0.5em]
&\geq \dfrac{1}{\mathbb{E} \big[ \widehat{V}_{T, L}^{\prime} \big( v - \epsilon_{T, L}^{*}(v) \big) \bigm| V_{T, L}(v), \widehat{V}_{T, L}^{\prime}(v) \big]}
\label{eq: proof-theorem-1.14}
\\[0.5em]
&= \dfrac{1}{\mathbb{E} \big[ \widehat{V}_{T, L}^{\prime} \big( v - \epsilon_{T, L}^{*}(v) \big) \bigm| \widehat{V}_{T, L}^{\prime}(v) \big]}
\label{eq: proof-theorem-1.15}
\\[0.5em]
&= \Bigg( \widehat{V}_{T, L}^{\prime}(v) \cdot \dfrac{\mathbb{P} \big( \Phi_{T, L}(j_{0}) > v - \epsilon_{T, L}^{*}(v) \big)}{\mathbb{P} \big( \Phi_{T, L}(j_{0}) > v \big)} \Bigg)^{-1}
\label{eq: proof-theorem-1.16}
\end{align}
The first inequality follows from Jensen's inequality. The first equation holds because $\widehat{V}_{T, L}^{\prime} \big( v - \epsilon_{T, L}^{*}(v) \big)$ and $V_{T, L}(v)$ are conditionally independent given $\widehat{V}_{T, L}^{\prime}(v)$ and the last line follows from Lemma~\ref{lemma: 7}. Plugging \eqref{eq: proof-theorem-1.13} and \eqref{eq: proof-theorem-1.16} into \eqref{eq: proof-theorem-1.6} yields
\begin{equation}
\mathbb{E} \big[ H_{T, L} \big( v - \epsilon_{T, L}^{*}(v) \big) \bigm| \mathcal{F}_{v} \big] \geq H_{T, L}(v),
\label{eq: proof-theorem-1.17}
\end{equation}
i.e., $\lbrace H_{T, L}(v) \rbrace_{v \in \mathcal{V}}$, with $\mathcal{V} = \lbrace \Phi_{T, L}(j) : j = 1, \ldots, p \rbrace$, is a backward-running super-martingale with respect to the filtration $\mathcal{F}_{v}$.
\end{proof}

\subsection{Proof of Theorem~\ref{theorem: Dummy generation} (Dummy generation)}
\label{subsec: Proof of Theorem 2 (Dummy generation)}
\begin{proof}
Since
\begin{equation}
\mathbb{E} \big[ D_{n, l, m, k} \big] = \dfrac{1}{\Gamma_{n, m, k}} \cdot \sum\limits_{i = 1}^{n} \gamma_{i, m, k} \cdot \mathbb{E} \big[ \XKvar_{i, l, k} \big] = 0
\label{eq: proof dummy generation-1}
\end{equation}
and
\begin{equation}
\Var \big[ D_{n, l, m, k} \big] = \dfrac{1}{\Gamma_{n, m, k}^{2}} \cdot \sum\limits_{i = 1}^{n} \gamma^{2}_{i, m, k} \cdot \Var \big[ \XKvar_{i, l, k} \big] = 1,
\label{eq: proof dummy generation-2}
\end{equation}
the Lindeberg-Feller central limit theorem can be used to prove that $D_{n, l, m, k} \overset{d}{\to} D$, $D \sim \mathcal{N}(0, 1)$. In order to do this, we define
\begin{equation}
\QKvar_{i, l, m, k} \coloneqq \dfrac{\gamma_{i, m, k} \cdot \XKvar_{i, l, k}}{\Gamma_{n, m, k}},
\label{eq: proof dummy generation-3}
\end{equation}
and check whether it satisfies the Lindeberg condition, i.e., whether for every $\tau > 0$
\begin{equation}
\lim_{n \to \infty} \sum\limits_{i = 1}^{n} \mathbb{E} \Big[ \QKvar_{i, l, m, k}^{2} \cdot I \big( \big| \QKvar_{i, l, m, k} \big| > \tau \big) \Big] = 0
\label{eq: proof dummy generation-4}
\end{equation}
holds. Rewriting the Lindeberg condition using the definition of $\QKvar_{i, l, m, k}$ yields 
\begin{equation}
\lim_{n \to \infty} \sum\limits_{i = 1}^{n} \bigg( \dfrac{\gamma_{i, m, k}}{\Gamma_{n, m, k}} \bigg)^{2} \mathbb{E} \bigg[ \XKvar_{i, l, k}^{2} \cdot I \bigg( \big| \XKvar_{i, l, k} \big| > \dfrac{\tau \Gamma_{n, m, k}}{| \gamma_{i, m, k} |} \bigg) \bigg] = 0,
\label{eq: proof dummy generation-5}
\end{equation}
where $I(\cdot)$ denotes the indicator function, i.e., $I(A > B)$ is equal to one if $A > B$ and equal to zero if $A \leq B$. Since
\begin{equation}
\lim_{n \to \infty} \max_{1 \leq i \leq n} \bigg( \dfrac{\gamma_{i, m, k}}{\Gamma_{n, m, k}} \bigg)^{2} = 0
\label{eq: proof dummy generation-6}
\end{equation}
and
\begin{equation}
\lim_{n \to \infty} \min_{1 \leq i \leq n} \bigg( \dfrac{\Gamma_{n, m, k}}{| \gamma_{i, m, k} |} \bigg) \to \infty,
\label{eq: proof dummy generation-7}
\end{equation}
the Lindeberg condition is satisfied and the theorem follows.
\label{proof: Dummy generation}
\end{proof}

\begin{rmk}
Loosely speaking, Theorem~\ref{theorem: Dummy generation} states that regardless of the distribution from which the dummies are sampled, the dummy correlation variables follow the standard normal distribution as $n \rightarrow \infty$. That is, the distribution of the dummies has no influence on the resulting distribution of the dummy correlation variables. Since the realizations of the dummy correlation variables determine which dummies are included along the LARS solution path, we can conclude that the decisions of which variable enters next along the solution path is independent of the distribution of the dummies. Thus, the dummies can be sampled from any univariate probability distribution with finite expectation and variance to serve as flagged null variables within the \textit{T-Rex} selector.
\label{remark: dummy generation}
\end{rmk}

\subsection{Proof of Corollary~\ref{corollary: 1}}
\label{subsec: Proof of Corollary 1}
\begin{proof}
Similarly to Theorem~\ref{theorem: Dummy generation}, we consider the predictors $\x_{j} = [x_{1, j}, \ldots, x_{n, j}]$ as $n$ i.i.d. realizations of $X_{j}$, which can also be considered as one realization from each of the i.i.d. random variables $X_{1, j}, \ldots, X_{n, j}$. Replacing
\begin{enumerate}
    \item[(i)] $\XKvar_{i, l, k}$, $i \in \lbrace 1, \ldots, n \rbrace$, $l \in \Dum_{m, k}$, $k \in \lbrace 1, \ldots, K \rbrace$,
\end{enumerate}
in Theorem~\ref{theorem: Dummy generation} with
\begin{enumerate}
    \item[(ii)] $X_{i,j}$, $i \in \lbrace 1, \ldots, n \rbrace$, $j \in \Z_{m, k}$, 
\end{enumerate}
and using A-\ref{assumption: 1} (see Appendix~\ref{sec: General Assumptions}), the conditions in Theorem~\ref{theorem: Dummy generation} are satisfied. Thus, it follows that, as $n \rightarrow \infty$,
\begin{equation}
    D_{n, j, m, k} \overset{d}{\rightarrow} D, \quad D \sim \mathcal{N}(0, 1),
    \label{eq: convergence to standard normal distribution of null correlation variables}
\end{equation}
i.e., the null correlation variables $\lbrace G_{j, m, k} : j \in \Z_{m, k} \rbrace$ are identically distributed.\footnote{Note that, as in Theorem~\ref{theorem: Dummy generation}, the constant factor $\Gamma_{n, m, k}$ in $D_{n, j, m, k} = (1 / \Gamma_{n, m, k})G_{j, m, k}$ is equal for all $j \in \Z_{m, k}$ and does not affect the distribution of $G_{j, m, k}$.} Since the non-included null random variables $\lbrace X_{j} : j \in \Z_{m, k} \rbrace$ are independent of the true active variables and mutually independent, the null correlation variables are also independently distributed. Thus, in combination with Theorem~\ref{theorem: Dummy generation}, the null and dummy correlation variables $\lbrace G_{j, m, k} : j \in \Z_{m, k} \cup \Dum_{m, k} \rbrace$ are i.i.d.

We define 
\begin{equation}
    g^{*}(j) \coloneqq \underset{g \in ( \Z_{m, k} \cup \Dum_{m, k} ) \backslash \lbrace j \rbrace}{\arg\max} \big\lbrace | G_{g, m, k} | \big\rbrace,
    \label{eq: largest absolute correlation value LARS}
\end{equation}
i.e., the largest absolute correlation with the current residual among all non-included nulls and dummies (except for variable~$j$) in the $m$th \textit{LARS} step. Since in each step $m$, the \textit{LARS} algorithm includes the variable with the largest absolute correlation with the current residual, we have
\begin{align}
    \mathbb{P}(&\text{``}X_{j} \text{ included in $m$th step of $k$th RE''} \mid j \in \Z_{m, k} \cup \Dum_{m, k} ) 
    \\[-0.5em]
    &= \mathbb{P}\big( | G_{j, m, k} | \geq | G_{g^{*}(j), m, k} | \,\, \big| \,\, j \in \Z_{m, k} \cup \Dum_{m, k} \big).
    \label{eq: LARS selection probability null and dummies}
\end{align}
Summing up the probabilities in~\eqref{eq: LARS selection probability null and dummies} over all $j^{\prime} \in \Z_{m, k} \cup \Dum_{m, k}$ yields
\begin{align}
    1 &= \,\,\,\quad \smashoperator{\sum\limits_{j^{\prime} \in \Z_{m, k} \cup \Dum_{m, k}}} \mathbb{P}\big( | G_{j^{\prime}, m, k} | \geq | G_{g^{*}(j^{\prime}), m, k} | \,\, \big| \,\, j^{\prime} \in \Z_{m, k} \cup \Dum_{m, k} \big)
    \label{eq: sum of all selection probabilities of nulls and dummies - 1}
    \\
    &= | \Z_{m, k} \cup \Dum_{m, k} | 
    \\[0.5em]
    &\qquad \cdot \mathbb{P}\big( | G_{j, m, k} | \geq | G_{g^{*}(j), m, k} | \,\, \big| \,\, j \in \Z_{m, k} \cup \Dum_{m, k} \big)
    \label{eq: sum of all selection probabilities of nulls and dummies - 2}
\end{align}
for all $j \in \Z_{m, k} \cup \Dum_{m, k}$. The second line follows from the fact that the $G_{j, m, k}$'s are exchangeable because they are i.i.d. Exchangeability is meant in the sense that
\begin{align}
    \mathbb{P}\big(&| G_{j_{1}, m, k} | \geq | G_{g^{*}(j_{1}), m, k} | \,\, \big| \,\, j_{1} \in \Z_{m, k} \cup \Dum_{m, k} \big) 
    \\[1em]
    &= \mathbb{P}\big( | G_{j_{2}, m, k} | \geq | G_{g^{*}(j_{2}), m, k} | \,\, \big| \,\, j_{2} \in \Z_{m, k} \cup \Dum_{m, k} \big)
    \label{eq: exchangeability of the correlation values}
\end{align}
for all $j_{1}, j_{2} \in \Z_{m, k} \cup \Dum_{m, k}$.

Combining~\eqref{eq: LARS selection probability null and dummies} and~\eqref{eq: sum of all selection probabilities of nulls and dummies - 2}, we obtain
\begin{align}
    \mathbb{P}(&\text{``}X_{j} \text{ included in $m$th step of $k$th RE''} \mid j \in \Z_{m, k} \cup \Dum_{m, k} )
    \\[-1.5em]
     &= \dfrac{1}{| \Z_{m, k} \cup \Dum_{m, k} |}
    \label{eq: equal probability for nulls and dummies}
\end{align}
for all $j \in \Z_{m, k} \cup \Dum_{m, k}$.
\label{proof: corollary 1}
\end{proof}

\subsection{Proof of Theorem~\ref{theorem: T-Rex algorithm} (Optimality of Algorithm~\ref{algorithm: T-Rex})}
\label{subsec: Proof of Theorem 3 (Optimality of Algorithm 1)}
\begin{proof}
First, note that for all triples $(v, T, L)$ that satisfy $\widehat{\FDP}(v, T, L) \leq \alpha$, the objective functions in Step~\ref{algorithm: T-Rex Step 4} of Algorithm~\ref{algorithm: T-Rex} and in the optimization problem in~\eqref{eq: T-Rex relaxed optimization problem} are equivalent, i.e., $\big| \widehat{\A}_{L}(v, T) \big| = R_{T, L}(v)$. Thus, in order to prove that $(v^{*}, T^{*})$ is an optimal solution of~\eqref{eq: T-Rex relaxed optimization problem}, it must be shown that the set of feasible tuples obtained by the algorithm contains the feasible set of~\eqref{eq: T-Rex relaxed optimization problem}. This also proves that $(v^{*}, T^{*})$ is a feasible solution of~\eqref{eq: T-Rex optimization problem 1} and~\eqref{eq: T-Rex optimization problem 2} because the conditions of the optimization problems in~\eqref{eq: T-Rex optimization problem 1},~\eqref{eq: T-Rex optimization problem 2}, and~\eqref{eq: T-Rex relaxed optimization problem} are equivalent.

Since, ceteris paribus, $\widehat{\FDP}(v, T, L)$ is monotonically decreasing in $v$ and monotonically increasing in $T$, the minimum of $\widehat{\FDP}(v, T, L)$ is attained at $v = 1 - \Delta v$, $\Delta v = 1 / K$, for any $T = T_{\fin}$ that satisfies the inequalities $\widehat{\FDP}(v = 1 - \Delta v, T = T_{\fin}, L) \leq \alpha$ and $\widehat{\FDP}(v = 1 - \Delta v, T = T_{\fin} + 1, L) > \alpha$. All in all, and since $v = 1 - \Delta v$ asymptotically ($K \rightarrow \infty$) coincides with the supremum of the interval $[0.5, 1)$, the feasible set of~\eqref{eq: T-Rex relaxed optimization problem} can be rewritten as follows:
\begin{align}
&\big\lbrace (v, T) : \widehat{\FDP}(v, T, L) \leq \alpha \rbrace 
\\[0.2em]
&= \Big\lbrace (v, T) : v \in [0.5, 1 - \Delta v],\, T \in \lbrace 1, \ldots, T_{\fin} \rbrace,
\\[0.2em]
& \qquad \qquad \quad \widehat{\FDP}(v, T, L) \leq \alpha \Big\rbrace.
\label{eq: proof T-Rex algorithm-1}
\end{align}
Note that the $v$-grid in Algorithm~\ref{algorithm: T-Rex} is adapted to the number of random experiments $K$ and, therefore, all values of the objective function (i.e., $| \widehat{\A}_{L}(v, T) |$) that can be attained by off-grid solutions can also be attained by at least one on-grid solution. Therefore, we can replace the right side of Equation~\eqref{eq: proof T-Rex algorithm-1} by
\begin{align}
\Big\lbrace (v, T) : \, & v \in \lbrace  0.5, 0.5 + \Delta v, 0.5 + 2 \cdot \Delta v, \ldots, 1 - \Delta v \rbrace,
\\[0.2em]
& T \in \lbrace 1, \ldots, T_{\fin} \rbrace, \, \widehat{\FDP}(v, T, L) \leq \alpha \Big\rbrace.
\label{eq: proof T-Rex algorithm-1.1}
\end{align}
The ``while''-loop in Step~\ref{algorithm: T-Rex Step 3} of Algorithm~\ref{algorithm: T-Rex} is terminated when $T = T_{\fin} + 1$. Thus, the feasible set of the optimization problem in Step~\ref{algorithm: T-Rex Step 4} of Algorithm~\ref{algorithm: T-Rex} can be written as follows:
\begin{align}
\Big\lbrace (v, T) : \, & v \in \lbrace  0.5, 0.5 + \Delta v, 0.5 + 2 \cdot \Delta v, \ldots, 1 - \Delta v \rbrace,
\\[0.2em]
& T \in \lbrace 1, \ldots, T_{\fin} \rbrace, \, \widehat{\FDP}(v, T, L) \leq \alpha \Big\rbrace.
\label{eq: proof T-Rex algorithm-2}
\end{align}
Since~\eqref{eq: proof T-Rex algorithm-1.1} is equal to~\eqref{eq: proof T-Rex algorithm-2}, the theorem follows.
\label{proof: T-Rex algorithm}
\end{proof}

\section{Main Ingredients of the T-Rex Selector}
\label{sec: Main Ingredients of the T-Rex Selector}
%
\begin{figure*}[t]
  \centering
  		\scalebox{0.851}{
  			\includegraphics[width = 0.5\linewidth]{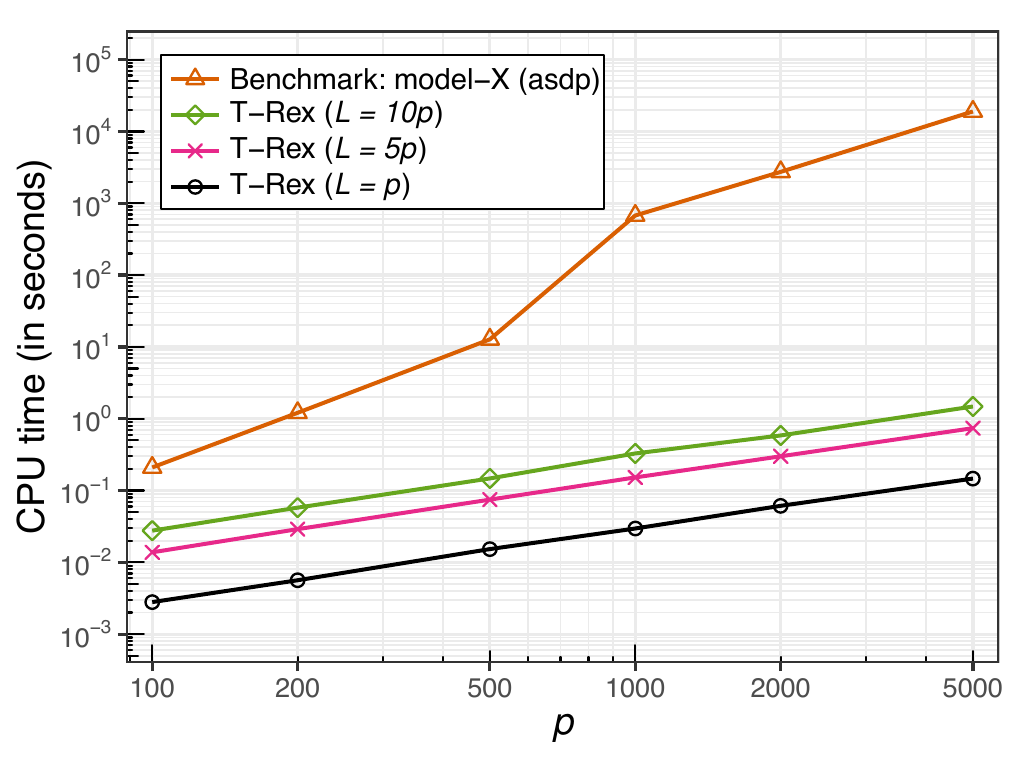}
  		}
 \caption[\textbf{Ingredient 1} - sampling dummies from the univariate standard normal distribution. The sequential computation time of generating one dummy matrix for the proposed \textit{T-Rex} selector is multiple orders of magnitude lower than the computation time of generating a knockoff matrix for the \textit{model-X} knockoff method, which is a current benchmark. Although the \textit{T-Rex} selector requires, e.g., $K = 20$ of such dummy matrices, its computation time is still multiple orders of magnitude lower than that of the \textit{model-X} knockoff method. For example, for $p = 5{,}000$ and $L = p$, the \textit{T-Rex} dummy generation process requires less than a second as compared to more than five hours for the \textit{model-X} knockoff method. The jump in computation time for the \textit{model-X} knockoff method between $p = 500$ and $p = 1{,}000$ is due to the suggestion of the authors to solve their proposed approximate semi-definite program (asdp) instead of their original semi-definite program for $p > 500$ in order to reduce the computation time required to generate \textit{model-X} knockoffs. Note that both axes are scaled logarithmically. Setup: $n = 300$, $MC = 955$.]{\textbf{Ingredient 1} - sampling dummies from the univariate standard normal distribution. The sequential computation time of generating one dummy matrix for the proposed \textit{T-Rex} selector is multiple orders of magnitude lower than the computation time of generating a knockoff matrix for the \textit{model-X} knockoff method, which is a current benchmark. For example, for $p = 5{,}000$ and $L = p$, the \textit{T-Rex} dummy generation process requires less than a second as compared to more than five hours for the \textit{model-X} knockoff method. Even taking into account that the \textit{T-Rex} selector requires, e.g., $K = 20$ of such dummy matrices, its sequential computation time is still multiple orders of magnitude lower than that of the \textit{model-X} knockoff method. The jump in computation time for the \textit{model-X} knockoff method between $p = 500$ and $p = 1{,}000$ is due to the suggestion of the authors to solve their proposed approximate semi-definite program (asdp) instead of their original semi-definite program for $p > 500$ in order to reduce the computation time required to generate \textit{model-X} knockoffs.\footnotemark Note that both axes are scaled logarithmically. Setup: $n = 300$, $MC = 955$.}
  \label{fig: Ingredient 1 - dummy generation}
\end{figure*}
\begin{figure*}[!htbp]
  \centering
     \captionsetup{width=0.47\linewidth}
  \subfloat[Number of selected variables vs. $T$. Setup: $n = 150$, $p = 300$, $p_{1} = 5$, $v = 0.8$, $L = p$, $K = 20$, $\text{SNR} = 1$, $MC = 500$.]{
  		\scalebox{0.779}{
  			\includegraphics[width = 0.413\linewidth, valign = t]{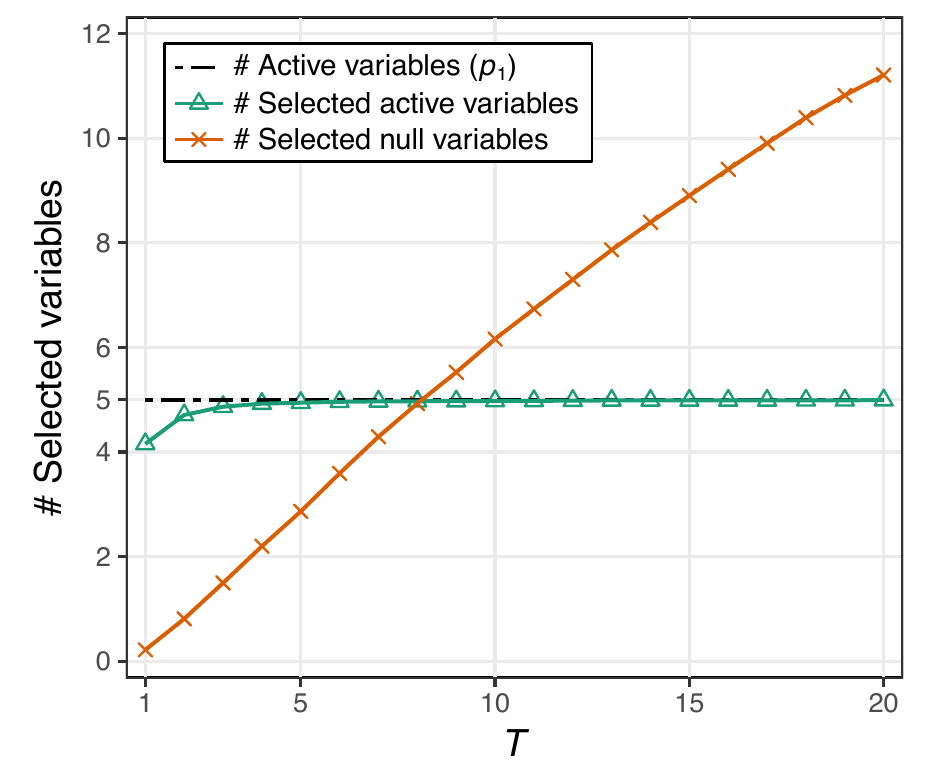}
  			\vphantom{\includegraphics[width = 0.404\linewidth, valign = t]{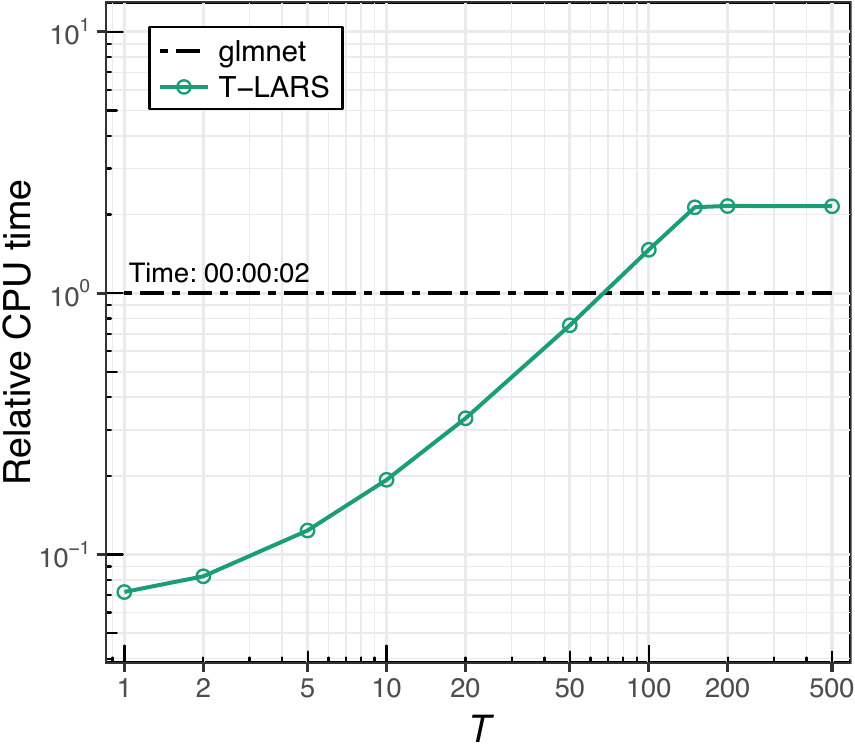}}
  		}
   		\label{fig: V_vs_T}
   }  
	\hspace{7em}
  \subfloat[The pathwise coordinate descent algorithm from the R package ``glmnet'' \cite{Friedman2010glmnet} (used to compute the \textit{Lasso} on a $\lambda$-grid with $500$ values) and the Terminating-\textit{LARS} (\textit{T-LARS}) algorithm from the R package ``tlars'' \cite{machkour2022tlars}. Note that both axes are scaled logarithmically. Setup: $n = 300$, $p = 5{,}000$, $p_{1} = 5$, $L = p$, $\text{SNR} = 1$, $MC = 955$.]{
  		\scalebox{0.779}{
  			\includegraphics[width = 0.404\linewidth, valign = t]{Figures/comp_time_LARS_vs_glmnet_p_5.000_with_tlars_package.pdf}
  		}
   		\label{fig: comp_time_LARS_vs_glmnet}
   }
      \captionsetup{width=1\linewidth}
  \caption{\textbf{Ingredient 2} - early terminating the solution paths of the random experiments. Figure (a) exemplifies that, on average, the number of selected active variables quickly increases towards the sparsity level $p_{1}$ (i.e., the number of active variables) and already for three included dummies almost all active variables are selected on average. However, the number of selected null variables also increases with increasing $T$. Figure (b) illustrates that for $p = 5{,}000$ and $L = p$, when terminated early, the Terminating-\textit{LARS} (\textit{T-LARS}) algorithm (a fundamental building block of the \textit{T-Rex} selector) is substantially faster than fitting the entire \textit{Lasso} solution path using the pathwise coordinate descent algorithm for $2p$ variables as it is done by the \textit{fixed-X} and \textit{model-X} knockoff methods. Although the \textit{T-Rex} selector needs to run the \textit{T-LARS} algorithm for, e.g., $K = 20$ random experiments within the \textit{T-Rex} selector, its sequential computation time is still comparable to that of a single run of ``glmnet'' in high-dimensional settings where $p$ is much larger than $n$. Moreover, the independent random experiments can be run in parallel on multicore computers to achieve a substantial reduction in computation time. The ``glmnet'' computation time is used as the reference computation time and its absolute value is given above the reference line (format: hh:mm:ss). Note that after $T = 150$ dummies are included the computation time of the \textit{T-LARS} algorithm does not increase further because the \textit{T-LARS} algorithm includes at most $\min\lbrace n, p + L \rbrace = n = 300$ variables and with $T = 150$ we can expect that, on average, also $150$ null variables plus the $5$ active variables are included.}
  \label{fig: Ingredient 2 - early stopping}
\end{figure*}
\begin{figure*}[!htbp]
  \centering
  		\scalebox{0.851}{
			\includegraphics[width = 0.5\linewidth]{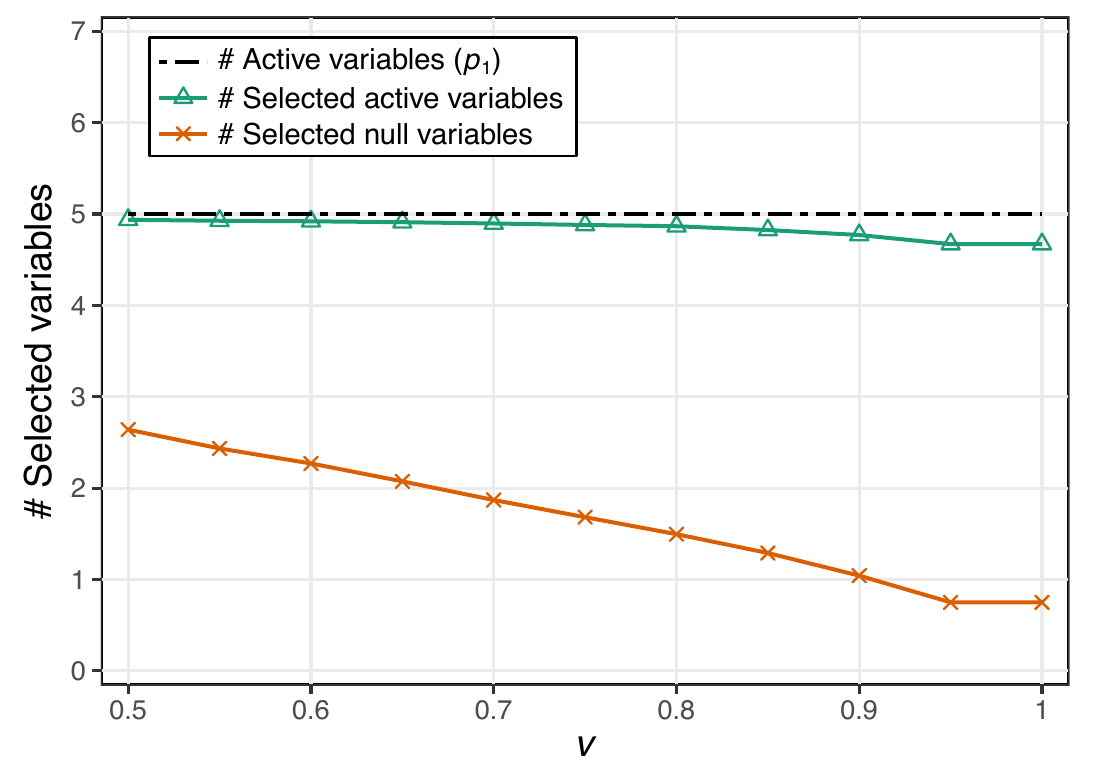}
  		}
  \caption{\textbf{Ingredient 3} - fusing the candidate sets based on their relative occurrences and a voting level $v \in [0.5, 1)$. The number of selected active variables remains high when increasing the voting level, while the number of selected null variables decreases faster with increasing $v$. Setup: $n = 150$, $p = 300$, $p_{1} = 5$, $T = 3$, $L = p$, $K = 20$, $\text{SNR} = 1$, $MC = 500$.}
  \label{fig: Ingredient 3- voting}
\end{figure*}

\newpage
\footnotetext{See the default parameters in the R package implementing the \textit{fixed-X} method and the \textit{model-X} method, which is available at \url{https://CRAN.R-project.org/package=knockoff} (last access: January 31, 2024).}
%
The following example helps to develop an intuition for the three main ingredients of the \textit{T-Rex} selector, which are
\begin{enumerate}[label=\arabic*., ref=\arabic*]
\item sampling dummies from the univariate standard normal distribution (see Figure~\ref{fig: Ingredient 1 - dummy generation}),
\item early terminating the solution paths of the random experiments (see Figure~\ref{fig: Ingredient 2 - early stopping}), and
\item fusing the candidate sets based on their relative occurrences and a voting level $v \in [0.5, 1)$ (see Figure~\ref{fig: Ingredient 3- voting}).
\end{enumerate}
In the example, we generate sparse high-dimensional data sets with $n$ observations and $p$ predictors and a response that is generated by the linear model in~\eqref{eq: linear model}. Further, $\beta_{j} = 1$ for active variables and $\beta_{j} = 0$ for null variables. The predictors are sampled from the standard normal distribution. The standard deviation~$\sigma$ is chosen such that the signal-to-noise ratio (SNR), which is given by $\Var[ \X\bbeta] \, / \, \sigma^{2}$, is equal to one.\footnote{Note that, in this case, $\Var$ denotes the sample variance operator.} The specific values of $n$, $p$, $p_{1}$ (i.e., the number of active variables), $v$, $T$, $L$, $K$, $\text{SNR}$, and $\text{MC}$ (i.e., the number of Monte Carlo realizations that the results are averaged over) are reported along with the discussion of the results in Figures~\ref{fig: Ingredient 1 - dummy generation}, \ref{fig: Ingredient 2 - early stopping}, and \ref{fig: Ingredient 3- voting}.
%
%
%

\cleardoublepage
\section{The Deflated Relative Occurrence}
\label{sec: The Deflated Relative Occurrence}
In order to provide an intuitive explanation of the deflated relative occurrence, we rewrite the expression as follows:
\begin{strip}
\begin{align}
&\Phi_{T, L}^{\prime}(j) = 
\\[-2em]
&\sum\limits_{t = 1}^{T}  \left( 1 - \dfrac{\dfrac{1}{L - (t - 1)}}{\dfrac{\sum\limits_{q \in \widehat{\A}(0.5)} \Delta \Phi_{t, L}(q)}{p - \sum\limits_{q = 1}^{p} \Phi_{t, L}(q)}} \right) \Delta \Phi_{t, L}(j)
= \sum\limits_{t = 1}^{T} \left( 1 - \dfrac{ \overbrace{\dfrac{t - (t - 1)}{L - (t - 1)}}^{\text{(i)}}}{\underbrace{\dfrac{\dfrac{1}{K} \sum\limits_{k = 1}^{K} \left( \sum\limits_{q \in \widehat{\A}(0.5)} \mathbbm{1}_{k} (q, t, L) - \sum\limits_{q \in \widehat{\A}(0.5)} \mathbbm{1}_{k}(q, t - 1, L) \right)}{p - \dfrac{1}{K} \sum\limits_{k = 1}^{K} \sum\limits_{q = 1}^{p} \mathbbm{1}_{k}(q, t, L)}}_{\text{(ii)}}} \right) \Delta \Phi_{t, L}(j).
\label{eq: reformulated path-adapted relative occurrence-2}
\end{align}
\end{strip}
The last equation follows by rewriting the expression in the denominator within the first expression using Definition~\ref{definition: relative occurrence}. In the last expression, each element of the sum consists of $\Delta \Phi_{t, L}(j)$ multiplied with what we call the \textit{deflation factor}. That factor is computed by subtracting from one the fraction of
\begin{enumerate}
\item[(i)] the number of included dummies at step $t$, which is always one, divided by the number of non-included dummies up until step $t - 1$ and

\item[(ii)] the average number of included candidates at step $t$ divided by the average number of non-included candidates up until step $t$.
\end{enumerate}
That is, the larger (smaller) the fraction of included candidates at step $t$ compared to the fraction of included dummies at step $t$, the more (less) weight is given to the change in relative occurrence in that step. Loosely speaking, if the number of non-included null variables and dummies is equal in step $t - 1$ of the $k$th random experiment, then allowing one more dummy to enter the solution path leads, on average, to the inclusion of one more null variable. Thus, if going from step $t - 1$ to $t$ leads to the inclusion of many variables, then still only one null variable is expected to be among them and, therefore, the deflation factor for that step is close to one.
\begin{rmk}
The reader might wonder whether the deflation factors affect not only the inflated $\Delta \Phi_{t, L}(j)$'s of the null variables but also those of the active variables. In the following, we shall give an intuitive explanation of why the deflation factors have only a negligible effect on the $\Delta \Phi_{t, L}(j)$'s of the active variables: Since usually most active variables enter the solution paths early, i.e., at low values of $t$ and because they are accompanied by very few null variables, the deflation factor is close to one. For this reason, the $\Delta \Phi_{t, L}(j)$'s of the active variables are relatively unaffected. With increasing values of $t$, the $\Delta \Phi_{t, L}(j)$'s of the active variables are close to zero, because for active variables the increases in relative occurrence are usually high for low values of $t$ and, consequently, low (or even zero) at higher values of $t$. Summarizing, the deflation factors have little or no effect on the $\Delta \Phi_{t, L}(j)$'s of the active variables because for low values of $t$ they are close to one and for large values of $t$ the $\Delta \Phi_{t, L}(j)$'s of the active variables are close to zero or zero.
\label{remark: deflation factor}
\end{rmk}

\section{Hyperparameter Choices for the Extended Calibration Algorithm}
\label{sec: Hyperparameter Choices for the Extended Calibration Algorithm}
In this appendix, we discuss the choices of the reference voting level $\tilde{v}$ and the maximum values of $L$ and $T$, namely $L_{\max}$ and $T_{\max}$ for the extended calibration algorithm in Algorithm~\ref{algorithm: Extended T-Rex}:
\begin{enumerate}[label=\arabic*., ref=\arabic*]
\item $\tilde{v} = 0.75$: The choice of $\tilde{v}$ is a compromise between the $50\%$ and $100\%$ voting levels. Setting $\tilde{v} = 0.5$ would require low values of $L$ to push $\widehat{\FDP}(v = \tilde{v}, T, L)$ below the target FDR level while setting $\tilde{v} = 1$ would require very high values of $L$. Thus, $\tilde{v} = 0.75$ is a compromise between tight FDR control and memory consumption. Note that the FDR control property holds for any choice of $\tilde{v} \in [0.5, 1)$.
\item $L_{\max} = 10 p$: In order to allow for sufficiently large values of $L$ such that tight FDR control is possible while not running out of memory, setting $L_{\max} = 10 p$ has proven to be a practical choice. Note that the FDR control property in Theorem~\ref{theorem: FDR control} holds for any choice of $L$. However, we can achieve tighter FDR control with larger values of $L$.
\item $T_{\max} = \lceil n / 2 \rceil$: As discussed for the \textit{T-LARS} algorithm in the caption of Figure~\ref{fig: Ingredient 2 - early stopping}, the \textit{LARS} algorithm includes at most $\min \lbrace n, p \rbrace$ variables and in high-dimensional settings ($p > n$), the maximum number of included variables in each random experiment is $n$. Since for $L = p$ we expect roughly as many null variables as dummies in very sparse settings, choosing $T_{\max} = \lceil n / 2 \rceil$ ensures that the \textit{LARS} algorithm could potentially run until (almost) the end of the solution path. In contrast, for $L = 10 p$ we expect $10$ times as many dummies as null variables in very sparse settings. Thus, for $L = p$ we allow the solution paths to potentially run until the end, although this might only happen in rare cases, while for $L = 2 p, \ldots, 10 p$ we restrict the run length. This is a compromise between a higher computation time and a higher TPR (i.e., power) that are both associated with larger values of $T_{\max}$.
\end{enumerate}

\section{Computational Complexity}
\label{sec: Computational Complexity}
The computational complexities of sampling dummies from the univariate standard normal distribution and fusing the candidate sets are negligible compared to the computational complexity of the utilized forward selection method. Therefore, it is sufficient to analyze the computational complexities of the early terminated forward selection processes. We restrict the following analysis to the \textit{LARS} algorithm~\cite{efron2004least}, which also applies to the Lasso~\cite{tibshirani1996regression}.\footnote{Since the \textit{Lasso} solution path can be computed by a slightly modified \textit{LARS} algorithm, the \textit{Lasso} and the \textit{LARS} algorithm have the same computational complexity.} The $\kappa$th step of the \textit{LARS} algorithm has the complexity $\mathcal{O} \big( (p - \kappa) \cdot n + \kappa^{2} \big)$, where the terms $(p - \kappa) \cdot n$ and $\kappa^{2}$ account for the complexity of determining the variable with the highest absolute correlation with the current residual (i.e., the next to be included variable) and the so-called equiangular direction vector, respectively. Replacing $p$ by $p + L$, since the original predictor matrix is replaced by the enlarged predictor matrix, and summing up the complexities of all steps until termination yields the computational complexity of the \textit{T-Rex} selector. First, we define the run lengths as the cardinalities of the respective candidate sets, i.e.,
\begin{equation}
\kappa_{T, L}(k) \coloneqq \big| \C_{k, L}(T) \big|, \quad k = 1, \ldots, K,
\label{eq: maximum run length forward selection process}
\end{equation}
and assume $L \geq p$. Then, the sum over all steps until the termination of the $k$th random experiment is given by
\begin{align}
\sum_{\kappa = 1}^{\kappa_{T, L}(k)} &\big( (p + L - \kappa) \cdot n + \kappa^{2} \big)
\\
&= n \cdot  \kappa_{T, L}(k) \cdot (p + L) - n\cdot \sum_{\kappa = 1}^{\kappa_{T, L}(k)} \kappa + \sum_{\kappa = 1}^{\kappa_{T, L}(k)} \kappa^{2} \\
&\leq n \cdot  \kappa_{T, L}(k) \cdot (p + L) + \big( \kappa_{T, L}(k) \big)^{3} \\
&\leq 2 \cdot n \cdot  \kappa_{T, L}(k) \cdot (p + L).
\end{align}
We can write $L = \lceil \eta \cdot p \rceil$, $\eta > 0$, and the expected run length can be upper bounded as follows:
\begin{equation}
\mathbb{E} \big[ \kappa_{T, L}(k) \big] \leq p_{1} + T + \mathbb{E} \big[ \Psi \big] = p_{1} + T + \dfrac{T}{L + 1} \cdot p_{0} \leq p_{1} + 2T,
\end{equation}
where the first equation follows from $\Psi \sim \NHG(p_{0} + L, p_{0}, T)$ and the second inequality holds because $L \geq p$. So, the expected computational complexity of one random experiment of the proposed \textit{T-Rex} selector is $\mathcal{O}(np)$. Although the theoretical FDR control result requires $K \rightarrow \infty$, as stated in Section~\ref{subsec: The T-Rex Selector: Methodology}, choosing $K \geq 20$ provides excellent empirical results and we did not observe any notable improvements for $K \geq 100$. Therefore, with fixed $K$ (e.g., $K = 20$), the overall expected computational complexity of the \textit{T-Rex} selector is $\mathcal{O}(np)$. The computational complexity of the original (i.e., non-terminated) \textit{LARS} algorithm in high-dimensional settings is $\mathcal{O}(p^{3})$. Thus, on average the high computational complexity of the \textit{LARS} algorithm does not carry over to the \textit{T-Rex} selector because within the \textit{T-Rex} selector the solution paths of the random experiments are early terminated. Moreover, the computational complexity of the \textit{T-Rex} selector is the same as that of the pathwise coordinate descent algorithm~\cite{Friedman2010glmnet}.

\section{General Assumptions}
\label{sec: General Assumptions}
It is important to note that existing theory for FDR control in high-dimensional settings, i.e., the \textit{model-X} knockoff methods~\cite{candes2018panning}, relies on an accurate estimation of the covariance matrix of the predictors, which is known to not be possible, in general, when $p \gg n$ (see, e.g., Figure~7 in~\cite{candes2018panning}). Further, the knockoff generation algorithm in~\cite{candes2018panning} is practically infeasible due to its exponential complexity in $p$ and the authors resort to second-order \textit{model-X} knockoffs for which no FDR control proof exists. In contrast, the \textit{T-Rex} selector does not rely on an accurate estimate of a high-dimensional covariance matrix and does not resort to an approximation of its theory to obtain a feasible algorithm.

Instead, to establish the FDR control theory for the \textit{T-Rex} selector, we will introduce two general and mild assumptions that are thoroughly verified in relevant use-cases and especially for non-Gaussian simulated genomics data using the software HAPGEN2~\cite{su2011hapgen2} (see Appendices~\ref{sec: Exemplary Numerical Verification of Assumptions} and~\ref{sec: Setup, Preprocessing, and Additional Results: Simulated Genome-Wide Association Study}).

Knockoff methods~\cite{barber2015controlling}, as well as many popular FDR-controlling methods (i.e.,~\cite{benjamini1995controlling,storey2004strong,gavrilov2009adaptive}) assume that the null $p$-values are i.i.d. and uniformly distributed between $0$ and $1$. In particular, to prove the FDR control property of the knockoff methods in~\cite{barber2015controlling,candes2018panning}, the authors assume that the null $p$-values
    \begin{enumerate}[label=\arabic*., ref=\arabic*]
        \item are i.i.d.,
        \label{standard assumption FDR control - 1}
        \item are independent of the $p$-values corresponding to the true active variables, and
        \label{standard assumption FDR control - 2}
        \item stochastically dominate a random variable following the uniform distribution with support between $0$ and $1$.
        \label{standard assumption FDR control - 3}
    \end{enumerate}
Since we do not use $p$-values, we make a different assumption and explain how our weaker assumption is implied by the aforementioned standard assumptions.

\begin{assumption}
Let $\A$ and $\Z$ be the index sets of the true active and the null variables, respectively, and let the candidate variables $X_{1}, \ldots, X_{p}$ be standardized (i.e., $\mathbb{E}[X_{j}] = 0$ and $\Var[X_{j}] = 1$ for all $j \in \lbrace 1, \ldots, p \rbrace$) and follow probability distributions with finite mean and variance. Then,
\begin{enumerate}
    \item[(i)] $X_{j}$ is independent of $\lbrace X_{g} : g \in \A \rbrace$ for all $j \in \Z$, i.e., the null variables are independent of the true active variables,
    \item[(ii)] $\lbrace X_{j} : j \in \Z \rbrace$ is a set of independent random variables, i.e., the nulls are mutually independent.
\end{enumerate}
\label{assumption: 1}
\end{assumption}

\begin{rmk}
Points~\ref{standard assumption FDR control - 1} and~\ref{standard assumption FDR control - 2} of the above standard assumption in FDR control theory state that the null $p$-values are i.i.d. and independent of the $p$-values corresponding to the true active variables. This can also be stated in terms of test statistics. That is, the test statistics corresponding to the null variables are i.i.d. and independent of the test statistics corresponding to the true active variables. Null $p$-values are defined by
\begin{equation}
    P_{j} = 1 - F_{0}(T_{j}),
    \label{eq: alternative formulation of p-values}
\end{equation}
where $P_{j}$ and $T_{j}$ are the null $p$-value and the null test statistic, respectively, corresponding to the $j$th null variable and $F_{0}(\cdot)$ is the distribution of the null test statistics~\cite{sackrowitz1999p}. From this definition of $p$-values, it is obvious that this assumption can be stated equivalently either in terms of $p$-values or test statistics, which is frequently done~\cite{benjamini1995controlling,benjamini2001control,storey2004strong,barber2015controlling}. As stated in~\cite{barber2015controlling} (p. 2075), especially in the case where the test statistics stem from the regression coefficient estimates $\hatbbeta = \big[ \hat{\beta}_{1} \, \cdots \, \hat{\beta}_{p} \big]^{\top} \sim \mathcal{N} \big( \bbeta, \sigma^{2} (\X^{\top}\X)^{-1} \big)$, the coefficient estimates (and the test statistics) are mutually independent if and only if $\X^{\top}\X$ is a diagonal matrix (i.e., orthogonal design). This implies that the null test statistics are mutually independent and independent of the test statistics corresponding to the true active variables if and only if the null variables are mutually independent and independent of the true active variables. Note that this is what we are stating in our A-\ref{assumption: 1}. Thus, the standard assumption in FDR control theory implies A-\ref{assumption: 1} and, since this implication does not require Point~\ref{standard assumption FDR control - 3} of the above standard assumption in FDR control theory, A-\ref{assumption: 1} is weaker.
\label{remark: null p values iid and independent of true active p values implies our assumption}
\end{rmk}
For a numerical verification of A-\ref{assumption: 1} in relevant use-cases and especially for non-Gaussian simulated genomics data using the software HAPGEN2~\cite{su2011hapgen2}, see Appendices~\ref{sec: Exemplary Numerical Verification of Assumptions} and~\ref{sec: Setup, Preprocessing, and Additional Results: Simulated Genome-Wide Association Study}.

\begin{rmk}
In the genomics literature, it is well-known that SNPs (i.e., variables) form groups of highly correlated SNPs. The biological phenomenon that leads to such dependency structures is called linkage disequilibrium~\cite{reich2001linkage} (see also Appendix~\ref{sec: Setup, Preprocessing, and Additional Results: Simulated Genome-Wide Association Study}. It is common in genomics research to use pruning methods to group SNPs and to keep only one representative SNP from each group and, thus, drastically reduce the dependencies among the SNPs before applying any variable selection procedure~(see, e.g.,~\cite{sesia2019gene} and references therein). Therefore, SNP pruning is a valid method to satisfy A-\ref{assumption: 1} in practice. When choosing the amount of pruning (i.e., the number of groups that the SNPs are grouped into) one must consider the trade-off between 
\begin{enumerate}
    \item the reduction of dependencies among SNPs (by creating few SNP groups) and
    \item the increase of the resolution of the to be detected regions on the genome (by creating many SNP groups).
\end{enumerate}
For details on how this trade-off is commonly tuned for GWAS, see Appendix~\ref{sec: Setup, Preprocessing, and Additional Results: Simulated Genome-Wide Association Study}.
\label{remark: SNP pruning allows to satisfy Assumption 1}
\end{rmk}

As shown in Figure~\ref{fig: T-Rex selector}, the estimator of the active set $\widehat{\A}(v)$ results from fusing the candidate sets $\C_{1, L}(T), \ldots, \C_{K, L}(T)$ based on a voting level that is applied to the relative occurrences of the candidate variables. Therefore, the number of selected null variables $V_{T, L}(v)$ is related to the distribution of the number of included null variables in the terminal step $t = T$. We state this relationship as an assumption:
\begin{assumption}
For any $v \in [0.5, 1)$, the number of selected null variables is stochastically dominated by a random variable following the negative hypergeometric distribution with parameters specified in Corollary~\ref{corollary: 1}, i.e.,
\begin{equation}
V_{T, L}(v) \overset{d}{\leq} \NHG(p_{0} + L, p_{0}, T).
\label{eq: assumption-2.1}
\end{equation}
\label{assumption: 2}
\end{assumption}
For a numerical verification of A-\ref{assumption: 2} in relevant use-cases and especially for non-Gaussian simulated genomics data using the software HAPGEN2~\cite{su2011hapgen2}, see Appendices~\ref{sec: Exemplary Numerical Verification of Assumptions} and~\ref{sec: Setup, Preprocessing, and Additional Results: Simulated Genome-Wide Association Study}.

The expression for $\widehat{V}_{T, L}^{\prime}(v)$ from Remark~\ref{remark: upper-bounded FDP} can be rewritten as follows:
\begingroup
\allowdisplaybreaks
\begin{align}
&\widehat{V}_{T, L}^{\prime}(v)
\\[0.5em]
&= \sum\limits_{t = 1}^{T} \dfrac{p - \sum_{q = 1}^{p} \Phi_{t, L}(q)}{L - (t - 1)} \cdot \dfrac{\sum_{j \in \widehat{\A}(v)} \Delta \Phi_{t, L}(j)}{\sum_{q \in \widehat{\A}(0.5)} \Delta \Phi_{t, L}(q)}
\label{eq: rewritten-V_T_L-2}
\\[0.5em]
&= \sum\limits_{t = 1}^{T} \dfrac{p_{0} - \sum_{q \in \Z} \Phi_{t, L}(q) + \overbracket{p_{1} - \textstyle\sum_{q \in \A} \Phi_{t, L}(q)}}{L - (t - 1)} 
\\[0.5em]
& \qquad \qquad \cdot \dfrac{\sum_{j \in \widehat{\A}^{\, 0}(v)} \Delta \Phi_{t, L}(j) + \overbracket{\textstyle\sum_{j \in \widehat{\A}^{\, 1}(v)} \Delta \Phi_{t, L}(j)}}{\sum_{q \in \widehat{\A}^{\, 0}(0.5)} \Delta \Phi_{t, L}(q) + \overbracket{\textstyle\sum_{q \in \widehat{\A}^{\, 1}(0.5)} \Delta \Phi_{t, L}(q)}}
\label{eq: rewritten-V_T_L-3}
\\[0.5em]
&\approx \sum\limits_{t = 1}^{T} \dfrac{p_{0} - \sum_{q \in \Z} \Phi_{t, L}(q)}{L - (t - 1)} \cdot \dfrac{\sum_{j \in \widehat{\A}^{\, 0}(v)} \Delta \Phi_{t, L}(j)}{\sum_{q \in \widehat{\A}^{\, 0}(0.5)} \Delta \Phi_{t, L}(q)}
\label{eq: rewritten-V_T_L-5}
\end{align}
\endgroup
The marked terms consider only the relative occurrences of the active variables. Recall that, assuming that the variable selection method is better than random selection, almost all active variables are selected early, i.e., terminating the \textit{T-Rex} selector after a small number of $T$ dummies have been included allows to select almost all active variables (see Figure~\ref{fig: V_vs_T}). Thus, the relative occurrences of the active variables are approximately one for a sufficient number of included dummies. In consequence, and since $\Delta \Phi_{t, L} = \Phi_{t, L} - \Phi_{t - 1, L}$, $t \in \lbrace 1, \ldots, T \rbrace$, the $\Delta \Phi_{t, L}$'s of the active variables are approximately zero for a sufficiently large $t$ and $T$. This motivates the assumption that the marked terms can be neglected.
\begin{assumption}
For sufficiently large $T \in \lbrace 1, \ldots, L \rbrace$ it holds that
\begin{align}
\widehat{V}_{T, L}^{\prime}(v)
= \sum\limits_{t = 1}^{T} \dfrac{p_{0} - \sum_{q \in \Z} \Phi_{t, L}(q)}{L - (t - 1)} \cdot \dfrac{\sum_{j \in \widehat{\A}^{\, 0}(v)} \Delta \Phi_{t,L}(j)}{\sum_{q \in \widehat{\A}^{\, 0}(0.5)} \Delta \Phi_{t,L}(q)}.
\label{eq: assumption-3.1}
\end{align}
\label{assumption: 3}
\end{assumption}
See Appendices~\ref{sec: Exemplary Numerical Verification of Assumptions} and~\ref{sec: Setup, Preprocessing, and Additional Results: Simulated Genome-Wide Association Study} for the motivation, technical details, and extensive numerical verifications of A-\ref{assumption: 3}.

\section{Exemplary Numerical Verification of A-\ref{assumption: 1}, A-\ref{assumption: 2}, and A-\ref{assumption: 3}}
\label{sec: Exemplary Numerical Verification of Assumptions}
In this section, A-\ref{assumption: 1}, A-\ref{assumption: 2}, and A-\ref{assumption: 3} from Appendix~\ref{sec: General Assumptions} are verified. The general setup for the exemplary numerical verification of all assumptions is as described in Section~\ref{subsec: Setup and Results}. The specific values of the generic high-dimensional simulation setting in Section~\ref{subsec: Setup and Results} and the parameters of the proposed \textit{T-Rex} selector and the proposed extended calibration algorithm in Algorithm~\ref{algorithm: Extended T-Rex}, i.e., the values of $n$, $p$, $p_{1}$, $v$, $T$, $L$, $K$, and $\text{SNR}$ are specified in the figure captions. All results are averaged over $MC = 500$ Monte Carlo realizations. An additional verification for our use-case of GWAS is provided in Appendix~\ref{sec: Setup, Preprocessing, and Additional Results: Simulated Genome-Wide Association Study}.
\subsection{Exemplary Numerical Verification of A-\ref{assumption: 1}}
\label{subsec: Exemplary Numerical Verification of Assumption 1}
\begin{figure*}[!htbp]
  \centering
  \subfloat[Histogram and theoretical distribution for $t = T = 20$. Note that the histogram is based on $K = 20$ random experiments for each of the $500$ Monte Carlo realizations.]{
  		\scalebox{1.0}{
  			\includegraphics[width=0.4\linewidth, valign = t]{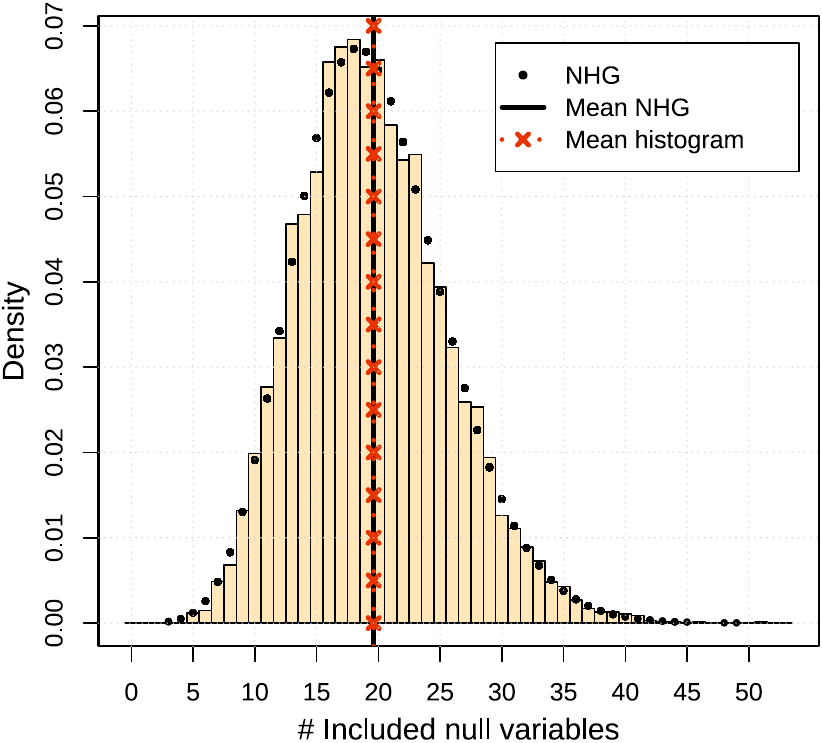}
  		}
   		\label{fig: NHG histogram}
   }
   \hspace{1em}
  \subfloat[Q-Q plot corresponding to Figure~(a).]{
  		\scalebox{1.0}{
  			\includegraphics[width=0.4\linewidth, valign = t]{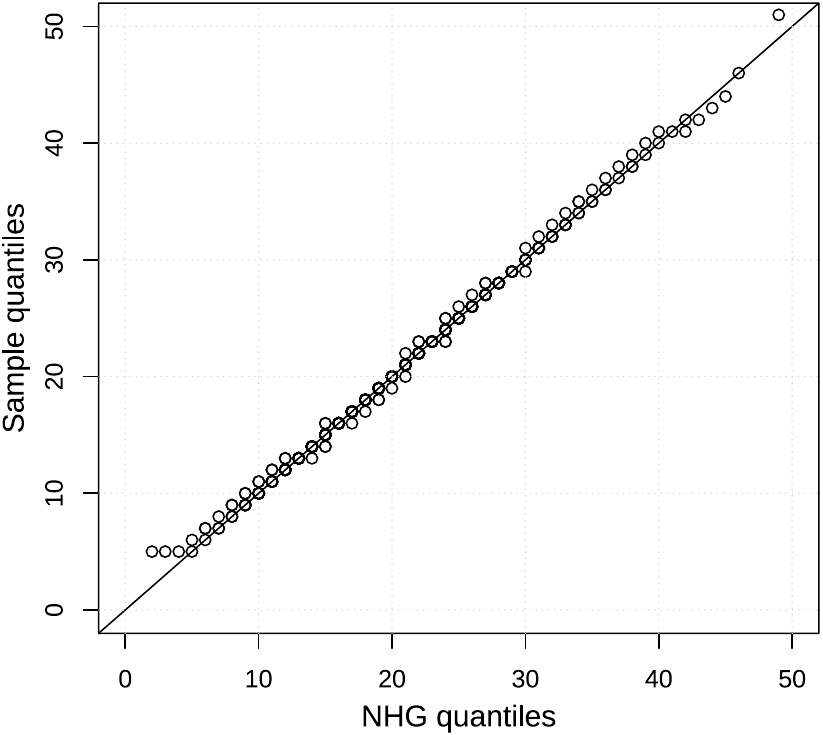}
  			\vphantom{\includegraphics[width=0.4\linewidth, valign = t]{Figures/NHG_histogram.pdf}}
  		}
   		\label{fig: NHG QQ-plot}
   }
  \caption{\textbf{Exemplary numerical verification of Corollary~\ref{corollary: 2} and A-\ref{assumption: 1}}: The histogram of the number of included null variables in Figure~(a) approximates the theoretical probability mass function (PMF). The expected value of a random variable following the negative hypergeometric distribution with the parameters specified in the last sentence of this caption is given by $T \cdot p_{0} \, / \, (L + 1) = 20 \cdot 290 \, / \, (300 + 1) \approx 19.27$, which fits the mean of the histogram. The Q-Q plot in Figure~(b) confirms that the number of included null variables follows the negative hypergeometric distribution. Setup: $n = 150$, $p = 300$, $p_{1} = 5$, $T = 20$, $L = p$, $K = 20$, $\text{SNR} = 1$, $MC = 500$.}
  \label{fig: NHG histogram and QQ-plot}
\end{figure*}

Figure~\ref{fig: NHG histogram} shows the histogram of the number of included null variables for $T = 20$ and for $500$ Monte Carlo replications consisting of $K = 20$ candidate sets each while Figure~\ref{fig: NHG QQ-plot} shows the corresponding Q-Q plot. The histogram closely approximates the probability mass function (PMF) of the negative hypergeometric distribution with the parameters specified in Corollary~\ref{corollary: 2}. Moreover, the points in the Q-Q plot closely approximate the ideal line. Thus, Figure~\ref{fig: NHG histogram and QQ-plot} provides an exemplary numerical verification of Corollary~\ref{corollary: 2} and, therewith, an implicit exemplary verification of A-\ref{assumption: 1}.
\subsection{Exemplary Numerical Verification of A-\ref{assumption: 2}}
\label{subsec: Exemplary Numerical Verification of Assumption 2}
\begin{figure}[h]
  \centering
  		\scalebox{0.9}{
  			\includegraphics[width=0.97\linewidth]{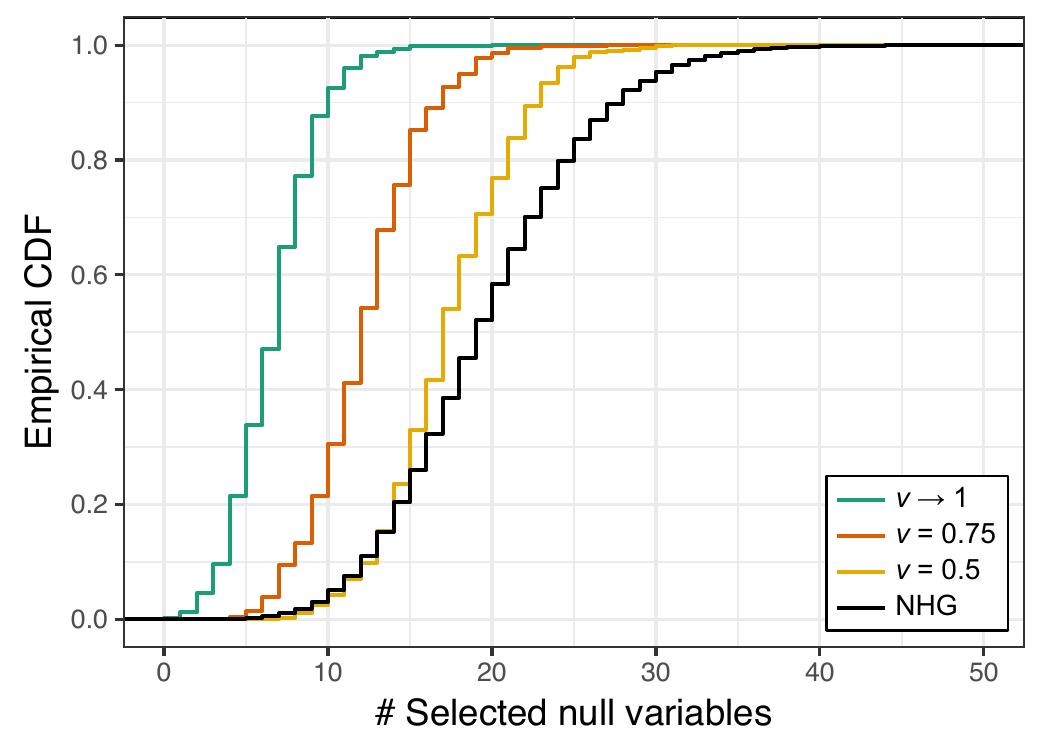}
  		}
  \caption{\textbf{Exemplary numerical verification of A-\ref{assumption: 2}}: For $v \geq 0.5$, a random variable following the negative hypergeometric distribution stochastically dominates the random variable $V_{T, L}(v)$ (i.e., the number of selected null variables) at almost all values of $V_{T, L}(v)$. Setup: $n = 150$, $p = 300$, $p_{1} = 5$, $T = 20$, $L = p$, $K = 20$, $\text{SNR} = 1$, $MC = 500$.}
  \label{fig: stochastic dominance cdf}
\end{figure}

Figure~\ref{fig: stochastic dominance cdf} shows the empirical cumulative distribution function (CDF) of $V_{T, L}(v)$ for $T = 20$ and different values of the voting level $v$ and the CDF of the negative hypergeometric distribution. The empirical CDFs are based on $500$ Monte Carlo replications. Already for a small number of random experiments, i.e., $K = 20$, the CDF of the negative hypergeometric distribution with its parameters being as specified in A-\ref{assumption: 2} lies below the empirical CDFs of $V_{T, L}(v)$ for all $v \geq 0.5$ at almost all values of $V_{T, L}(v)$. For values of $V_{T, L}(v)$ between $6$ and $12$, we observe that the CDF of the negative hypergeometric distribution lies slightly above the empirical CDF for $v = 0.5$. All in all, we conclude that a random variable following the negative hypergeometric distribution stochastically dominates $V_{T, L}(v)$ at almost all values and for all $v \geq 0.5$, which exemplarily verifies A-\ref{assumption: 2}.
\begin{figure*}[!htbp]
  \centering
  \subfloat[Approximations and true values for different choices of $v$ averaged over $500$ Monte Carlo realizations.]{
  		\scalebox{0.9}{
  			\includegraphics[width=0.4\linewidth]{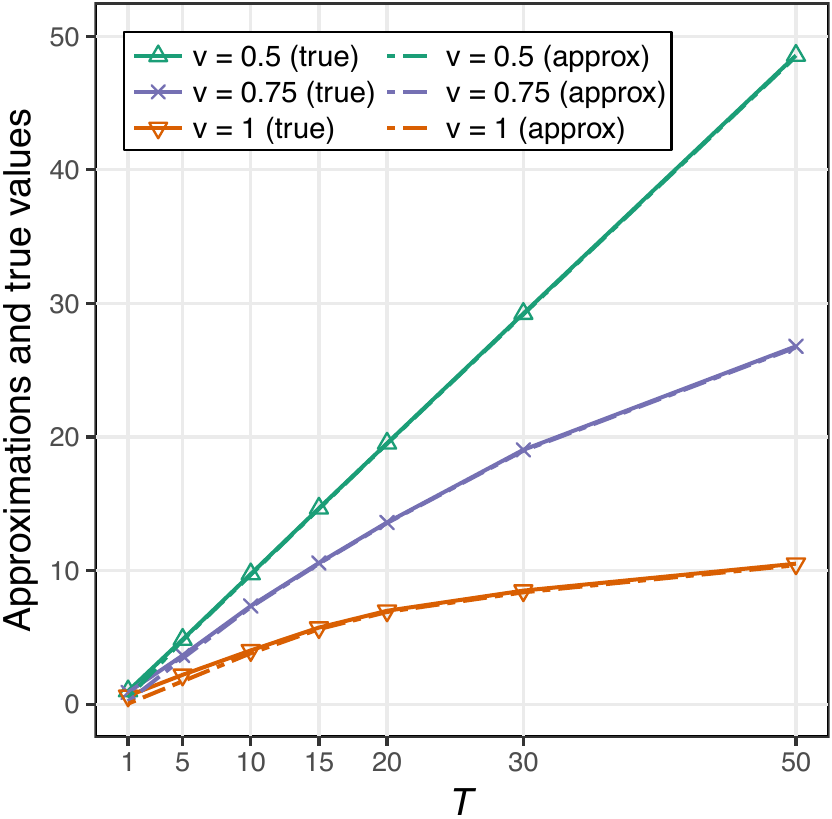}
  		}
   		\label{fig: relative_error_vs_v}
   }
     \hspace{1em}
  \subfloat[Approximations and true values for different choices of $T$ averaged over $500$ Monte Carlo realizations.]{
  		\scalebox{0.9}{
  			\includegraphics[width=0.4\linewidth]{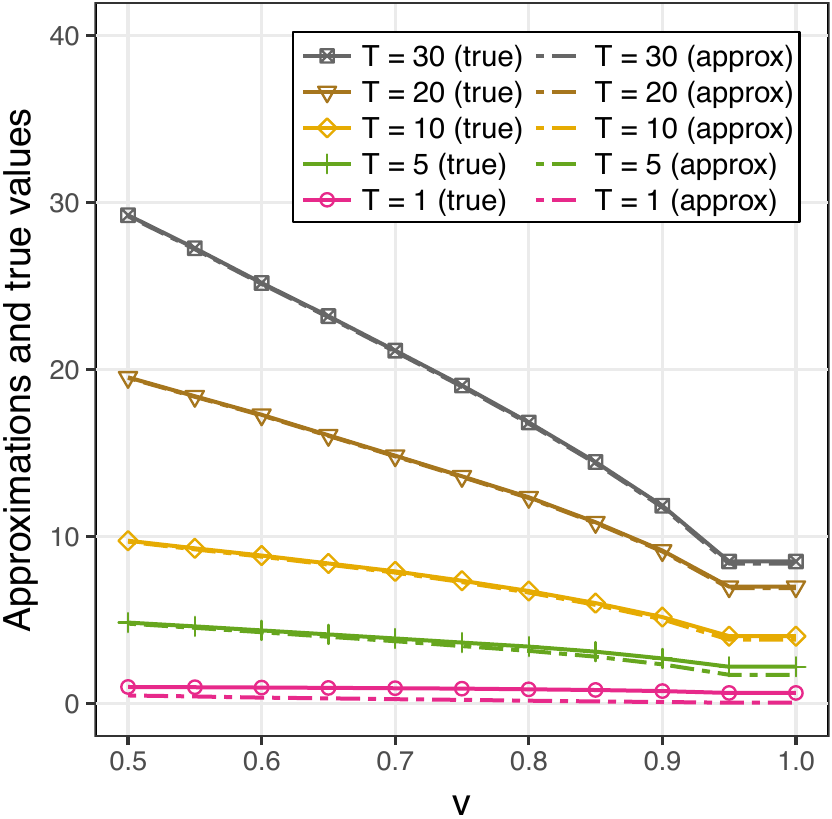}
  		}
   		\label{fig: relative_error_vs_v}
   }
   \\
  \subfloat[Box plots of approximations and true values corresponding to the lines for $v = 0.75$ in Figure (a).]{
  		\scalebox{0.9}{
  			\includegraphics[width=0.4\linewidth]{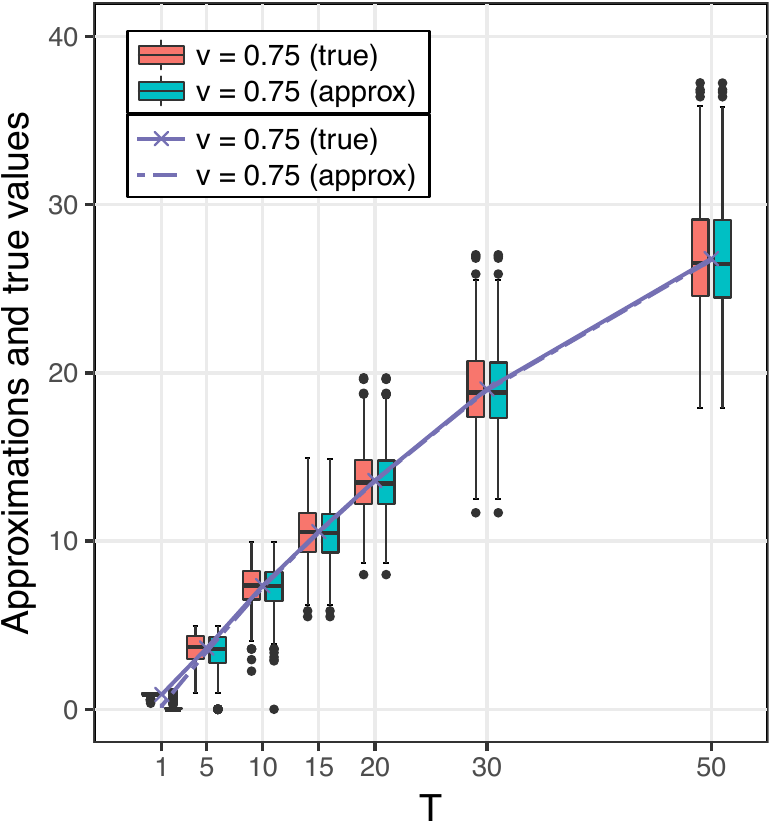}
  		}
   		\label{fig: relative_error_vs_v}
   }
     \hspace{1em}
  \subfloat[Box plots of approximations and true values corresponding to the lines for $T = 10$ in Figure (b).]{
  		\scalebox{0.9}{
  			\includegraphics[width=0.405\linewidth]{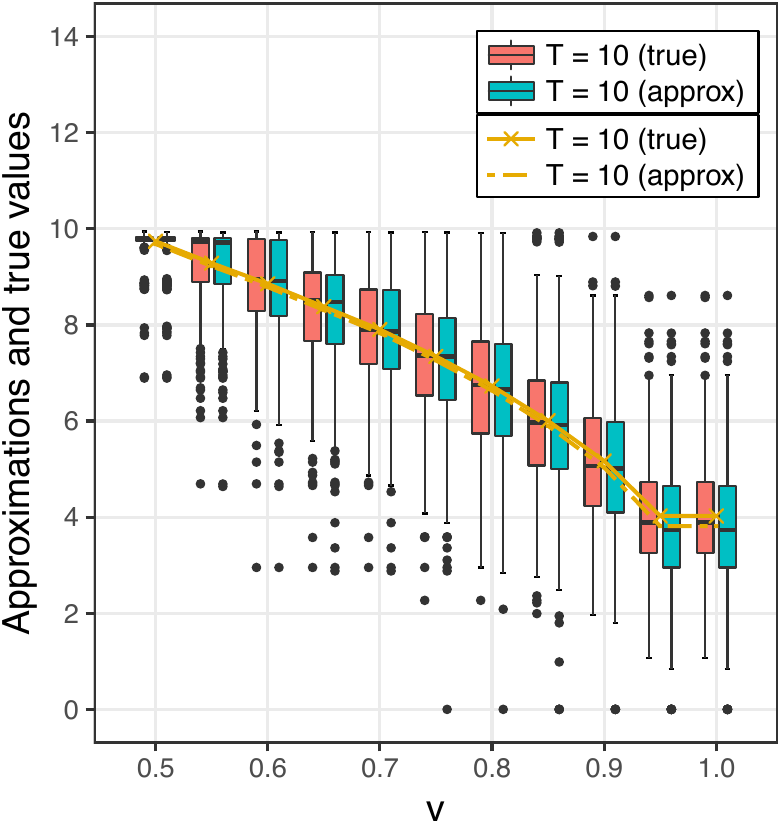}
  		}
   		\label{fig: relative_error_vs_v}
   }
  \caption{\textbf{Exemplary numerical verification of A-\ref{assumption: 3}}: In Figures (a) and (b), we see that the approximations and the true values are almost identical for different values of $v$ and $T$. The corresponding box plots in Figures (c) and (d) show that also the distributions of approximations and true values are very similar. Setup: $n = 150$, $p = 300$, $p_{1} = 5$, $L = p$, $K = 20$, $\text{SNR} = 1$, $MC = 500$.}
  \label{fig: relative_error_vs_T_and_v}
\end{figure*}
\subsection{Exemplary Numerical Verification of A-\ref{assumption: 3}}
\label{subsec: Exemplary Numerical Verification of Assumption 3}
An exemplary numerical verification of A-\ref{assumption: 3} is given in Figure~\ref{fig: relative_error_vs_T_and_v}, where we see that approximations and true values are almost identical for different choices of $v$ and $T$.

\section{Benchmark Methods for FDR Control}
\label{sec: Benchmark Methods for FDR Control}
As mentioned in Section~\ref{sec: Numerical Simulations}, the benchmark methods in low-dimensional settings (i.e., $p \leq n$) are the Benjamini-Hochberg (\textit{BH}) method~\cite{benjamini1995controlling}, the Benjamini-Yekutieli (\textit{BY}) method~\cite{benjamini2001control}, and the \textit{fixed-X} knockoff methods~\cite{barber2015controlling}, while the \textit{model-X} knockoff methods~\cite{candes2018panning} are the benchmarks in high-dimensional settings (i.e., $p > n$). These methods are briefly described and discussed in the following.

\subsection{The BH and the BY Method}
\label{subsec: The BH and the BY Method}
For low-dimensional sparse regression, we can formulate the null hypotheses $H_{j}: \beta_{j} = 0$, $j = 1, \ldots, p$ with associated $p$-values $P_{1}, \ldots, P_{p}$. Thus, when a variable is selected, we can interpret this as the rejection of the corresponding null hypothesis in favor of the alternative hypothesis. The \textit{BH} method and the \textit{BY} method were designed to control the FDR at the target level $\alpha \in [0, 1]$ for multiple hypothesis testing based on $p$-values. For all variables in the sparse regression setting, the $p$-values are computed and sorted in an ascending order. Then, the estimate of the number of active variables $\widehat{p}_{1}(\alpha)$ is determined by finding the largest $p$-value that does not exceed a threshold depending on $\alpha$ by solving
\begin{equation}
\widehat{p}_{1}(\alpha) = \max\bigg\lbrace m : P_{m} \leq \dfrac{m}{p \cdot c(p)}\cdot \alpha \bigg\rbrace,
\label{eq: BH and BY}
\end{equation}
where $c(p) = 1$ for the \textit{BH} method and $c(p) = \sum_{j = 1}^{p} 1/j \approx \ln(p) + \gamma$ for the \textit{BY} method with $\gamma \approx 0.577$ being the Euler-Mascheroni constant. If no such $\widehat{p}_{1}(\alpha)$ exists, then no hypothesis is rejected. Otherwise, the variables corresponding to the $\widehat{p}_{1}(\alpha)$ smallest $p$-values are selected. The \textit{BH} method requires independent hypotheses or, at least, a so-called positive regression dependency among the candidates to guarantee FDR control at the target level. In contrast, the \textit{BY} method provably controls the FDR at the target level and does not require independent hypotheses or any assumptions regarding the dependency among the hypotheses. However, the \textit{BY} method is more conservative than the \textit{BH} method, i.e., it achieves a considerably lower power than the \textit{BH} method at the same target FDR level.

\subsection{The fixed-X and the model-X Methods}
\label{subsec: The Fixed-X and the Model-X Method}
The \textit{fixed-X} knockoff method is a relatively new method for controlling the FDR in sparse linear regression settings. Since it requires $n \geq 2 p$ observations, it is not suitable for high-dimensional settings. The method generates a knockoff matrix $\XK$ consisting of $p$ knockoff variables and appends it to the original predictor matrix. Unlike for our proposed \textit{T-Rex} selector, the knockoff variables are designed to mimic the covariance structure of $\X$. Further, they are designed to be, conditional on the original variables, independent of the response. Hence, the knockoff variables act as a control group and when a knockoff variable enters the active set before its original counterpart it provides some evidence against this variable being a true positive. 

The predictor matrix of, e.g., the \textit{Lasso} optimization problem in~\eqref{eq: Lasso} is then replaced by $[\X \,\, \XK]$ and the $\lambda$-values corresponding to the first entry points of the original and knockoff variables are extracted from the solution path resulting in $Z_{j} = \sup\lbrace \lambda : \hat{\beta}_{j} \neq 0\,\, \text{first time} \rbrace$ and $\accentset{\circ}{Z}_{j} = \sup\lbrace \lambda : \hat{\beta}_{j+p} \neq 0\,\, \text{first time} \rbrace$, $j = 1,\ldots,p$. The authors suggest to use the test statistics
\begin{equation}
W_{j} = (Z_{j} \vee \accentset{\circ}{Z}_{j}) \cdot \text{sign}(Z_{j} - \accentset{\circ}{Z}_{j}), \quad j = 1, \ldots, p,
\label{eq: test statistic knockoff selector}
\end{equation}
and to determine the threshold 
\begin{equation}
\tau = \min \bigg \lbrace \tau^{\prime} \in \mathcal{W} : \dfrac{b + \big| \lbrace j : W_{j} \leq - \tau^{\prime} \rbrace \big|}{\big| \lbrace j : W_{j} \geq \tau^{\prime} \rbrace \big| \vee 1} \leq \alpha \bigg \rbrace,
\label{eq: threshold knockoff selector}
\end{equation}
where $\mathcal{W} = \lbrace |W_{j}| : j=1, \ldots, p \rbrace\backslash\lbrace 0 \rbrace$. Note that this is only one of the test statistics that were proposed by the authors. In general, many other test statistics obeying a certain sufficiency and anti-symmetry property are suitable for the knockoff method. In our simulations, we stick to the test statistic in~\eqref{eq: test statistic knockoff selector}. In~\eqref{eq: threshold knockoff selector}, $b = 0$ yields the knockoff method and $b = 1$ the more conservative (higher threshold $\tau$) knockoff+ method. Finally, only those variables whose test statistics exceed the threshold are selected, which gives us the selected active set $\widehat{\mathcal{A}} = \lbrace j : W_{j} \geq \tau \rbrace$. The knockoff+ method controls the FDR at the target level $\alpha$ and the knockoff method controls a modified version of the FDR. The advantage of the knockoff over the knockoff+ method is that it is less conservative and will, generally, have a higher power at the cost of controlling only a related quantity but not the FDR.

The \textit{model-X} knockoff method was proposed as an extension to the \textit{fixed-X} knockoff method for high-dimensional settings~\cite{candes2018panning}. It does not require any knowledge about the conditional distribution of the response given the explanatory variables $Y | X_{1},\ldots, X_{p}$ but needs to know the distribution of the covariates $(X_{i1}, \ldots, X_{ip})$, $i = 1,\ldots,n$. The difference to the deterministic design of \textit{fixed-X} knockoffs is that \textit{model-X} knockoffs need to be designed probabilistically by sequentially sampling each knockoff predictor $\xK_{j}$, $j = 1, \ldots p$, from the conditional distribution of $X_{j} | X_{-j}, \accentset{\circ}{X}_{1:j-1}$, where $X_{-j}$ is the set of all explanatory variables except for $X_{j}$ and $\accentset{\circ}{X}_{1:j-1} \coloneqq \lbrace \accentset{\circ}{X}_{1}, \ldots, \accentset{\circ}{X}_{j-1} \rbrace$. However, the authors state that determining a new conditional distribution for each knockoff predictor and sampling from it turned out to be complicated and computationally very expensive~\cite{candes2018panning}. The only case in which \textit{model-X} knockoffs can be easily constructed by sampling from the Gaussian distribution with a certain mean vector and covariance matrix is when the covariates follow the Gaussian distribution. For all other distributions of the covariates, especially when $p$ is large, the authors consider an approximate construction of \textit{model-X} knockoffs which yields the so-called second-order \textit{model-X} knockoffs. Unfortunately, however, there is no proof that FDR control is achieved with second-order \textit{model-X} knockoffs. Nevertheless, in our simulations we consider these knockoffs. Moreover, for $p > 500$ we consider the approximate semidefinite program (asdp) instead of the original semidefinite program that needs to be solved to construct second-order \textit{model-X} knockoffs. This is the default choice in the R package accompanying the \textit{fixed-X} and \textit{model-X} papers.\footnote{The R package containing the implementations of the \textit{fixed-X} and the \textit{model-X} methods is available at \url{https://CRAN.R-project.org/package=knockoff} (last access: January 31, 2024).}

\section{Additional Simulation Results}
\label{sec: Additional Simulation Results}
\begin{figure*}[!htb]
  \centering
  \subfloat[Setup: $n = 300$, $p = 100$, $p_{1} = 10$, $T_{\max} = \lceil n/2 \rceil$, $L_{\max} = 10p$, $K = 20$, $MC = 955$.]{
  		\scalebox{0.85}{
  			\includegraphics[width=0.45\linewidth]{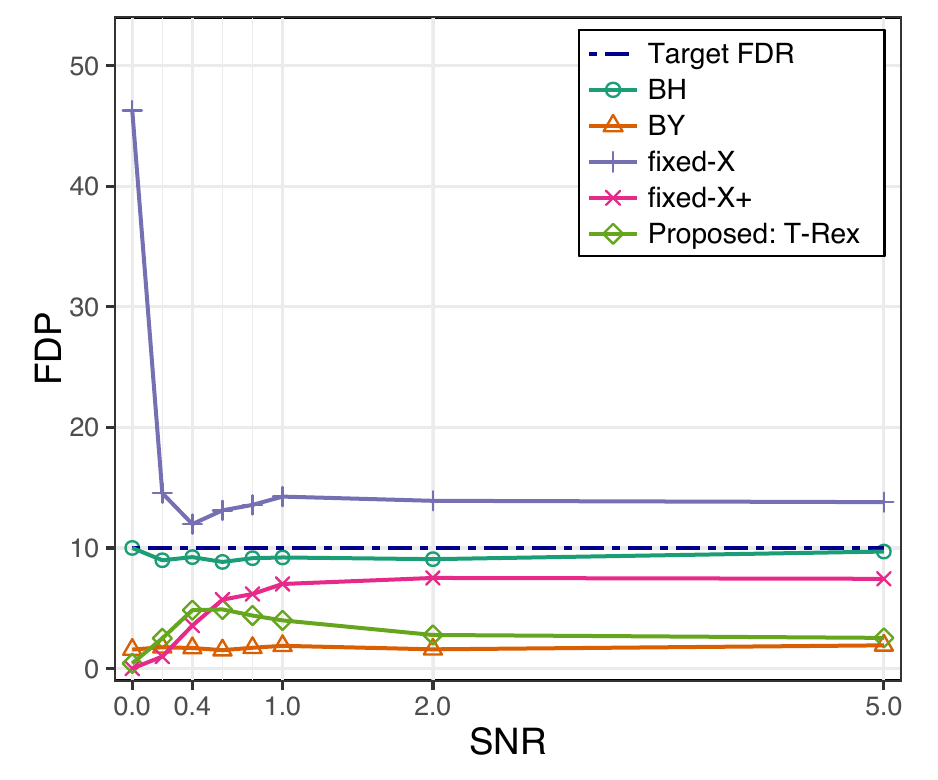}
  		}
   		\label{fig: FDP_vs_SNR_p_100_Optimal_T_L}
   }
     \hspace{1em}
  \subfloat[Setup: Same as in Figure~(a).]{
  		\scalebox{0.85}{
  			\includegraphics[width=0.4580\linewidth]{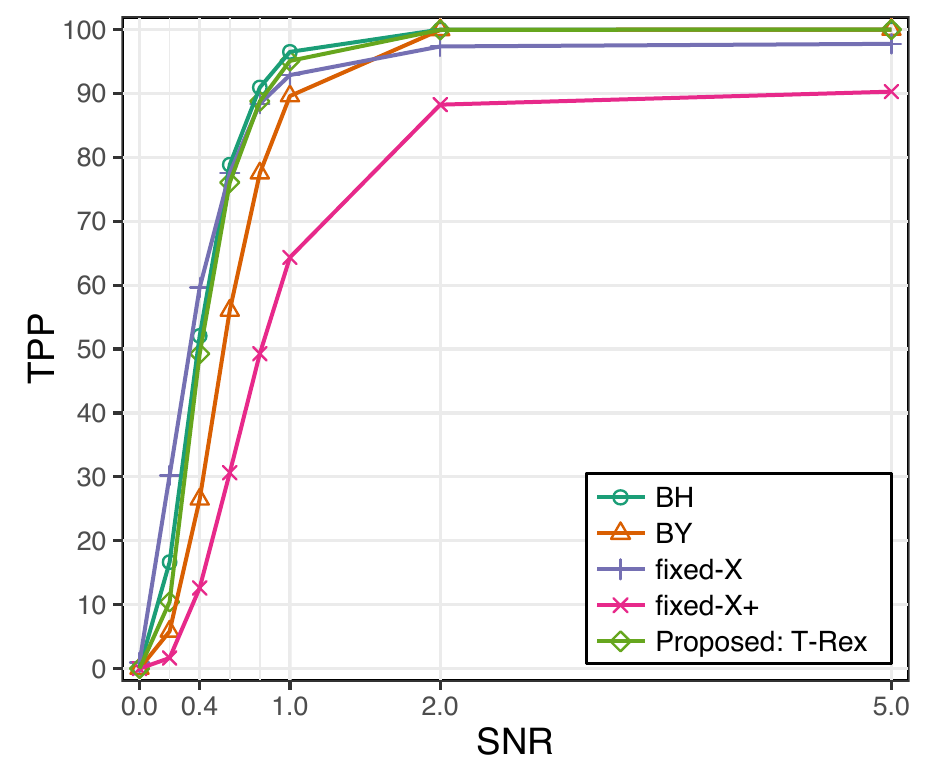}
  		}
   		\label{fig: TPP_vs_SNR_p_100_Optimal_T_L}
   }
\\
    \subfloat[Setup: $n = 300$, $p = 100$, $T_{\max} = \lceil n/2 \rceil$, $L_{\max} = 10p$, $K = 20$, $\text{SNR} = 1$, $MC = 955$.]{
  		\scalebox{0.85}{
  			\includegraphics[width=0.45\linewidth]{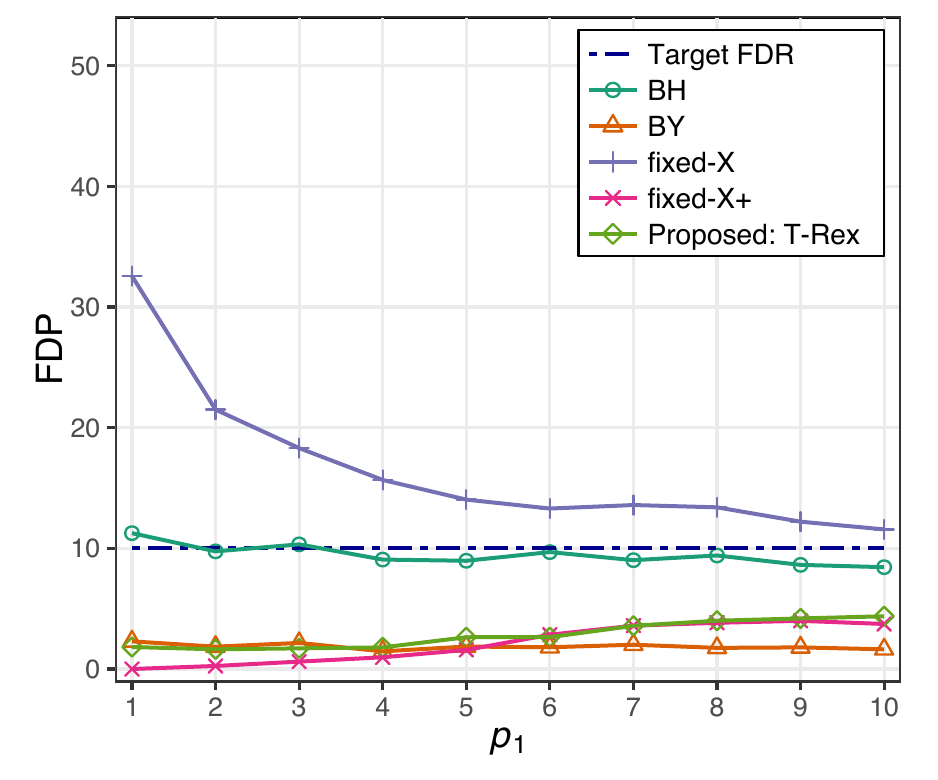}
  		}
   		\label{fig: FDP_vs_p1_p_100_Optimal_T_L}
   }  
	  \hspace{1em}
  \subfloat[Setup: Same as in Figure~(c).]{
  		\scalebox{0.85}{
  			\includegraphics[width=0.4575\linewidth]{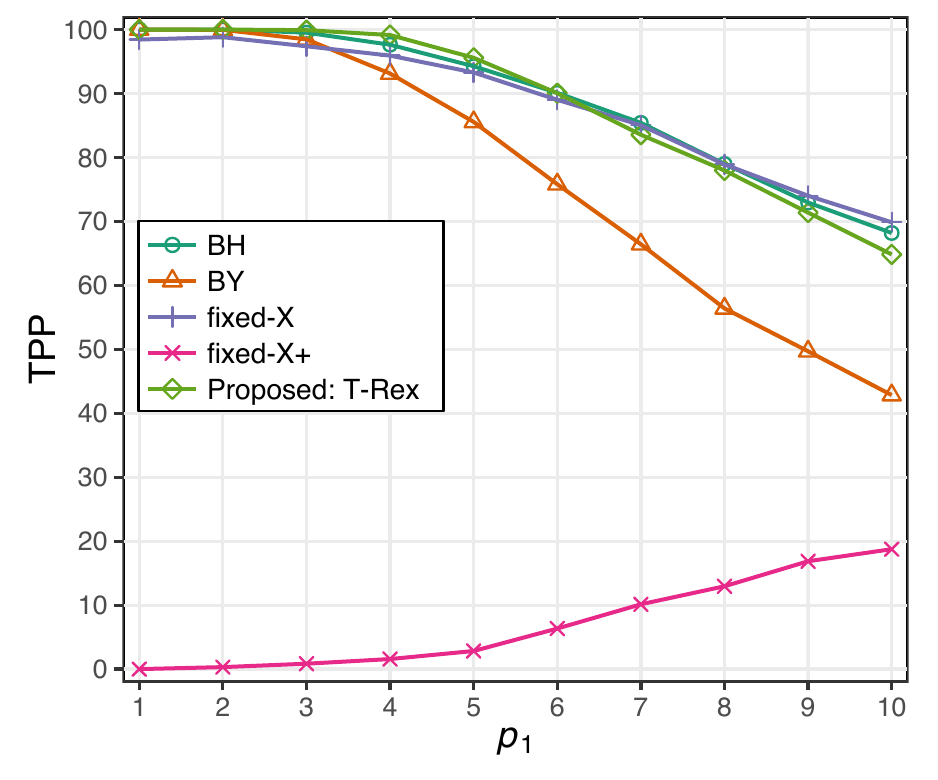}
  		}
   		\label{fig: TPP_vs_p1_p_100_Optimal_T_L}
   }
  \caption{\textbf{General}: The \textit{fixed-X} knockoff method fails to control the FDR. In terms of power, the proposed \textit{T-Rex} selector outperforms the \textit{fixed-X} knockoff method, the \textit{fixed-X} knockoff+ method, and the \textit{BY} method and shows a comparable performance to the \textit{BH} method. \textbf{Details}: \textbf{(a)}~All methods except for the \textit{fixed-X} knockoff method control the FDR at a target level of $10\%$ for the whole range of SNR values. The \textit{fixed-X} knockoff method fails to control the FDR and performs poorly at low SNR values. \textbf{(b)}~As expected, the TPR (i.e., power) increases with respect to the SNR. It is remarkable that the TPP (i.e., power) of the proposed \textit{T-Rex} selector is comparable to that of the \textit{BH} method, although the FDR of the \textit{T-Rex} selector is less than half of the achieved FDR of the \textit{BH} method (see Figure~(a)). The high power of the \textit{fixed-X} knockoff method cannot be interpreted as an advantage because the method does not control the FDR. \textbf{(c)}~The proposed \textit{T-Rex} selector, the \textit{fixed-X} knockoff+ method, and the \textit{BY} method control the FDR at a target level of $10\%$, while the \textit{BH} method exceeds the target level for some low values of $p_{1}$ and the curve of the \textit{fixed-X} knockoff method never falls below the target level. \textbf{(d)}~Among the methods that control the FDR for all considered values of $p_{1}$, the proposed \textit{T-Rex} selector has the highest power. It is remarkable that the TPP (i.e., power) of the proposed \textit{T-Rex} selector is comparable to that of the \textit{BH} method, although the FDR of the \textit{T-Rex} selector is approximately only half of the achieved FDR of the \textit{BH} method (see Figure~(c)).}
  \label{fig: sweep p1 and sweep snr plots p = 100}
\end{figure*}
For the sake of completeness, we present additional simulation results for the classical low-dimensional setting, i.e., $p \leq n$. The data is generated as described in Section~\ref{subsec: Setup and Results}. The specific values of the generic simulation setting in Section~\ref{subsec: Setup and Results} and the parameters of the proposed \textit{T-Rex} selector and the proposed extended calibration algorithm in Algorithm~\ref{algorithm: Extended T-Rex}, i.e., the values of $n$, $p$, $p_{1}$, $T_{\max}$, $L_{\max}$, $K$, and $\text{SNR}$ are specified in the captions of Figure~\ref{fig: sweep p1 and sweep snr plots p = 100}. All results are averaged over $955$ Monte Carlo realizations. The simulations were conducted using the R packages \textit{TRexSelector}~\cite{machkour2022TRexSelector} and \textit{tlars}~\cite{machkour2022tlars}.

Summarizing in brief, the proposed \textit{T-Rex} selector controls the FDR at the target level of $10\%$ while, in terms of power, outperforming the \textit{fixed-X} knockoff method, the \textit{fixed-X} knockoff+ method, and the \textit{BY} method and showing a comparable performance to the \textit{BH} method. A detailed discussion of the simulation results is given in the captions of Figure~\ref{fig: sweep p1 and sweep snr plots p = 100} and its subfigures.
\section{Setup, Preprocessing, and Additional Results: Simulated Genome-Wide Association Study}
\label{sec: Setup, Preprocessing, and Additional Results: Simulated Genome-Wide Association Study}
\begin{figure*}[!htbp]
  \centering
  \subfloat[FDP box plots.]{
  		\scalebox{0.85}{
  			\includegraphics[width=0.45\linewidth]{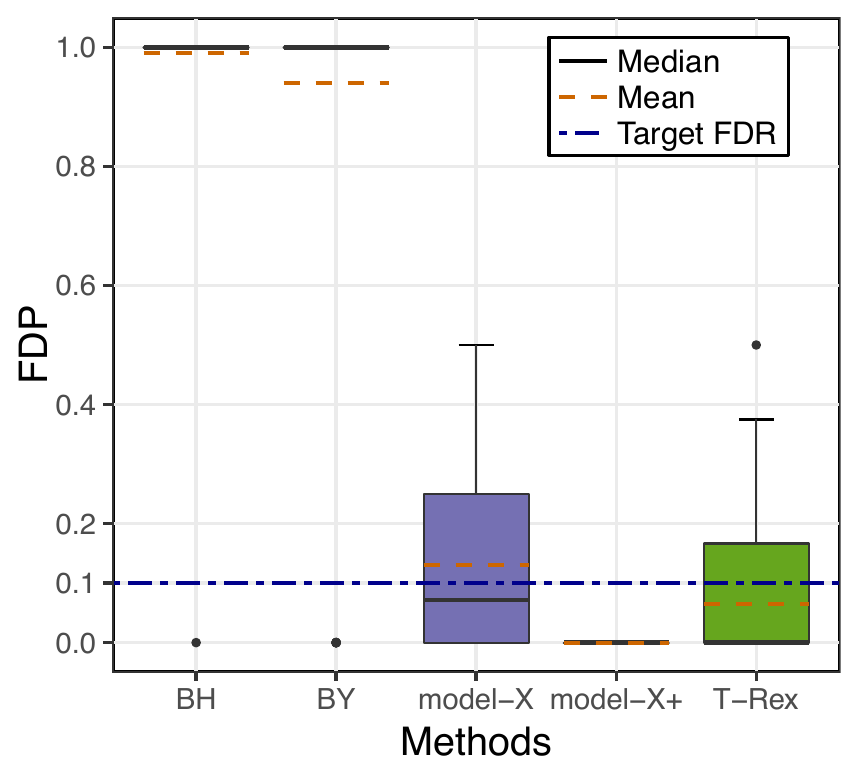}
  		}
   		\label{fig: FDP_simulted_GWAS_with_TRexSelector_package}
   }
   \hspace{1em}
  \subfloat[TPP box plots.]{
  		\scalebox{0.85}{
  			\includegraphics[width=0.45\linewidth]{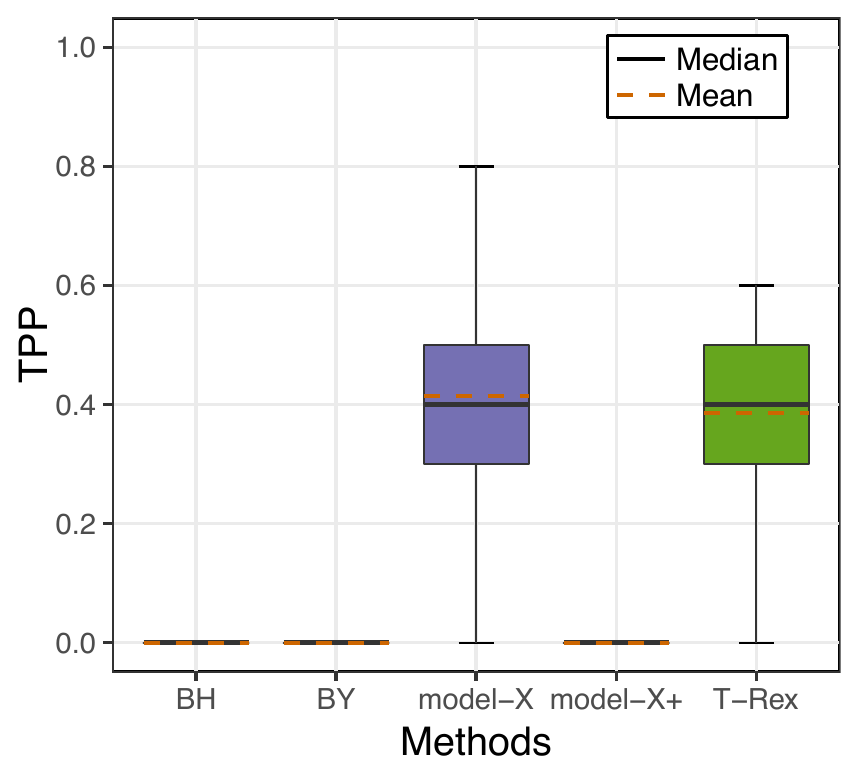}
  		}
   		\label{fig: TPP_simulted_GWAS_with_TRexSelector_package}
   }
  \caption{The proposed \textit{T-Rex} selector is the only method that has an average FDP below the target FDR level and that has a non-zero power. Note that the FDP can be different across the realizations and even for FDR controlling methods it is not necessarily below the target level for every realization. We use box plots to visualize the distribution of the results and give the reader a sense of how the FDP and TPP (i.e., power) vary around the mean.}
  \label{fig: simulated GWAS with TRexSelector package}
\end{figure*}
This appendix provides additional details on the setup of the simulated genome-wide association study (GWAS) in Section~\ref{sec: Simulated Genome-Wide Association Study} and the preprocessing of the data, presents additional results, and verifies A-\ref{assumption: 1}, A-\ref{assumption: 2}, and A-\ref{assumption: 3} on simulated genomics data.
\subsection{Setup}
\label{subsec: Setup Appendix GWAS}
\begin{figure}[!htb]
  \centering
  		\scalebox{1}{
  			\includegraphics[width=0.98\linewidth]{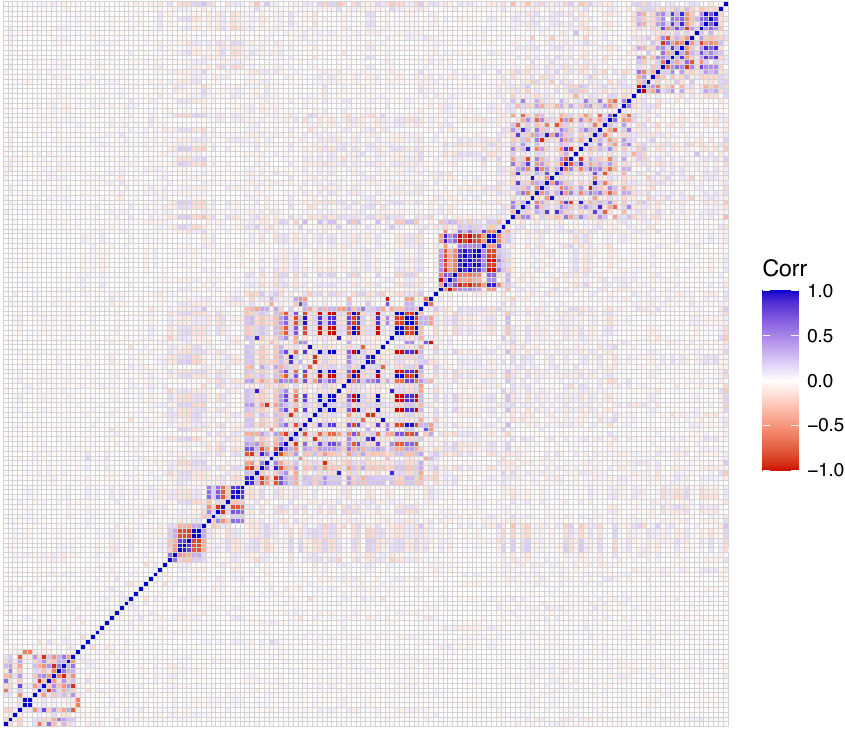}
  		}
   		\label{fig: correlation_heatmap_hapgen_150_snps_10_disease_loci}
 \caption{The heatmap visualizes the correlation matrix of the first $150$ out of $3{,}000$ SNPs (containing $10$ disease SNPs, i.e., true active variables) that were generated using the software HAPGEN2~\cite{su2011hapgen2}.}
  \label{fig: correlation heatmap hapgen}
\end{figure}

The genotype matrix, i.e., the matrix $\X$ containing the SNPs as columns consists of groups of highly correlated SNPs. This is due to a phenomenon called linkage disequilibrium~\cite{reich2001linkage}. In order to visualize this phenomenon and understand the implications it has on the data structure, we have generated $3{,}000$ SNPs using the genomics software HAPGEN2. That is, we have fed real world haplotypes from the International HapMap project (phase 3)~\cite{international2010integrating} into the software HAPGEN2. The software takes into account biological characteristics of genomics principles to simulate realistic genotypes (i.e., predictor matrix $\X$) with known ground truth. This data contains groups of highly correlated variables. Figure~\ref{fig: correlation heatmap hapgen} visualizes the correlation matrix of the first $150$ SNPs in $\X$. We can observe the dependency structure among the predictors/SNPs that form groups of highly correlated predictors.\\

In GWAS, our goal is not to find specific SNPs/variables that are associated with a disease of interest but rather to find the groups of highly correlated SNPs/variables that point to the broader locations on the genome that are associated with the disease of interest. Therefore, in genomics research, it is a standard procedure to apply a preprocessing method called \textit{SNP pruning} before applying any variable selection method (see, e.g.,~\cite{sesia2019gene}). The main idea behind \textit{SNP pruning} is to cluster the SNPs into groups of highly correlated SNPs using a dendrogram and to select one representative from each group of highly correlated SNPs. After this procedure has been carried out, we are left with an SNP matrix whose dimension is reduced and that exhibits only weak dependencies among the representative SNPs.

For the simulated GWAS, we generated $100$ data sets satisfying the specifications in Section~\ref{subsec: GWAS Setup} using the software HAPGEN2~\cite{su2011hapgen2}. According to the authors, HAPGEN2 uses the time of the current day in seconds to set the seed of the random number generator, and, therefore multiple simulations should not be started very close in time to avoid identical results. Therefore we have generated the data sets sequentially and since generating a single data set took roughly six minutes, a sufficient time period between the starts of consecutive simulations was allowed.\footnote{The data sets were generated on a compute node of the Lichtenberg High-Performance Computer of the Technische Universität Darmstadt that consists of two ``Intel\textsuperscript{®} Xeon\textsuperscript{®} Platinum 9242 Processors'' with 96 cores and 384 GB RAM (DDR4-2933) in total.}

\subsection{Preprocessing and Additional Results}
\label{subsec: Preprocessing and Additional Results Appendix GWAS}
The preprocessing is carried out as suggested in~\cite{sesia2019gene} and on the accompanying website.\footnote{URL: \url{https://web.stanford.edu/group/candes/knockoffs/tutorials/gwas_tutorial.html} (last access: January 31, 2024).} That is, SNPs with a minor allele frequency or call rate lower than $1\%$ and $95\%$, respectively, are removed. Additionally, SNPs that violate the Hardy-Weinberg disequilibrium with a cutoff of $10^{-6}$ are removed. Since proximate SNPs are highly correlated, the remaining SNPs are clustered using SNP pruning that ensures that there exist no absolute sample correlations above $0.75$ between any two SNPs belonging to different clusters. The resulting average number of clusters is $8211$ while the minimum and maximum numbers of clusters are $8120$ and $8326$, respectively. Each cluster is represented by the strongest cluster representative which is selected by computing the marginal $p$-values using the Cochran-Armitage test based on $20\%$ of the data and picking the SNP with the smallest $p$-value. The marginal $p$-values that will be plugged into the \textit{BH} method and the \textit{BY} method are also computed using the Cochran-Armitage test but with the full data set.

In addition to the averaged results of the simulated GWAS in Section~\ref{sec: Simulated Genome-Wide Association Study} in the main paper, Figure~\ref{fig: simulated GWAS with TRexSelector package} shows how the FDP and TPP vary around the mean using box plots.

\begin{figure*}[!htbp]
  \centering
  \subfloat[Histogram and theoretical distribution for $t = T = 3$. Note that the histogram is based on $K = 20$ random experiments for each of the $100$ HAPGEN2 genomics data sets.]{
  		\scalebox{1.0}{
  			\includegraphics[width=0.43\linewidth, valign = t]{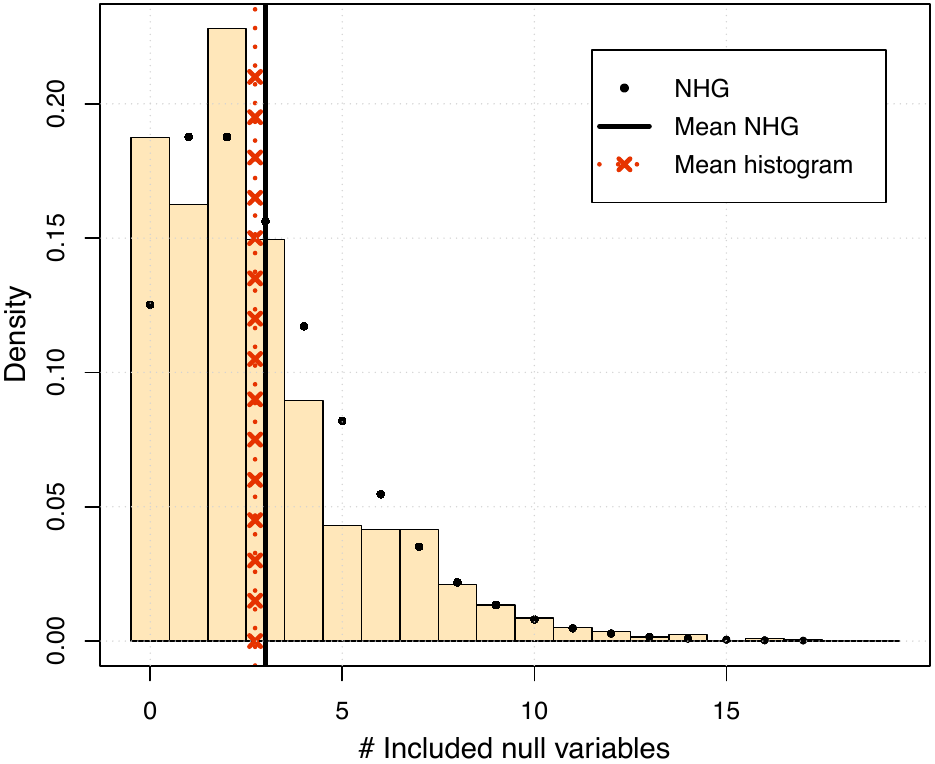}
  		}
   		\label{fig: NHG histogram _HAPGEN2_data}
   }
   \hspace{1em}
  \subfloat[Q-Q plot corresponding to Figure~(a).]{
  		\scalebox{1.0}{
  			\includegraphics[width=0.4\linewidth, valign = t]{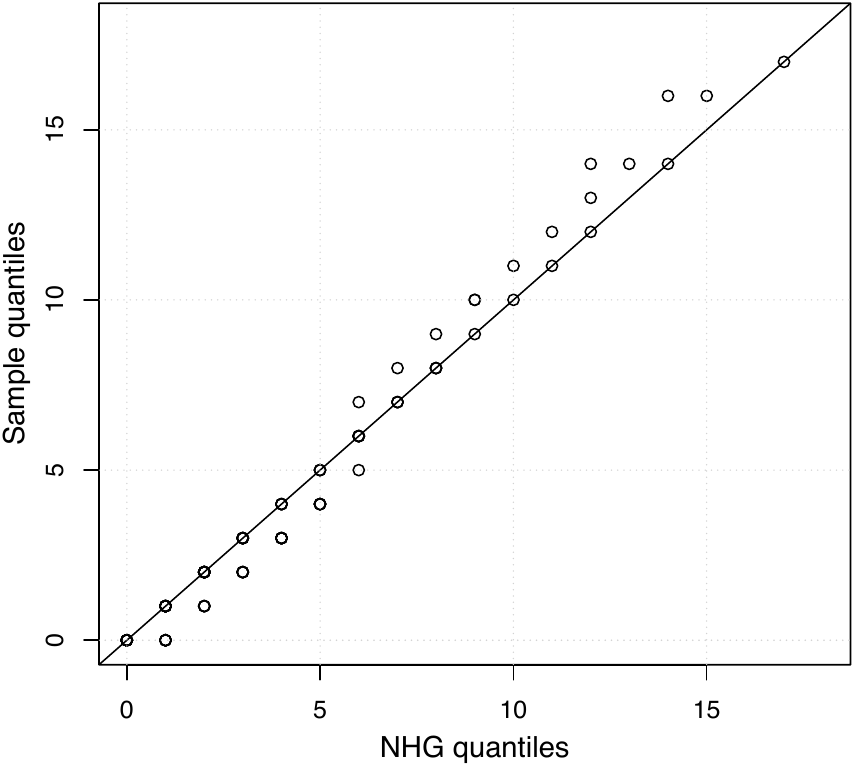}
  			\vphantom{\includegraphics[width=0.43\linewidth, valign = t]{Figures/NHG_histogram_HAPGEN2_data.pdf}}
  		}
   		\label{fig: NHG QQ-plot _HAPGEN2_data}
   }
  \caption{\textbf{Exemplary numerical verification of Corollary~\ref{corollary: 2} and A-\ref{assumption: 1} for HAPGEN2 genomics data}: The histogram of the number of included null variables in Figure~(a) approximates the theoretical probability mass function (PMF). The expected value of a random variable following the negative hypergeometric distribution with the parameters specified in the last sentence of this caption is given by $T \cdot p_{0} \, / \, (L + 1) = 3 \cdot 8{,}110 \, / \, (8{,}120 + 1) \approx 2.996$, which fits the mean of the histogram. The Q-Q plot in Figure~(b) confirms that the number of included null variables follows the negative hypergeometric distribution. Setup after preprocessing: $n = 1{,}000$, $p = 8{,}120$, $p_{1} = 10$, $T = 3$, $L = p$, $K = 20$.}
  \label{fig: NHG histogram and QQ-plot _HAPGEN2_data}
\end{figure*}

\begin{figure}[h]
  \centering
  		\scalebox{0.95}{
  			\includegraphics[width=0.97\linewidth]{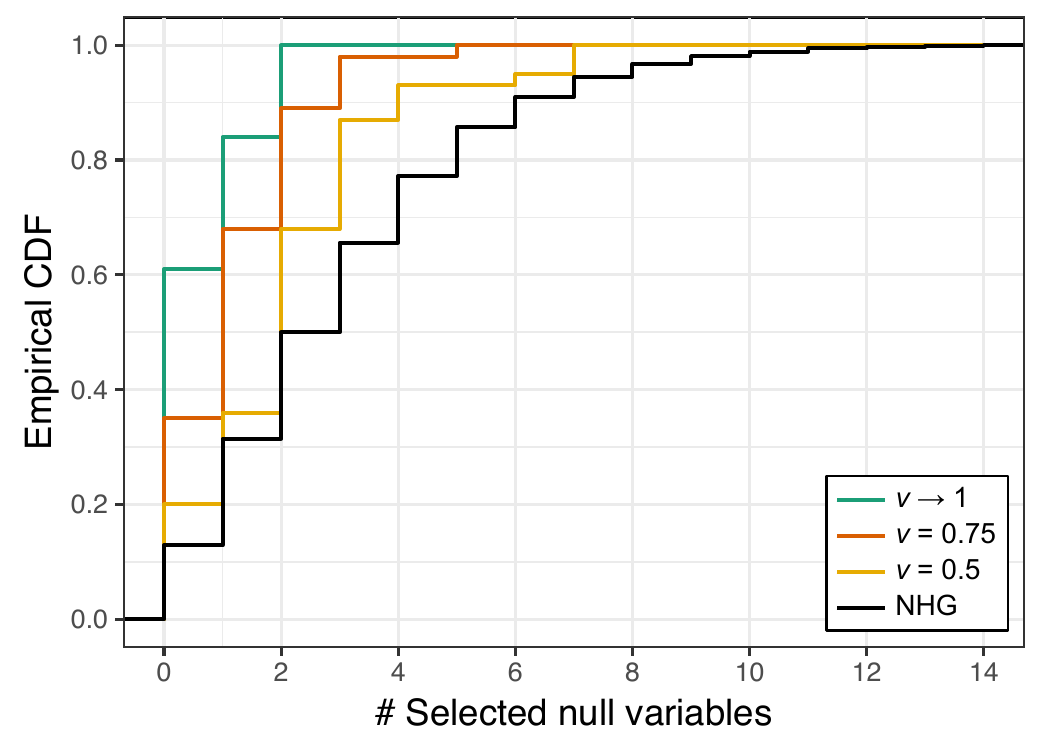}
  		}
  \caption{\textbf{Exemplary numerical verification of A-\ref{assumption: 2} for HAPGEN2 genomics data}: For $v \geq 0.5$, a random variable following the negative hypergeometric distribution stochastically dominates the random variable $V_{T, L}(v)$ (i.e., the number of selected null variables) at all values of $V_{T, L}(v)$. Setup after preprocessing: $n = 1{,}000$, $p = 8{,}120$, $p_{1} = 10$, $T = 3$, $L = p$, $K = 20$.}
  \label{fig: stochastic dominance cdf _HAPGEN2_data}
\end{figure}

\begin{figure*}[!htbp]
  \centering
  \subfloat[Approximations and true values for different choices of $v$ averaged over $100$ HAPGEN2 genomics data sets.]{
  		\scalebox{0.9}{
  			\includegraphics[width=0.4\linewidth]{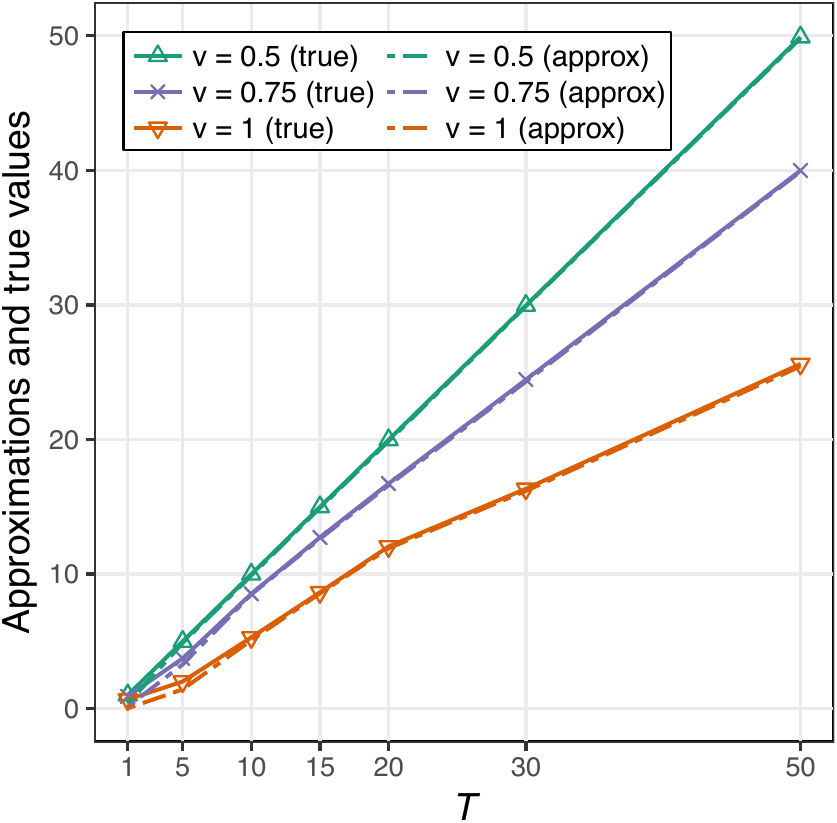}
  		}
   		\label{fig: relative_error_vs_v _HAPGEN2_data}
   }
     \hspace{1em}
  \subfloat[Approximations and true values for different choices of $T$ averaged over $100$ HAPGEN2 genomics data sets.]{
  		\scalebox{0.9}{
  			\includegraphics[width=0.4\linewidth]{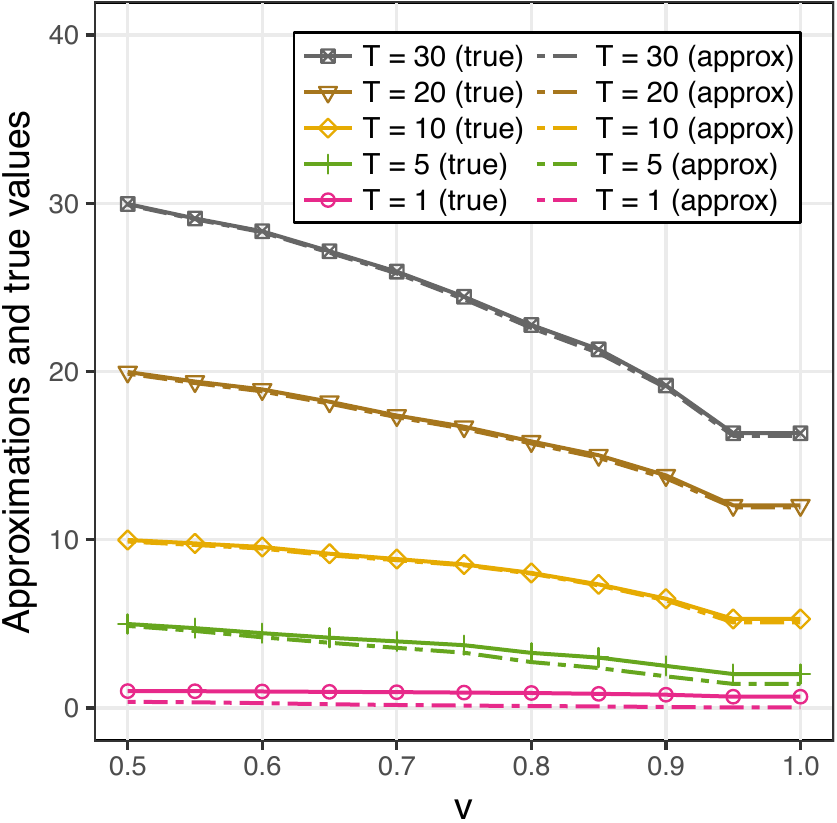}
  		}
   		\label{fig: relative_error_vs_v _HAPGEN2_data}
   }
   \\
  \subfloat[Box plots of approximations and true values corresponding to the lines for $v = 0.75$ in Figure (a).]{
  		\scalebox{0.9}{
  			\includegraphics[width=0.4\linewidth]{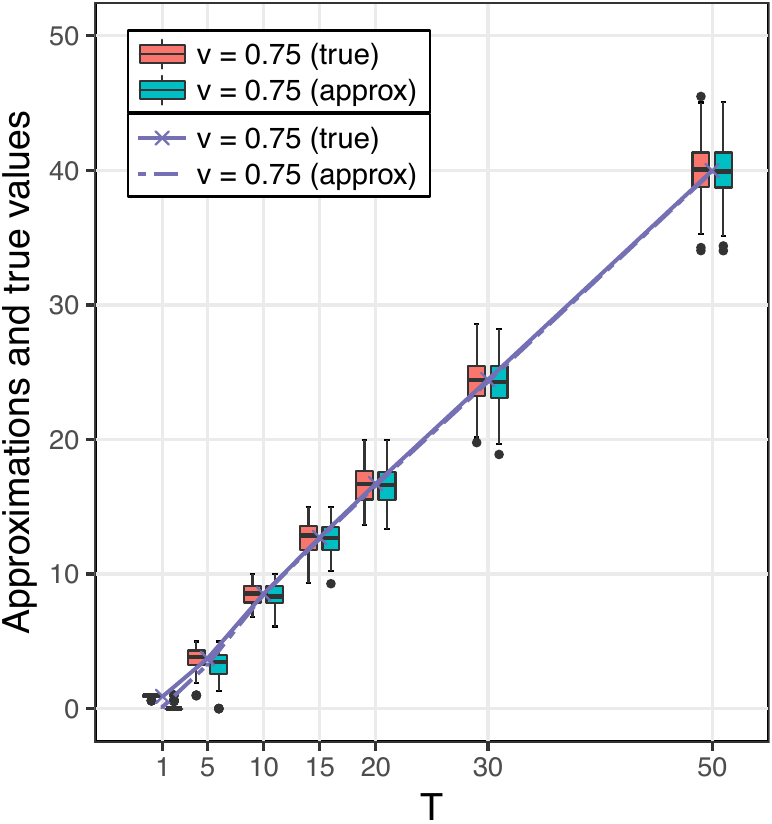}
  		}
   		\label{fig: relative_error_vs_v _HAPGEN2_data}
   }
     \hspace{1em}
  \subfloat[Box plots of approximations and true values corresponding to the lines for $T = 10$ in Figure (b).]{
  		\scalebox{0.9}{
  			\includegraphics[width=0.405\linewidth]{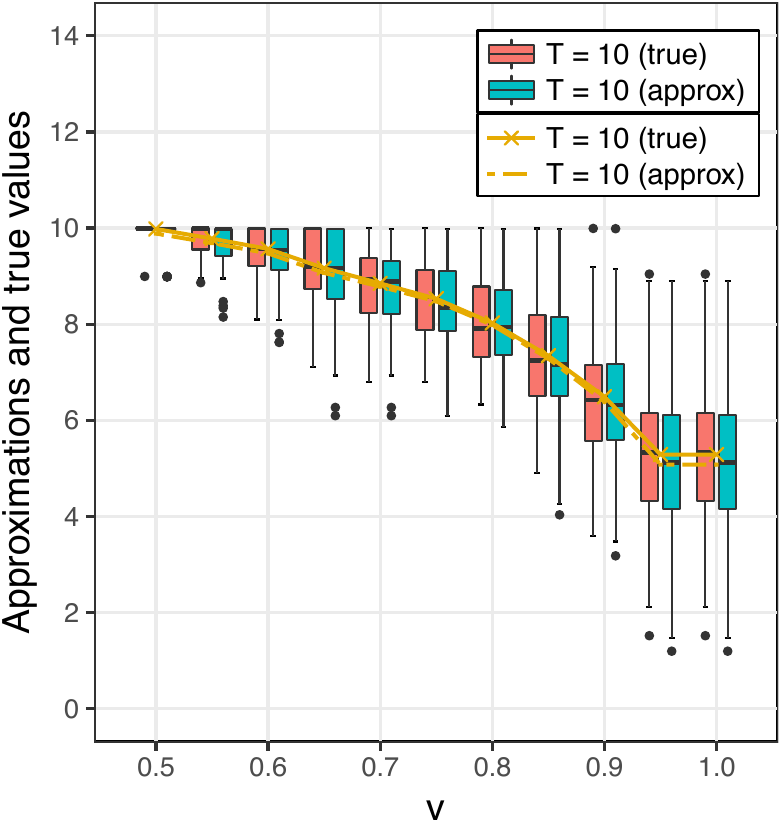}
  		}
   		\label{fig: relative_error_vs_v _HAPGEN2_data}
   }
  \caption{\textbf{Exemplary numerical verification of A-\ref{assumption: 3} for HAPGEN2 genomics data}: In Figures (a) and (b), we see that the approximations and the true values are almost identical for different values of $v$ and $T$. The corresponding box plots in Figures (c) and (d) show that also the distributions of approximations and true values are very similar. Setup after preprocessing: $n = 1{,}000$, $p = 8{,}120$, $p_{1} = 10$, $L = p$, $K = 20$.}
  \label{fig: relative_error_vs_T_and_v _HAPGEN2_data}
\end{figure*}
\subsection{Verification of A-\ref{assumption: 1}, A-\ref{assumption: 2}, and A-\ref{assumption: 3} on HAPGEN2 Genomics Data}
\label{subsec: Verification of Assumptions on HAPGEN2 Genomics Data}
Figures~\ref{fig: NHG histogram and QQ-plot _HAPGEN2_data},~\ref{fig: stochastic dominance cdf _HAPGEN2_data}, and~\ref{fig: relative_error_vs_T_and_v _HAPGEN2_data} show that for the genomics data analyzed in Section~\ref{sec: Simulated Genome-Wide Association Study} in the main paper and with the preprocessing (i.e., SNP pruning, etc.) described above, A-\ref{assumption: 1}, A-\ref{assumption: 2}, and A-\ref{assumption: 3} are surprisingly well satisfied. For our verifications here, we have only made one necessary minor adjustment to the preprocessing described in the previous section. The reason is that for each of the $100$ data sets, that have been generated using HAPGEN2~\cite{su2011hapgen2}, the SNP pruning procedure outputs pruned SNP sets with slightly different sizes. For the verification of the assumptions, it is necessary to have a constant number of SNPs. Therefore, we have removed very few randomly selected SNPs from all sets in order to match the size of the smallest SNP set, which contains $8{,}120$ out of originally $20{,}000$ SNPs after the preprocessing.

\section{Illustration of Theorem~\ref{theorem: Dummy generation} (Dummy Generation)}
\label{sec: Illustration of Theorem 2 (Dummy Generation)}
\begin{figure*}[!htbp]
  \centering
	\hspace*{-0.8em}
  \subfloat[Gaussian dummies.]{
  		\scalebox{1}{
  			\includegraphics[width=0.245\linewidth]{Figures/FDP_vs_SNR_p_1000_Optimal_T_L.pdf}
  		}
   		\label{fig: FDP_vs_SNR_p_1000_Optimal_T_L}
   }
	\hspace*{-1.4em}
  \subfloat[Gaussian dummies.]{
  		\scalebox{1}{
  			\includegraphics[width=0.245\linewidth]{Figures/TPP_vs_SNR_p_1000_Optimal_T_L.pdf}
  		}
   		\label{fig: TPP_vs_SNR_p_1000_Optimal_T_L}
   }
	\hspace*{-1.4em}
    \subfloat[Gaussian dummies.]{
  		\scalebox{1}{
  			\includegraphics[width=0.245\linewidth]{Figures/FDP_vs_p1_p_1000_Optimal_T_L.pdf}
  		}
   		\label{fig: FDP_vs_p1_p_1000_Optimal_T_L}
   }
	\hspace*{-1.4em}
  \subfloat[Gaussian dummies.]{
  		\scalebox{1}{
  			\includegraphics[width=0.245\linewidth]{Figures/TPP_vs_p1_p_1000_Optimal_T_L.pdf}
  		}
   		\label{fig: TPP_vs_p1_p_1000_Optimal_T_L}
   }   
   \\
   \vspace{0.2cm}
	\hspace*{-0.8em}
  \subfloat[Uniform dummies.]{
  		\scalebox{1}{
  			\includegraphics[width=0.245\linewidth]{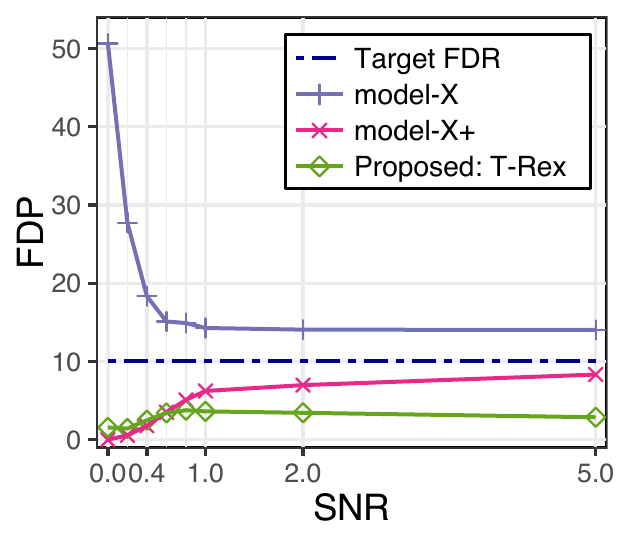}
  		}
   		\label{fig: FDP_vs_SNR_p_1000_Optimal_T_L_Unif_Dummies}
   }
	\hspace*{-1.4em}
  \subfloat[Uniform dummies.]{
  		\scalebox{1}{
  			\includegraphics[width=0.245\linewidth]{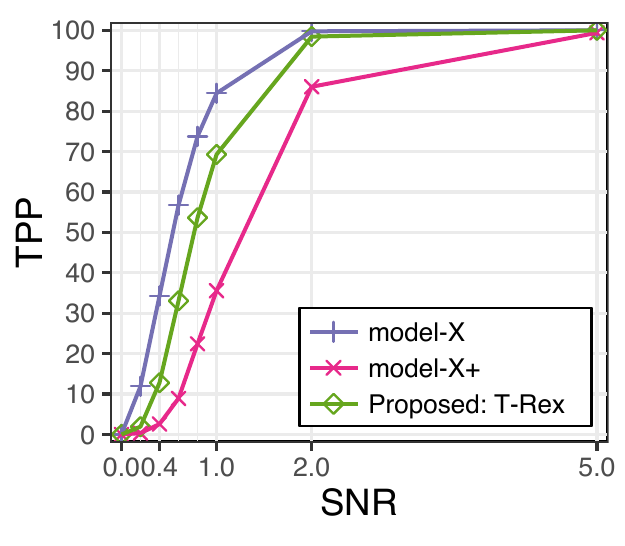}
  		}
   		\label{fig: TPP_vs_SNR_p_1000_Optimal_T_L_Unif_Dummies}
   }
	\hspace*{-1.4em}
    \subfloat[Uniform dummies.]{
  		\scalebox{1}{
  			\includegraphics[width=0.245\linewidth]{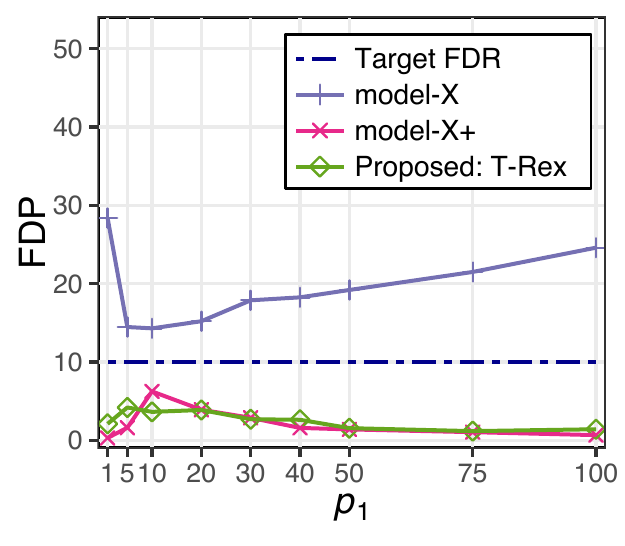}
  		}
   		\label{fig: FDP_vs_p1_p_1000_Optimal_T_L_Unif_Dummies}
   }
	\hspace*{-1.4em}
  \subfloat[Uniform dummies.]{
  		\scalebox{1}{
  			\includegraphics[width=0.245\linewidth]{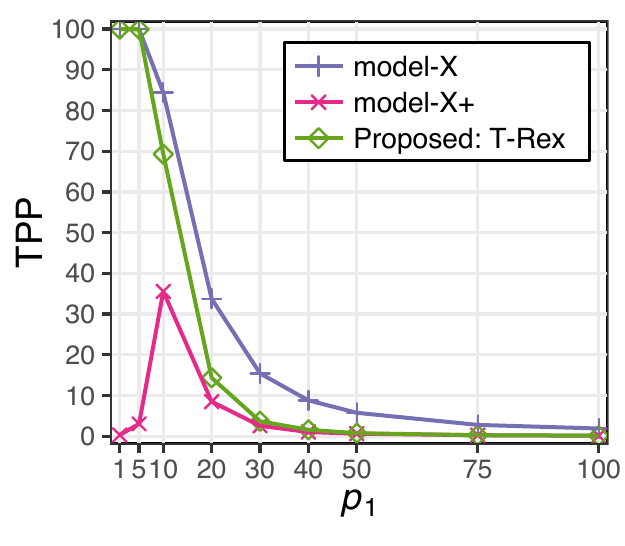}
  		}
   		\label{fig: TPP_vs_p1_p_1000_Optimal_T_L_Unif_Dummies}
   }
   \\
   \vspace{0.2cm}
   \hspace*{-0.8em}
  \subfloat[Student $t$ dummies.]{
  		\scalebox{1}{
  			\includegraphics[width=0.245\linewidth]{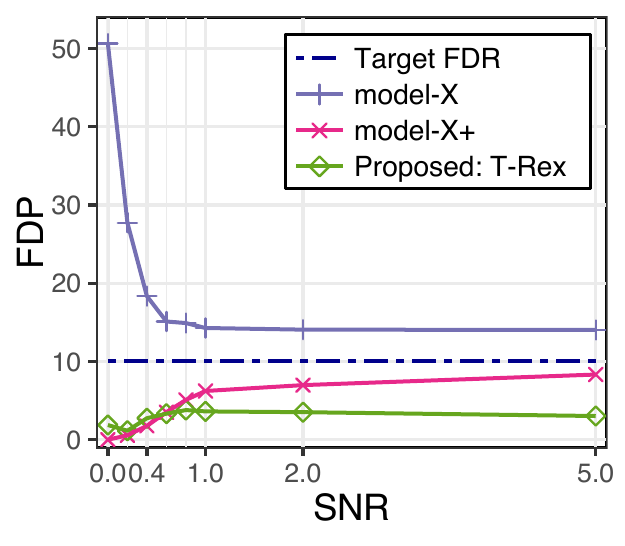}
  		}
   		\label{fig: FDP_vs_SNR_p_1000_Optimal_T_L_tDistr_Dummies}
   }
	\hspace*{-1.4em}
  \subfloat[Student $t$ dummies.]{
  		\scalebox{1}{
  			\includegraphics[width=0.245\linewidth]{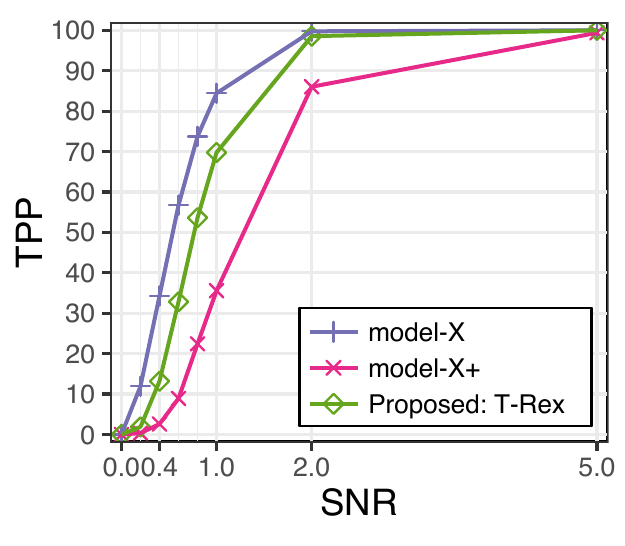}
  		}
   		\label{fig: TPP_vs_SNR_p_1000_Optimal_T_L_tDistr_Dummies}
   }
	\hspace*{-1.4em}
    \subfloat[Student $t$ dummies.]{
  		\scalebox{1}{
  			\includegraphics[width=0.245\linewidth]{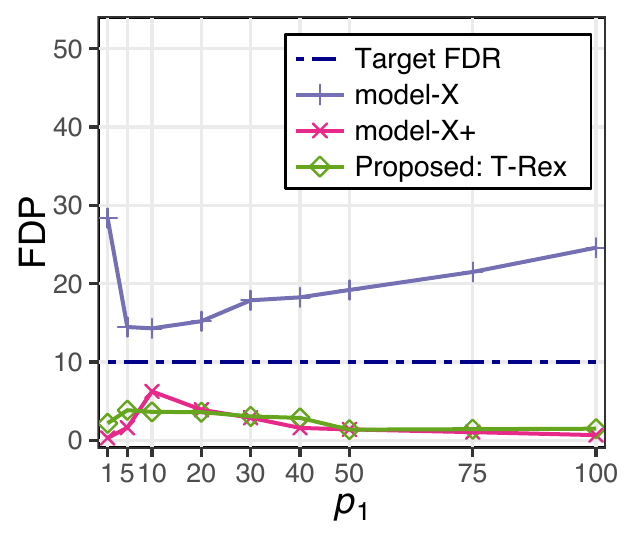}
  		}
   		\label{fig: FDP_vs_p1_p_1000_Optimal_T_L_tDistr_Dummies}
   }
	\hspace*{-1.4em}
  \subfloat[Student $t$ dummies.]{
  		\scalebox{1}{
  			\includegraphics[width=0.245\linewidth]{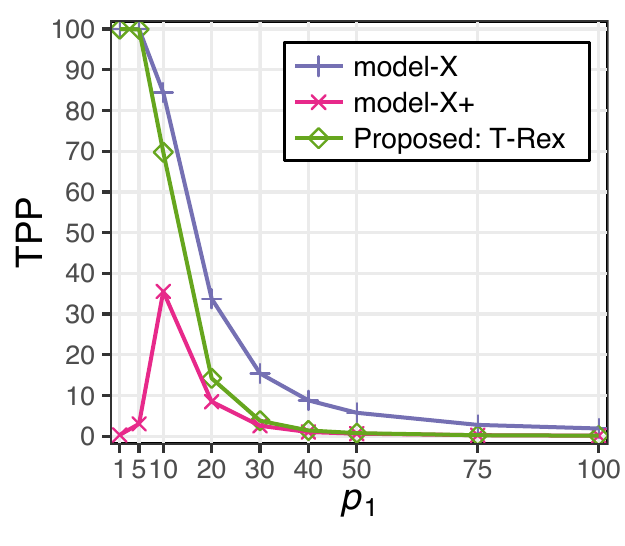}
  		}
   		\label{fig: TPP_vs_p1_p_1000_Optimal_T_L_tDistr_Dummies}
   }
   \\
   \vspace{0.2cm}
	\hspace*{-0.8em}
  \subfloat[Gumbel dummies.]{
  		\scalebox{1}{
  			\includegraphics[width=0.245\linewidth]{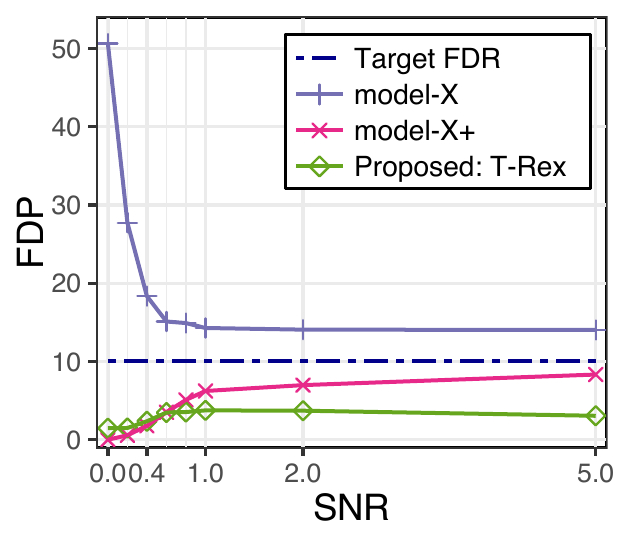}
  		}
   		\label{fig: FDP_vs_SNR_p_1000_Optimal_T_L_Gumbel_Dummies}
   }
	\hspace*{-1.4em}
  \subfloat[Gumbel dummies.]{
  		\scalebox{1}{
  			\includegraphics[width=0.245\linewidth]{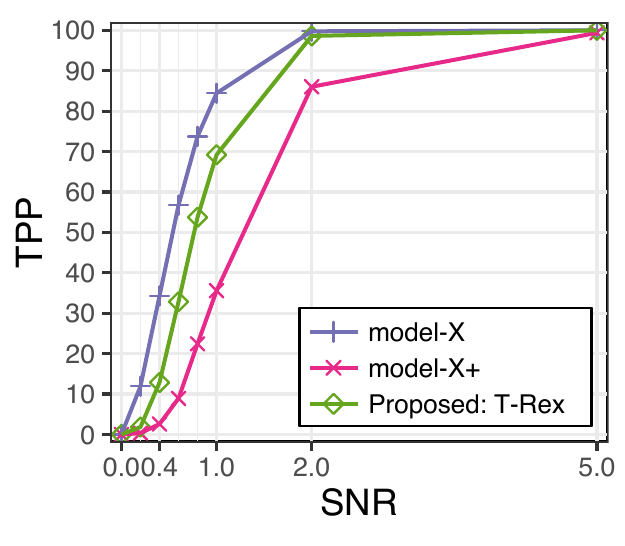}
  		}
   		\label{fig: TPP_vs_SNR_p_1000_Optimal_T_L_Gumbel_Dummies}
   }
	\hspace*{-1.4em}
    \subfloat[Gumbel dummies.]{
  		\scalebox{1}{
  			\includegraphics[width=0.245\linewidth]{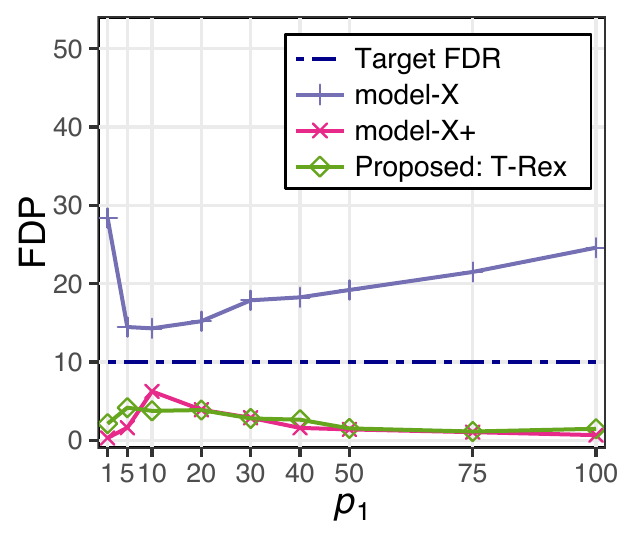}
  		}
   		\label{fig: FDP_vs_p1_p_1000_Optimal_T_L_Gumbel_Dummies}
   }
	\hspace*{-1.4em}
  \subfloat[Gumbel dummies.]{
  		\scalebox{1}{
  			\includegraphics[width=0.245\linewidth]{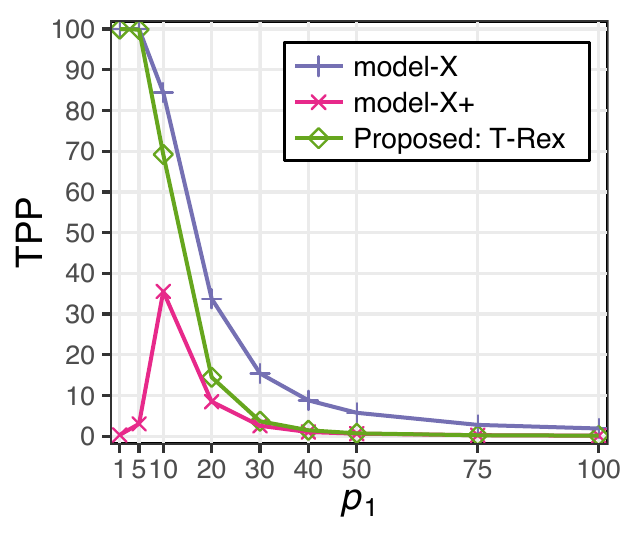}
  		}
   		\label{fig: TPP_vs_p1_p_1000_Optimal_T_L_Gumbel_Dummies}
   }
  \caption{\textbf{Illustration of Theorem~\ref{theorem: Dummy generation} (Dummy generation)}: The average FDP and TPP of the \textit{T-Rex} selector remain almost unchanged regardless of the distribution that the dummies are sampled from: (a) - (d) standard normal distribution, (e) - (h) uniform distribution with support between $0$ and $100$, (i) - (l) Student's $t$-distribution with $3$ degrees of freedom, (m) - (p) Gumbel distribution with its location and scale being $0$ and $1$, respectively. Setup: $n = 300$, $p = 1{,}000$, $p_{1} = 10$, $T_{\max} = \lceil n/2 \rceil$, $L_{\max} = 10p$, $K = 20$, $\text{SNR} = 1$, $MC = 955$.}
  \label{fig: sweep p1 and sweep snr plots p = 1000 _Optimal_T_L_Dummies}
\end{figure*}

Theorem~\ref{theorem: Dummy generation} is an asymptotic result that, loosely speaking, tells us that the FDR control property of the \textit{T-Rex} selector remains intact regardless of the distribution that the dummies are sampled from. In order to exemplify the somehow surprising results of Theorem~\ref{theorem: Dummy generation}, we have conducted simulations to show that the FDR control property of the \textit{T-Rex} selector remains intact for dummies sampled from the standard normal, uniform, $t\text{-}$, and Gumbel distribution, while the original predictors are sampled from the standard normal distribution. In Figure~\ref{fig: sweep p1 and sweep snr plots p = 1000 _Optimal_T_L_Dummies}, we see that the results remain almost unchanged regardless of the choice of the dummy distribution.

\begin{figure*}[!htbp]
  \centering
  \subfloat[$t(3)$ distributed $\X$.]{
		\scalebox{0.6}{
  			\includegraphics[width=0.49\linewidth]{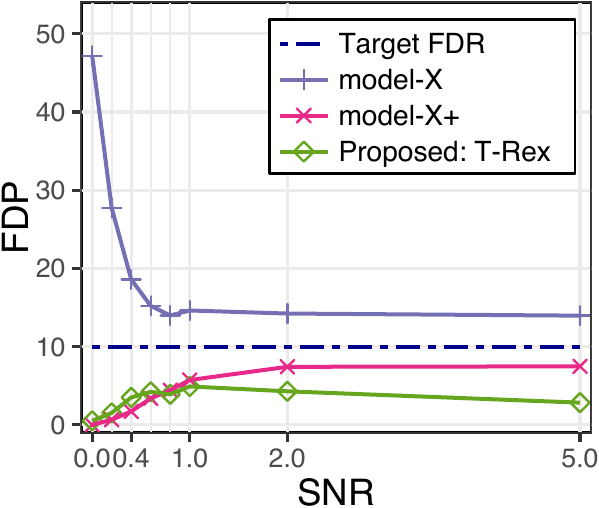}
  		}
   		\label{fig: FDP_vs_SNR_p_1000_Optimal_T_L_tDistr_df_3_X}
   }
  \subfloat[$t(3)$ distributed $\X$.]{
		\scalebox{0.6}{
  			\includegraphics[width=0.49\linewidth]{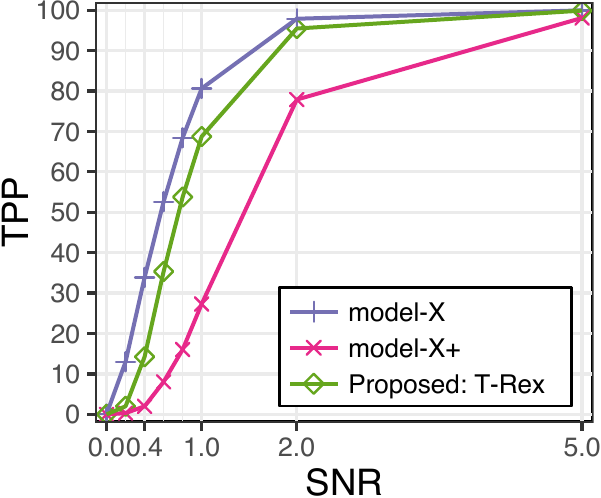}
  		}
   		\label{fig: TPP_vs_SNR_p_1000_Optimal_T_L_tDistr_df_3_X}
   }
   \\
    \subfloat[$t(2.1)$ distributed $\X$.]{
		\scalebox{0.6}{
  			\includegraphics[width=0.49\linewidth]{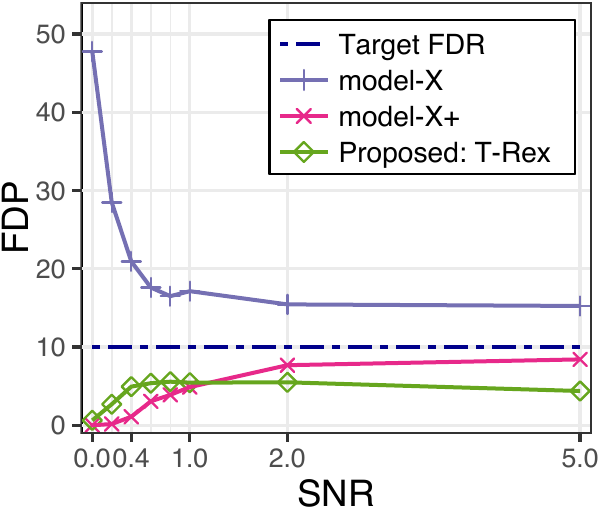}
  		}
   		\label{fig: FDP_vs_SNR_p_1000_Optimal_T_L_tDistr_df_2_1_X}
   }
  \subfloat[$t(2.1)$ distributed $\X$.]{
		\scalebox{0.6}{
  			\includegraphics[width=0.49\linewidth]{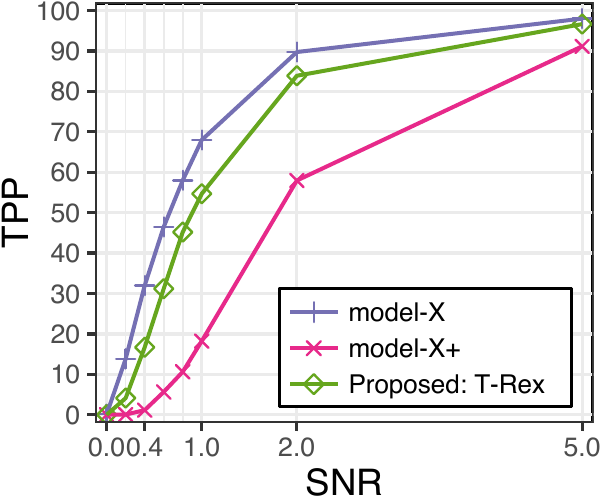}
  		}
   		\label{fig: TPP_vs_SNR_p_1000_Optimal_T_L_tDistr_df_2_1_X}
   }
   \\
    \subfloat[$t(2.01)$ distributed $\X$.]{
		\scalebox{0.6}{
  			\includegraphics[width=0.49\linewidth]{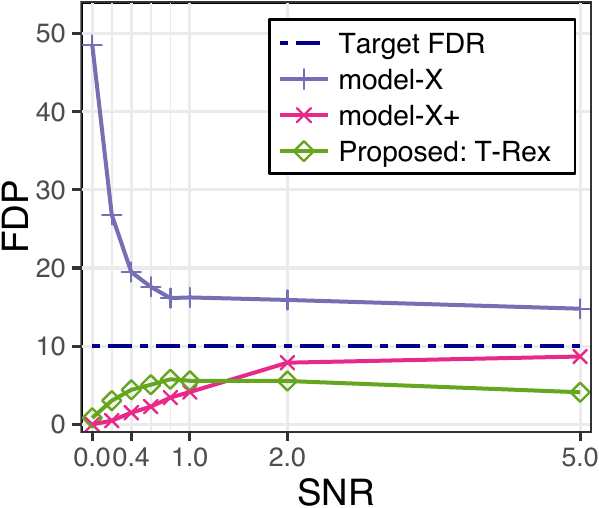}
  		}
   		\label{fig: FDP_vs_SNR_p_1000_Optimal_T_L_tDistr_df_2_01_X}
   }
  \subfloat[$t(2.01)$ distributed $\X$.]{
		\scalebox{0.6}{
  			\includegraphics[width=0.49\linewidth]{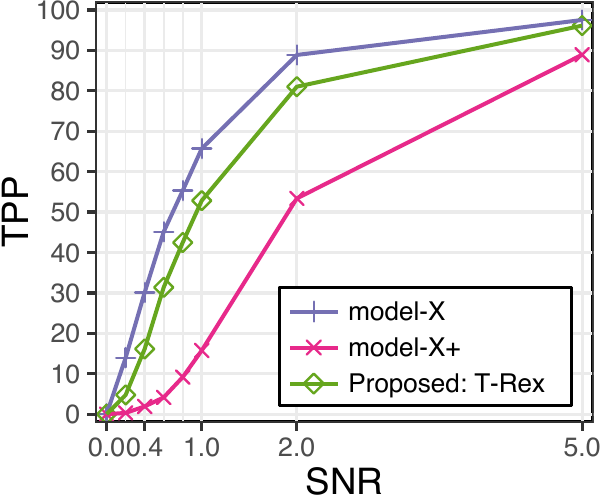}
  		}
   		\label{fig: TPP_vs_SNR_p_1000_Optimal_T_L_tDistr_df_2_01_X}
   }
   \\
    \subfloat[Gumbel$(0, 1)$ distributed $\X$.]{
		\scalebox{0.6}{
  			\includegraphics[width=0.49\linewidth]{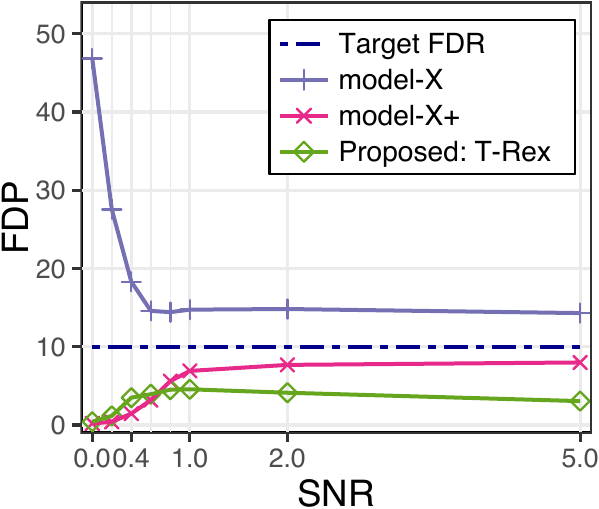}
  		}
   		\label{fig: FDP_vs_SNR_p_1000_Optimal_T_L_Gumbel_X}
   }
  \subfloat[Gumbel$(0, 1)$ distributed $\X$.]{
		\scalebox{0.6}{
  			\includegraphics[width=0.49\linewidth]{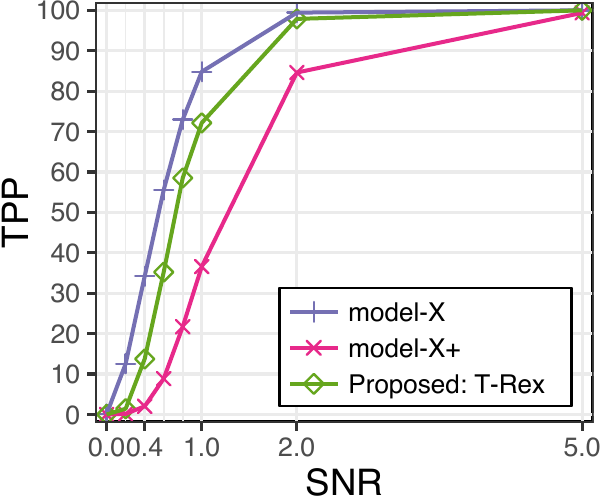}
  		}
   		\label{fig: TPP_vs_SNR_p_1000_Optimal_T_L_Gumbel_X}
   }
  \caption{\textbf{Average FDP and TPP in the case of non-Gaussian predictors in $\X$}: The FDR is controlled by the \textit{T-Rex} selector and the \textit{model-X} knockoff+ method while the \textit{model-X} knockoff method does not control the FDR. The predictors in $\X$ were sampled from (a) - (f) the Student's $t$ distribution with $3$, $2.1$, and $2.01$ degrees of freedom (i.e., $t(3)$, $t(2.1)$, and $t(2.01)$) and (g) - (h) the Gumbel distribution with location and scale being zero and one (i.e., Gumbel$(0, 1)$), respectively. The response was generated according to the linear model in~\eqref{eq: linear model}. Setup: $n = 300$, $p = 1{,}000$, $p_{1} = 10$, $T_{\max} = \lceil n/2 \rceil$, $L_{\max} = 10p$, $K = 20$, $MC = 955$.}
  \label{fig: sweep snr plots p = 1000 _Optimal_T_L_tDistr_df_3_df_2_01_Gumbel_X}
\end{figure*}

In order to also verify that the FDR control property holds for different distributions (with finite mean and variance) of the original predictors, we have conducted simulations in which the dummies are sampled from a standard normal distribution, while the original predictors are sampled from non-Gaussian heavy-tailed (i.e., Student's $t(3)$, $t(2.1)$, and $t(2.01)$) and skewed (i.e., Gumbel$(0, 1)$) distributions. Figure~\ref{fig: sweep snr plots p = 1000 _Optimal_T_L_tDistr_df_3_df_2_01_Gumbel_X} shows that, regardless of the mismatch between the distribution of the original variables and the dummies, the FDR control property holds for all these different distributions.
%
%
%

\section{Robustness of The T-Rex Selector}
\label{sec: Robustness of The T-Rex Selector}
\begin{figure*}[!htbp]
  \centering
	\hspace*{-0.8em}
  \subfloat[]{
  		\scalebox{1}{
  			\includegraphics[width=0.245\linewidth]{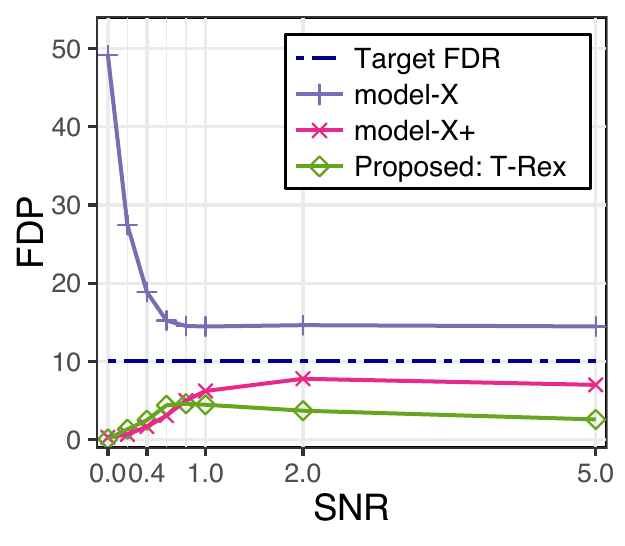}
  		}
   		\label{fig: FDP_vs_SNR_p_1000_Optimal_T_L_tDistr_noise}
   }
	\hspace*{-1.4em}
  \subfloat[]{
  		\scalebox{1}{
  			\includegraphics[width=0.245\linewidth]{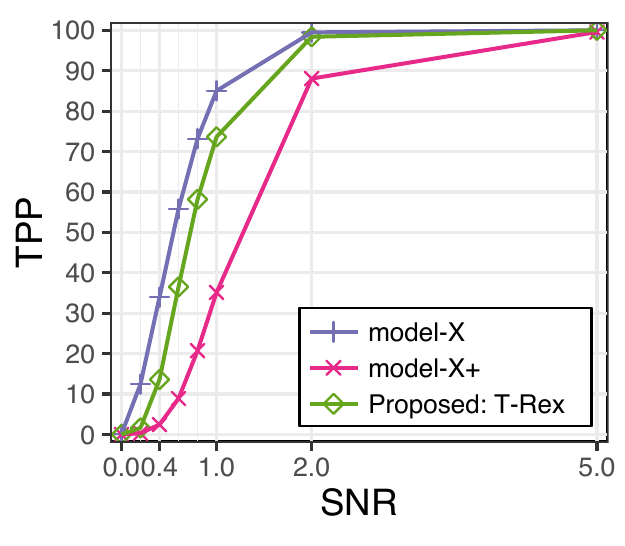}
  		}
   		\label{fig: TPP_vs_SNR_p_1000_Optimal_T_L_tDistr_noise}
   }
	\hspace*{-1.4em}
    \subfloat[]{
  		\scalebox{1}{
  			\includegraphics[width=0.245\linewidth]{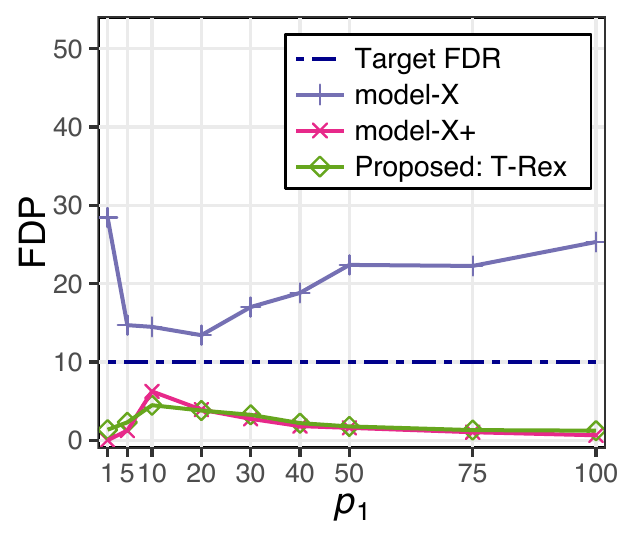}
  		}
   		\label{fig: FDP_vs_p1_p_1000_Optimal_T_L_tDistr_noise}
   }
	\hspace*{-1.4em}
  \subfloat[]{
  		\scalebox{1}{
  			\includegraphics[width=0.245\linewidth]{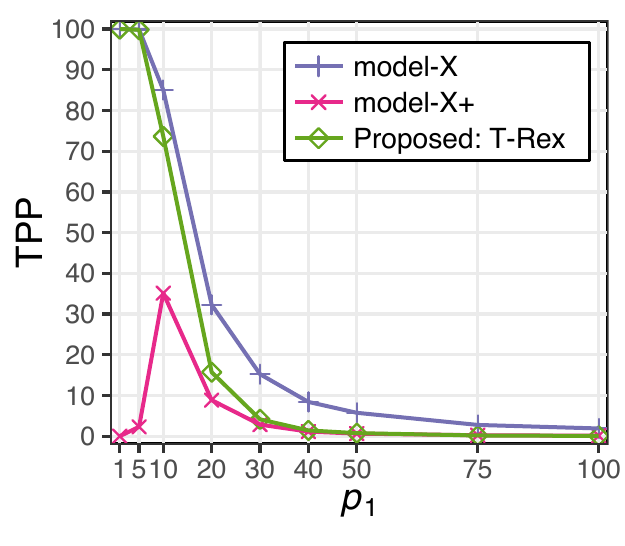}
  		}
   		\label{fig: TPP_vs_p1_p_1000_Optimal_T_L_tDistr_noise}
   }
  \caption{\textbf{Average FDP and TPP in the case of non-Gaussian noise}: The results are similar to the results of the Gaussian noise case in Figure~\ref{fig: sweep p1 and sweep snr plots p = 1000}. That is, all considered methods appear to be robust against deviations from the Gaussian noise assumption for the case of heavy-tailed ($t$-distributed) noise. The predictors in $\X$ were sampled from a univariate standard normal distribution and the response was generated according to the linear model in~\eqref{eq: linear model} with the noise vector $\epsilon$ being sampled from the $t$-distribution with $3$ degrees of freedom. Setup: $n = 300$, $p = 1{,}000$, $p_{1} = 10$, $T_{\max} = \lceil n/2 \rceil$, $L_{\max} = 10p$, $K = 20$, $\text{SNR} = 1$, $MC = 955$.}
  \label{fig: sweep p1 and sweep snr plots p = 1000 _Optimal_T_L_tDistr_noise}
\end{figure*}

In this appendix, we investigate the robustness of the proposed \textit{T-Rex} selector in the presence of non-Gaussian noise. We have conducted simulations with heavy-tailed noise following the t-distribution with three degrees of freedom. Figure~\ref{fig: sweep p1 and sweep snr plots p = 1000 _Optimal_T_L_tDistr_noise} shows that the proposed method performs well, even in the presence of heavy-tailed noise and, most importantly, maintains its FDR control property.
%
%

\bibliographystyle{IEEEtran_newLineInBibURL}
\IEEEtriggeratref{9}
\bibliography{bibliography}

\vfill

\typeout{get arXiv to do 4 passes: Label(s) may have changed. Rerun}